\documentclass{article}
\usepackage{ifthen}

\newboolean{isarxiv}
\setboolean{isarxiv}{true} 

\newboolean{isfinal}
\setboolean{isfinal}{true}

\ifthenelse{\boolean{isarxiv}}{
\PassOptionsToPackage{numbers,sort&compress}{natbib}  
\PassOptionsToPackage{pdfborder={0 0 0}}{hyperref}
\usepackage[preprint]{neurips_2025}
\setcitestyle{numbers,sort&compress}
}{
\ifthenelse{\boolean{isfinal}}{
\PassOptionsToPackage{numbers,sort&compress}{natbib}  
\PassOptionsToPackage{pdfborder={0 0 0}}{hyperref}
\usepackage[final]{neurips_2025}
\setcitestyle{numbers,sort&compress}
}{
}
}

\usepackage[hyphens]{url} 

\usepackage[utf8]{inputenc} 
\usepackage[T1]{fontenc}    
\usepackage{hyperref}       
\usepackage{url}            
\usepackage{booktabs}       
\usepackage{caption}
\usepackage{amsfonts}       
\usepackage{nicefrac}       
\usepackage{microtype}      
\usepackage{xcolor}         
\usepackage{cancel}
\usepackage{mathtools}
\usepackage{relsize} 
\usepackage[percent]{overpic} 

\usepackage{algorithm}
\usepackage{algorithmic}
\providecommand{\algorithmname}{Algorithm}
\floatname{algorithm}{\algorithmname}
\makeatletter
\newcommand{\RETURN}{\STATE \textbf{return} }
\makeatother

\newif\iflocal
\localfalse 

\IfFileExists{local}{
  \localtrue
}{}

\usepackage{microtype}
\usepackage{graphicx}
\usepackage{subcaption}

\usepackage{tabularx}              
\usepackage{ltablex}

\usepackage{amsmath,bm}
\usepackage{amssymb}
\usepackage{mathtools}
\usepackage{amsthm}

\usepackage[normalem]{ulem}
\usepackage{enumitem}

\usepackage[capitalize,noabbrev]{cleveref}

\usepackage{tikz}
\usetikzlibrary{arrows, fit, bending, positioning, calc, backgrounds,knots,decorations.pathreplacing}
\usepackage{adjustbox}
\usepackage{scalefnt}
\newcommand{\scalefactor}{0.6}



\DeclareMathSizes{4}{4}{3}{3}

\DeclareMathSizes{4}{4}{3}{3}

\ExplSyntaxOn
\cs_new:Npn \SF #1
  {
    \fp_to_decimal:n { #1 * \scalefactor } cm
  }
\cs_new:Npn \SFp #1
  {
    \fp_to_decimal:n { #1 * \scalefactor } pt
  }
\ExplSyntaxOff

\tikzset{mRectStyle/.style = {draw, rectangle, fill=green!20, rounded corners,minimum width=\SF{0.5}, minimum height=\SF{0.5}}}
\tikzset{mCircStyle/.style = {draw,circle,fill=white,inner sep=\SF{0.05},
      minimum width=\SF{0.85}, minimum height=\SF{0.85}}}      
\tikzset{mHardRectStyle/.style = {draw,rectangle,fill=white,inner sep=\SF{0.05},
      minimum width=\SF{0.85}, minimum height=\SF{0.85}}}      
\tikzset{mNoBorder/.style = {fill=white,inner sep=\SF{0.05},
      minimum width=\SF{0.4}, minimum height=\SF{0.4}}}           
\tikzset{mTensorStyle/.style = {inner sep=2pt}}           
\tikzset{
  mArrowStyle/.style={
     -{latex}, 
     shorten >=0.5mm
  }
}
\tikzset{mBraceTxtShift/.style = {below=4pt}}    

\tikzset{diagram brace/.style={decorate,
           decoration={brace,mirror,amplitude=6pt}}}

\theoremstyle{plain}
\newtheorem{theorem}{Theorem}[section]
\newtheorem{proposition}[theorem]{Proposition}
\newtheorem{lemma}[theorem]{Lemma}
\newtheorem{corollary}[theorem]{Corollary}
\theoremstyle{definition}
\newtheorem{definition}{Definition}[section]
\newtheorem{assumption}{Assumption}[section]
\theoremstyle{remark}
\newtheorem{remark}[theorem]{Remark}

\usepackage[textsize=tiny]{todonotes}

\usepackage{amsmath,amsfonts,bm,amsthm}
\usepackage{booktabs}

\def\eqref#1{equation~\ref{#1}}

\def\1{\bm{1}}

\DeclareMathAlphabet{\mathsfit}{\encodingdefault}{\sfdefault}{m}{sl}
\SetMathAlphabet{\mathsfit}{bold}{\encodingdefault}{\sfdefault}{bx}{n}

\def\sN{{\mathbb{N}}}

\def\sR{{\mathbb{R}}}

\newcommand{\E}{\mathbb{E}}

\newcommand{\R}{\mathbb{R}}

\newcommand{\KL}{D_{\mathrm{KL}}}
\newcommand{\Var}{\mathrm{Var}}

\DeclareMathOperator*{\argmax}{arg\,max}
\DeclareMathOperator*{\argmin}{arg\,min}

\newcommand{\appropto}{\mathrel{\vcenter{
  \offinterlineskip\halign{\hfil$##$\cr
    \propto\cr\noalign{\kern2pt}\sim\cr\noalign{\kern-2pt}}}}}

\newcommand{\madeplus}{$\text{MADE}^+$}
\newcommand{\target}{\pi}
\newcommand{\targetunnorm}{\eta}
\newcommand{\prior}{p}  
\newcommand{\likelihood}{p}  

\newcommand{\sattargetunnorm}{\tilde{\eta}}
\newcommand{\totalnormconst}{\mc{Z}}
\newcommand{\transdsupport}{\mc{X}}
\newcommand{\transform}{T}
\newcommand{\indivtransform}{\tau}
\newcommand{\catmdl}{\zeta}
\newcommand{\catmdltarget}{\target_\catmdl}
\newcommand{\catmdlsupport}{\mc{C}}
\newcommand{\catmdlparam}{\vimkp^\catmdl}
\newcommand{\neuralnet}{\operatorname{NN}}
\newcommand{\refdist}{\nu}
\newcommand{\data}{\mc{D}}
\newcommand{\mdl}{m}
\newcommand{\mdlspace}{\mc{M}}
\newcommand{\cvars}[1][\mdl]{\vct{\theta}_{#1}}

\newcommand{\refvarcoord}{z}
\newcommand{\refvars}{\vct{z}}

\newcommand{\auxvars}{\vct{u}}

\newcommand{\vimkp}{\psi} 
\newcommand{\vipp}{\phi} 
\newcommand{\vimkq}{q_\vimkp}
\newcommand{\satvipq}{\tilde{q}_\vipp}
\newcommand{\vipq}{q_\vipp}

\newcommand{\igthreshold}{\beta_{\mathrm{IG}(\vimkp)}}
\newcommand{\modeltomask}{A} 
\newcommand{\masktoapmask}{B} 
\newcommand{\ctm}{C} 
\newcommand{\ap}{\vct{\rho}} 
\newcommand{\staticap}{\ap^{\mathrm{Id}}} 
\newcommand{\vct}[1]{\boldsymbol{#1}} 
\newcommand{\mm}[1]{\mathbf{#1}} 
\newcommand{\Ex}{\E}

\newcommand{\dmax}{{d_{\mathrm{max}}}}
\newcommand{\dmdl}{{d_{\mdl}}}

\newcommand{\method}{CoSMIC}
\newcommand{\dkl}{D_{\mathrm{KL}}}
\newcommand{\normal}{\mc{N}}

\newcommand{\contextvars}{\vct{\xi}}

\newcommand{\lrperm}{\triangleleft}

\newcommand{\controlvariate}{\varsigma}

\newcommand*{\vim}{q}               
\newcommand*{\pspace}{\mc{P}}       
\newcommand*{\vpspace}{\Phi}        
\newcommand*{\vimkpspace}{\Psi}
\newcommand*{\objective}{f}         
\newcommand*{\observable}{\tilde{\objective}}
\newcommand*{\iterIdx}{t}           
\newcommand*{\observation}{y}       
\newcommand*{\obsnoise}{\epsilon}   
\newcommand*{\gp}{\mc{GP}}     
\newcommand*{\kernel}{\kappa}            
\newcommand*{\Kernel}{\mm{K}}            

\newcommand*{\mKernel}{\mm{K}}
\newcommand*{\nbatch}{B}            
\newcommand*{\norm}[1]{\lVert #1 \rVert}    
\newcommand*{\transpose}{\top}        
\newcommand*{\eye}{\mm{I}}                 
\newcommand*{\ucb}{u}
\newcommand*{\gpMean}{\mu}
\newcommand*{\entropy}{\mathrm{H}}
\newcommand*{\nIterations}{N}   

\newcommand*{\stepsize}{\lambda}
\newcommand*{\domain}{\mc{S}}       
\newcommand*{\location}{x}          
\newcommand*{\locations}{\mc{X}}
\newcommand*{\anyfunction}{g}       

\newcommand*{\anyconstant}{C}         
\newcommand*{\obscount}{N}
\newcommand*{\history}{\mathfrak{H}}
\newcommand*{\bound}{b}
\newcommand*{\batch}{\mc{B}}
\newcommand*{\obsIdx}{n}

\newcommand*{\bigo}{\mc{O}}
\newcommand*{\nmodels}{M}
\newcommand*{\indic}[1]{\mathbb{I}[#1]} 

\DeclareMathAlphabet{\pazocal}{OMS}{zplm}{m}{n}
\newcommand\mc{\pazocal}
\DeclareMathOperator*{\Prob}{\mathbb{P}} 

\renewcommand{\cal}[1]{\mathcal{#1}}

\newcommand{\bb}[1]{\mathbb{#1}}
\newcommand{\gvn}{\, | \,}

\definecolor{red}{rgb}{0.81, 0.09, 0.13}

\let\SQRT\sqrt
\renewcommand{\sqrt}[2][]{\ensuremath{\ifx\\#1\\ \SQRT{#2} \else \SQRT[#1]{#2} \fi \;}}

\DeclareMathAlphabet{\pazocal}{OMS}{zplm}{m}{n}

\newcommand{\stars}[1]{\ifcase#1 \or \(\star\) \or \(\star\star\) \or \(\star\star\star\) \or \(\star\star\star\star\) \or \(\star\star\star\star\star\) \fi}

\newcommand{\rebuttalred}[1]{\textcolor{red}{#1}}

\makeatletter
\newcommand{\MLPDbsize}{B}                 
\newcommand{\MLPDnodes}{N_{d}}                 
\newcommand{\MLPDsamples}{n}               
\newcommand{\MLPDhid}{H}                   

\newcommand{\MLPDData}{\vct{X}}    
   
\newcommand{\MLPDPerm}{\mm{P}}        
\newcommand{\MLPDEdges}{\mm{U}}       
\newcommand{\MLPDAdj}{\mm{A}}         
\newcommand{\MLPDParamVec}{\vct{\theta}}      
\newcommand{\MLPDThetaJ}[1]{\vct{\theta}^{(#1)}} 

\newcommand{\MLPDrelu}{\operatorname{ReLU}}           
\newcommand{\MLPDPermmap}{\varpi}                     
\newcommand{\MLPDparents}[1]{\operatorname{pa}_{\MLPDAdj}\!\left(#1\right)} 
\newcommand{\MLPDctxmask}{\ctm}                       
\newcommand{\MLPDModel}{\mdl}                         
\newcommand{\MLPfunc}[1]{f_{#1}}                      
\newcommand{\MLPDuvec}[1]{\vct u_{#1}}            

\newcommand{\MLPDNoiseStd}{\sigma}                    
\newcommand{\MLPDNoiseVar}{\MLPDNoiseStd^{2}}         
\newcommand{\MLPDEdgeProb}{\rho_{\operatorname{Edge}}}
\newcommand{\MLPDPriorScale}{\sigma_{0}}              
\newcommand{\MLPDSparsityPen}{\lambda}                
\newcommand{\MLPDBiasFlag}{\beta}                     

\makeatother

\title{Variational Transdimensional Inference}
\author{%
Laurence Davies$^{1*}$ \quad Dan Mackinlay$^{2\textasciicircum}$ \quad Rafael Oliveira$^{2\textasciicircum}$ \quad Scott A.~Sisson$^{1\$}$\\
$^1$University of New South Wales \quad $^2$CSIRO Data61\\
\texttt{$^{*}$laurence@latentlogic.com.au,  $^{\$}$scott.sisson@unsw.edu.au}\\
\texttt{$\textasciicircum$\{dan.mackinlay,rafael.dossantosdeoliveira\}@data61.csiro.au}
}

\begin{document}

\maketitle

\begin{abstract}
The expressiveness of flow-based models combined with stochastic variational inference (SVI) has expanded the application of optimization-based Bayesian inference to highly complex problems. However, despite the importance of multi-model Bayesian inference for problems defined on a transdimensional joint model and parameter space, such as Bayesian structure learning and model selection, flow-based SVI has been limited to problems defined on a fixed-dimensional parameter space. We introduce \method\, normalizing flows (COntextually-Specified Masking for Identity-mapped Components), an extension to neural autoregressive conditional normalizing flow architectures that enables use of a single flow-based variational density for inference over a transdimensional (multi-model) conditional target distribution. We propose a combined stochastic variational transdimensional inference (VTI) approach to training \method\, flows using ideas from Bayesian optimization and Monte Carlo gradient estimation. Numerical experiments show the performance of VTI on challenging problems that scale to high-cardinality model spaces.
\end{abstract}

\section{Introduction}\label{sec:introduction}
%
Variational inference via stochastic optimization~\citep{PaisleyVariational2012,HoffmanStochastic2013}
has surged in interest since the introduction of normalizing flows~\citep{RezendeVariational2015}. 
Flow-based densities 
can be used for a variety of downstream tasks, such as importance sampling~\citep{RezendeVariational2015}, simulation-based inference~\citep{Papamakarios2019Sequential, ZammitMangion2024Neural}, adaptive Markov chain Monte Carlo (MCMC) \citep{Gabrie2022Adaptive}, and generative modeling~\citep{Kingma2013AutoEncoding}.
While many 
existing approaches only consider continuous supports, there is a growing interest in applications where the support is either discrete or discretely indexed~\citep{Diluvi2024Mixed}.
One such application concerns a \textit{target} transdimensional probability distribution
$\target$
with support 
$\transdsupport=\smash{\mathop{\textstyle\bigcup}\nolimits_{\mdl\in\mdlspace}(\{\mdl\}\times\Theta_\mdl)}$,
where
$\mdlspace$ is a finite discrete {\em index} set, $\Theta_\mdl\subseteq\mathbb{R}^{d_\mdl}$, and 
the dimension $d_\mdl$ of
$\Theta_\mdl$ may vary with $\mdl$. Hence $\transdsupport$ is a {\em transdimensional} space~\citep{green1995reversible, SissonTransdimensional2005,fan2024reversible}.
Such spaces arise 
in Bayesian model
inference, where 
$\Theta_\mdl$ correspond to model {\em parameters}, and $\mdl\in\mdlspace$ is a {\em model index}.
Discrete indices parameterize many practical inference 
problems, including variable selection \citep{fan2024reversible}, mixtures-of-regressions,  learning directed acyclic graphs (DAGs) from data \citep{thompson2025prodag},  phylogenetic tree topology search \cite{everitt2020smctrans}, mixture-component inference \cite{das2014ttmcmc}, geoscientific inversion \cite{Sambridge2013}, and change-point models \cite{green1995reversible}. 
This article is concerned with estimating the target distribution $\pi$ with associated density function $\target(\mdl,\cvars)$,
$\cvars\in\Theta_\mdl$,
whose dimension depends on $\mdl$.
For simplicity we refer to $\target(\mdl,\cvars)$ and related functions as {\em density} functions, even though they are not 
continuous.
Typically, this density is only available in a conditional unnormalized form, $\targetunnorm(\cvars\gvn\mdl)=\totalnormconst_\mdl\target(\cvars\gvn\mdl)$, where $\totalnormconst_\mdl=\smash{\int_{\Theta_\mdl}\targetunnorm(\cvars\gvn\mdl)d\cvars}$.
The factorization $\targetunnorm(\mdl,\cvars)=\targetunnorm(\cvars\gvn \mdl)\target(\mdl)$ implies there is a discrete target probability mass function
over models, $\target(\mdl)=\totalnormconst_\mdl\totalnormconst^{-1}$, where $\totalnormconst=\smash{\sum_{\mdl\in\mdlspace}\totalnormconst_\mdl}$.
Estimation of $\targetunnorm(\mdl,\cvars)$ then becomes estimation of both $\targetunnorm(\cvars\mid\mdl)$ and $\target(\mdl)$.

In the presence of a likelihood function $\likelihood(\data\gvn\mdl,\cvars)$ for data $\data$, and priors $\prior(\cvars\mid\mdl)$ and $\prior(\mdl)$, the target distribution is defined by the
$\data$-conditional
transdimensional
posterior $\target(\mdl,\cvars\gvn \data)\propto\likelihood(\data\gvn\mdl,\cvars)\prior(\cvars|\mdl)\prior(\mdl)$.
In the context of \textit{variational Bayesian inference}~(VI; see \citep{Jordan1999Introduction,Blei2017Variational})
approximation of the
transdimensional
posterior $\target(\mdl,\cvars\gvn \data)$
has not been addressed in generality.
%
Such a scheme would approximate some
unnormalized
target density $\targetunnorm(\mdl,\cvars\gvn \data)=\totalnormconst\target(\mdl,\cvars\gvn \data)$ by 
choosing 
parameters  $\vipp\in\mathbb{R}^{n_\vipp},\vimkp\in\mathbb{R}^{n_\vimkp}$ of a tractable variational density family $q_{\vimkp,\vipp}(\mdl,\cvars)=\vipq(\cvars\mid\mdl)\vimkq(\mdl)$ to minimize
%
%
\begin{align}
     \vimkp^*, \vipp^* := \argmin_{\vimkp,\vipp} \mc{L}(\vimkp,\vipp), 
\qquad
    \mc{L}(\vimkp,\vipp) =  \dkl(q_{\vimkp,\vipp}||\targetunnorm),\label{eq:abstractloss}
\end{align}
where 
$\dkl$
is the Kullback-Leibler (KL) divergence.
There are two 
impediments to
constructing such a variational approximation:
(i) defining and optimizing $q_\vipp$ as $\cvars$ may vary in dimension conditional on $\mdl$, and (ii) the inference of $q_\vimkp$ for discrete latent variables $\mdl$ during the optimization of $q_\vipp$, a non-stationary objective as $\vipp\rightarrow\vipp^*$ and $\vimkp\rightarrow\vimkp^*$ are interdependent.



\noindent{\bf Background: Flow-based models for stochastic variational inference:}
\citet{RezendeVariational2015} showed that 
a \textit{normalizing flow} for $q_\vipp$ (with fixed $\mdl$)
is able to approximate many challenging fixed-dimensional distributions that are not well approximated by common parametric families.
A normalizing flow is defined by a diffeomorphism $\transform_\vipp:\mathbb{R}^d\rightarrow\mathbb{R}^d$ between two random vectors $\vct{\theta}\sim q$ and $\refvars\sim\refdist_d$, such that their distributions $q$ and $\refdist_d$ are absolutely continuous with respect to a $d$-dimensional Lebesgue measure, have well-defined densities $q(\vct{\theta})$ and $\refdist_d(\refvars)$ respectively, and can be related by 
$\refvars=\transform_\vipp(\vct{\theta})$ so that $q(\vct{\theta} ; \vipp)=\refdist_d(\transform_{\vipp}(\vct{\theta}))|\det\nabla \transform_{\vipp}(\vct{\theta})|$, $\vct{\theta}\in\bb R^d$.
As is
typical of normalizing flow-based models, we refer to  $\refdist_d$ as the \emph{reference} distribution and assume it factorizes into a product of $d$ identical marginal distributions $\refdist_d=\refdist\otimes\dots\otimes\refdist=\otimes_d\refdist$.
Construction of $\smash{\transform_\vipp}$ is typically achieved by defining $d$ 
bijective, univariate functions $\smash{\indivtransform_{\ap_i}:\mathbb{R}\mapsto\mathbb{R}}$, $\smash{z_i=\indivtransform_{\ap_i}(\theta_i)}$ for $\smash{i\in\{1,\dots,d\}}$.
The parameters $\ap_i=\neuralnet_\vipp(\vct{\theta}_{\backslash i})$ for the $i^{\mathrm{th}}$ transformation are determined by a neural network $\neuralnet_\vipp$ such that $\ap_i$ is not dependent on $\theta_i$, so that the inverse $\smash{\indivtransform_{\ap_i}^{-1}(\,\cdot\,)}$ can  be calculated without requiring 
inversion of $\neuralnet_\vipp$.
This independency remains if the neural network $\neuralnet_\vipp$ is  \textit{autoregressive} with respect to the inputs $\theta_1,\dots,\theta_d$ \citep{PapamakariosMasked2017}. Benefits of autoregressive flows are their higher-overall expressiveness and efficiency in the variational inference setting versus e.g.~coupling flows \citep{Coccaro2024Comparison}. For these reasons, 
this paper employs
autoregressive $\neuralnet_\vipp$. 
%
%
A \textit{conditional} normalizing flow extends
is a natural extension of a normalizing flow  
with a conditioning variate, $\contextvars$, passed as a contextual input to the  $\neuralnet_\vipp$, such that $\ap_i=\neuralnet_\vipp(\vct{\theta}_{\backslash i};\contextvars)$. 
Applications include classification, where $\contextvars$ is an index, or
likelihood estimation \citep{WinklerLearning2019} where $\contextvars$ 
encodes
the parameters of the likelihood function.

The MADE 
encoder \citep{GermainMADE2015} enables 
autoregressive neural flow architectures, which can be coupled with any $\indivtransform$ such as affine~\citep{PapamakariosMasked2017} and spline~\citep{Durkan2019Neural} 
transformations.
The cost of an autoregressive flow depends on the direction. In the forward (sampling) direction, it evaluates each dimension sequentially, for a time complexity of $\mathcal{O}(d)$. In the the \emph{inverse}  (likelihood) direction, computation can be parallel. The \textit{inverse autoregressive flow} (IAF) \citep{KingmaImproving2016} reverses this dependence,   setting $\vct{\theta}=\transform_\vipp(\refvars)$, yielding the variational density $\smash{q_\vipp(\cvars)=\refdist_d(\transform_\vipp^{-1}(\vct{\theta}))|\det\nabla \transform_\vipp^{-1}(\vct{\theta})|=\refdist_d(\refvars)|\det\nabla \transform_\vipp(\refvars)|^{-1}}$.


\noindent{\bf Contributions:}
We introduce \method\, ({\em COntextually-Specified Masking for Identity-mapped Components}) flows, a widely applicable and simple modification to conditional neural flow architectures (Section \ref{sec:method}). \method\, flows fundamentally expand the use cases for normalizing flows to encompass variational inference applications, so that a single flow-based variational density can be used for variational inference over a transdimensional (multi-model) target distribution. In effect, this extends the reparameterization trick exploited by IAF-based VI to the transdimensional setting. 
In Section \ref{sec:modelweights}, we demonstrate the efficacy of \method\, transformations within a novel {\em variational transdimensional inference} (VTI) framework with two implementations. The first builds upon principles of Bayesian optimization~\citep{SrinivasGaussian2012}, and the second uses Monte Carlo gradient estimation~\citep{MohamedMonte2020}.
We also provide a theoretical analysis of VTI approximation error bounds under a Gaussian process surrogate, and convergence guarantees for the marginal model distribution under convergent optimization steps.
%
%
%
Finally, we demonstrate the applicability of VTI to problems with model spaces that cannot be easily enumerated within the memory limitations of current computing architectures. In particular, 
Section \ref{sec:examples} explores problems in Bayesian robust variable selection~\citep{OHaraReview2009} and 
Bayesian causal discovery~\citep{Heckerman2006Bayesian}. \footnote{PyTorch CUDA code for all experiments is available at \href{https://github.com/daviesl/avti}{https://github.com/daviesl/avti}.}

\section{Formulating a transdimensional variational density}\label{sec:method}
\begin{figure}[t]
    \centering
\includegraphics[width=\textwidth]{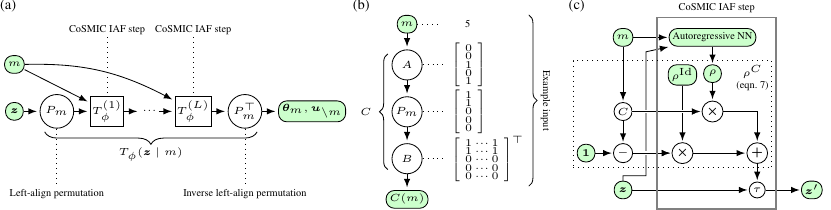}
\caption{(a) \method{} flow composition, (b) Context-to-mask map, (c) A single \method{} IAF step.}
    \label{fig:cosmic-iaf-step}
\end{figure}

Rather than constructing a variational density separately for each model $\mdl\in\mdlspace$, it is preferable to construct a single density on the transdimensional support $\transdsupport$.
To account for the 
varying dimension of $\cvars$,
we adopt the \textit{dimension saturation} approach of \citet{BrooksEfficient2003}, where the dimension of the parameter space conditional on each model is unified across all models. This is achieved by augmenting the space of model-conditional parameters with auxiliary variables $\auxvars\sim\refdist$,
as discussed below.
We use
$\backslash\mdl$ to identify auxiliary variables of dimension $\dmax-\dmdl$, where $\dmax:=\max_\mdl \{\dmdl\}$.
We define the saturated support 
$(\cvars,\auxvars_{\backslash\mdl})\in\Theta_\mdl\times\mc{U}_\mdl\subseteq\mathbb{R}^{\dmax}$,
with
unnormalized, dimension-saturated,
conditional target density
\begin{align}
    \sattargetunnorm(\cvars,\auxvars_{\backslash\mdl}\gvn\mdl) &= \targetunnorm(\cvars\gvn\mdl)
    \refdist_{\backslash\mdl}
    (\auxvars_{\backslash\mdl})\label{eq:saturatedtarget}.
\end{align}
Defined on the same augmented support is the family of saturated variational densities
\begin{align}
    \tilde{q}_{\vimkp,\vipp}(\mdl,\cvars,\auxvars_{\backslash\mdl})&=\satvipq(\cvars,\auxvars_{\backslash\mdl}\gvn \mdl)q_\vimkp(\mdl) ,\label{eq:cosmic-mixture}
\end{align}
where, noting the availability of a transport $(\cvars,\auxvars_{\backslash\mdl})=\transform_\vipp(\refvars\gvn \mdl)$, $\refvars\in \mc{U}^{d_{\mathrm{max}}}$, we define the IAF
\begin{align}
    \satvipq(\cvars,\auxvars_{\backslash\mdl}\gvn \mdl) &:=
    \refdist_{\dmax}(\transform^{-1}_\vipp(\cvars,\auxvars_{\backslash\mdl}\gvn \mdl))
    \left|\det\nabla \transform^{-1}_\vipp(\cvars,\auxvars_{\backslash\mdl}\mid\mdl)\right|, \nonumber \\
    &= \refdist_{\dmax}(\refvars)\left|\det\nabla \transform_\vipp(\refvars\mid \mdl)\right|^{-1}.
    \label{eq:cosmiciaf}
\end{align}
Our goal is to construct the IAF so that \eqref{eq:cosmiciaf} factorizes into active and \emph{i.i.d.}~auxiliary parts, i.e.
\begin{equation}
\satvipq\!\bigl(\cvars,\auxvars_{\setminus\mdl}\mid\mdl\bigr)
   \;=\;
   q_\vipp\!\bigl(\cvars\mid\mdl\bigr)\;
   \refdist_{d_{\setminus\mdl}}\!\bigl(\auxvars_{\setminus\mdl}\bigr),
\label{eq:cosmic-factorisation-target}
\end{equation}
and to exploit this factorization in the construction of a transdimensional loss function. To achieve this factorization, we define the following notation. Let 
$\modeltomask_i : \mdlspace \!\to\! \{0,1\}$ 
flag whether latent coordinate $i$ appears in model $\mdl$,  
and let $\masktoapmask_i : \{0,1\} \!\to\! \{0,1\}^{|\ap_i|},\;
\masktoapmask_i(b)=(b,\ldots,b)$,
broadcast this bit to the corresponding parameter block.
Their composition
$\ctm_i := \masktoapmask_i \circ \modeltomask_i : 
\mdlspace \!\to\! \{0,1\}^{|\ap_i|}$ 
therefore activates \emph{exactly} the autoregressive parameters $\ap_i$
needed by $\indivtransform_{\ap_i}(\refvars^{(i)})$ under model $\mdl$.
Concatenating the blocks gives the global context-to-mask map  
(see Figure \ref{fig:cosmic-iaf-step}(b) for a visualization):
\begin{align}
  \ctm(\mdl) := \bigl(\ctm_1(\mdl),\ldots,\ctm_{\dmax}(\mdl)\bigr)
  \in \{0,1\}^{|\ap|}, 
  \qquad |\ap| = \sum_{i=1}^{\dmax} |\ap_i|.\label{eq:ctmconcat}
\end{align}
Similarly, $\modeltomask$ and $\masktoapmask$ denote the respective coordinate-concatenated maps similar in form to \eqref{eq:ctmconcat}.
After a fixed left–align permutation aligning latents with $\cvars$,
Proposition~\ref{prop:cosmic-factorisation} proves this factorization is
\emph{exact} for any autoregressive network $\neuralnet_\vipp$ that
parametrizes the transport $\transform_\vipp$.

Recalling the univariate bijective maps of the inverse autoregressive flow 
as $\smash{\indivtransform_{\ap_i} : \mathbb{R} \mapsto \Theta_i}$ for $\smash{i=1,\dots,\dmax}$, we assume the existence of a \textit{static} point $\smash{\staticap}$ such that $\smash{\indivtransform_{\staticap}(z)=z}$ for all $z\in\mathbb{R}$, i.e., the transform becomes the identity map at $\staticap$. For example, a simple affine transformation (scale and location shift) is $\smash{\theta=\indivtransform_{\ap_i}(z)=\rho^{(0)} + \rho^{(1)}z}$, where $\smash{\ap_i=(\rho^{(0)},\rho^{(1)})}$. In this case, the static point is $\smash{\staticap=(0,1)}$ as then $\theta=z$. 
We can then construct
a simple mechanism for ``choosing'' between $\ap_i$ and $\staticap$ for each individual transform $\indivtransform$, $i=1,\dots,\dmax$, via the convex combination
\begin{align}
    \ap_i^\ctm = (\vct{1} - \ctm_i(\mdl)) \staticap + \ctm_i(\mdl) \ap_i,\,\,\,\,\mdl\in\mdlspace.\label{eqn:cosmic-rho-m}
\end{align}
Each coordinate-wise transform then becomes $\smash{\cvars^{(i)}=\indivtransform_{\ap_i^\ctm}(\refvars^{(i)})}$, $\smash{i\in\{1,\dots,\dmax\}}$.
That is, the transformation parameters become a context-dependent composition of the elements of $\ap_i$ and the static point $\staticap$ (Figure \ref{fig:cosmic-iaf-step}(c)). 
%
%
A composition of transforms parametrized according to \eqref{eqn:cosmic-rho-m} is a
\emph{Contextually-Specified Masking for Identity-mapped Components (\method{})}
normalizing flow.
\begin{lemma}\label{lemma:identity-map}
For a \method{} transform $(\cvars,\auxvars_{\setminus\mdl})=\transform_\vipp(\refvars_\mdl,\refvars_{\setminus\mdl})$, $\auxvars_{\setminus\mdl}=\refvars_{\setminus\mdl}$ $\forall\mdl\in\mdlspace$.
\end{lemma}
\begin{proposition}
\label{prop:cosmic-factorisation}
Fix $\mdl\in\mdlspace$.
Let $P_\mdl$ be the permutation matrix that places the coordinates
indexed by $I(\mdl)$ (from the proof of  \cref{lemma:identity-map}) before those in $I^{c}(\mdl)$ while preserving the
original order inside each group.  Define the 
left-align-permuted flow
$\transform_\vipp^{\lrperm}:=P_\mdl^{-1}\!\circ\transform_\vipp\circ P_\mdl$
and the corresponding density
$\smash{\satvipq^{\lrperm}(\cvars,\auxvars_{\setminus\mdl})
  =\refdist_{\dmax}(\refvars)\,
   |\det\nabla\transform_\vipp^{\lrperm}(\refvars\mid\mdl)|^{-1}}$, $\smash{\refvars=\transform_\vipp^{\lrperm,-1}(\cvars,\auxvars_{\setminus\mdl})}$. 
Redefine $\ctm:=\ctm^{\lrperm}=\masktoapmask\circ P_\mdl\circ\modeltomask$.
Then (a) $\satvipq^{\lrperm}(\cvars,\auxvars_{\setminus\mdl})$ factorizes as per \eqref{eq:cosmic-factorisation-target} with the substitution $\satvipq:=\satvipq^{\lrperm}$,
and (b) the marginal $\vipq(\cvars\mid\mdl)$ is consistent.
\end{proposition}
From here on, we use the notational convenience $\transform_\vipp:=\transform_\vipp^{\lrperm}$ and $q_\vipp := q_\vipp^{\lrperm}$ to denote the composition of transforms and associated variational density that include the left-align permutation $P_\mdl$ required by Proposition \ref{prop:cosmic-factorisation}. We also write the partitioning $\refvars=(\refvars_\mdl,\refvars_{\backslash\mdl})$ as explicitly obtained by $[\refvars_\mdl\,\refvars_{\setminus\mdl}]^\top=P_\mdl \refvars$. By construction, $\refdist_\dmax=\refdist_{d_\mdl}\otimes\refdist_{d_{\setminus\mdl}}$, i.e. $\refdist_\dmax(\refvars)=\refdist_{d_\mdl}(\refvars_\mdl)\refdist_{d_{\setminus\mdl}}(\refvars_{\setminus\mdl})$.

\begin{corollary}
\label{prop:cancellation}
Given Lemma \ref{lemma:identity-map} and Proposition \ref{prop:cosmic-factorisation}, then
\begin{align}
    \frac{\refdist_{\dmax}(\refvars)\left|\det\nabla \transform_\vipp(\refvars\gvn\mdl)\right|^{-1}}{\sattargetunnorm(\transform_{\vipp}(\refvars\gvn\mdl)\gvn\mdl)}
    &=
    \frac{\refdist_{d_\mdl}(\refvars_\mdl)\left|\det\nabla \transform_\vipp(\refvars\gvn\mdl)\right|^{-1}}{\targetunnorm(\cvars\gvn\mdl)}
    :=h_\vipp(\refvars\mid\mdl),\label{eq:desiredcancellation}
\end{align}
and, substituting $\ell(\mdl;\vipp):=\Ex_{\refvars\sim \refdist_\dmax}\left[\log h_\vipp(\refvars\mid\mdl)\right]$, the loss 
in \eqref{eq:abstractloss} becomes
\begin{align}
    \mc{L}(\vimkp,\vipp)
    &= \Ex_{\mdl\sim q_{\vimkp}}\left[
    \ell(\mdl;\vipp) - \log  \prior(\mdl) + \log q_\vimkp(\mdl)
    \right].\label{eq:varloss_by_mixture}
\end{align}
\end{corollary}
%
%

Proposition \ref{prop:cosmic-factorisation} states that, conditional on model $\mdl$, the \method{} IAF (Figure \ref{fig:cosmic-iaf-step}) achieves the factorization of the saturated-space variational approximation in \eqref{eq:cosmic-factorisation-target}. From Corollary \ref{prop:cancellation}, this means that when computing the loss function in \eqref{eq:abstractloss}, the ratio of the dimension-saturated variational density $\tilde{q}$ and conditional target $\tilde{\eta}$ (\eqref{eq:desiredcancellation}, LHS) which are both $\dmax$-dimensional and which involve the auxiliary variables, collapses down to a direct comparison only on the $d_\mdl$-dimensional model specific densities $q_\phi(\boldsymbol{\theta}_\mdl\mid m)$ and $\eta(\boldsymbol{\theta}_\mdl\mid m)$ (\eqref{eq:desiredcancellation}, RHS), and without the involvement of any auxiliary variables. 
That is, the \method{} flow enables the IAF to calculate on a fixed-dimensional space, while permitting the model-specific comparison within the loss function to operate on the natural $d_\mdl$-dimensional space.

The implementation of a \method{} inverse autoregressive flow step $\transform_i$ as part of a composition of transforms $\transform_L\circ\dots\circ\transform_1$ is visualized in Figure \ref{fig:cosmic-iaf-step}(a). Individual architectures for affine and rational quadratic spline transforms~\citep{Durkan2019Neural} and compositions are described in Appendix~\ref{apdx:flowarchitecture}.
\section{Formulating a model weights distribution}\label{sec:modelweights}

Formulating and estimating $\vimkq$ is not as straightforward as that of $\vipq$ because the discrete random variables $\mdl\sim\vimkq$ are not automatically linked to the density parameters $\vimkp$ by automatic differentiation. This problem naturally lends itself to methods developed in black-box variational inference \citep{Ranganath2014Black, Titsias2014Doubly, Wingate2013Automated} and multi-armed bandits \citep{BubeckRegret2012}, as described below. 
%
%
%
The representation of $\mdl$ is any discrete random variable on a finite space $\mdlspace$. Writing the true distribution of $\mdl$ as $\target_\mdl$, a finite $\mdlspace$ implies the existence of a categorical distribution $\catmdltarget$ which is bijectively equivalent to $\target_\mdl$. The random variables $\catmdl\sim\catmdltarget$ exist on the finite support $\catmdl\in\catmdlsupport\subset\mathbb{N}$, thus $|\catmdlsupport|=|\mdlspace|$. This property is used by the surrogate-based approach described in Section \ref{sec:surrogatemodelweightdensity}.We formalize this concept via \autoref{prop:unique_categorical_representation}. 
We consider two approaches to model $\vim_\vimkp$. Firstly, we derive a non-parametric surrogate-based approach which comes equipped with theoretical convergence guarantees and is applicable to model spaces $\mdlspace$ of low cardinality. We then present an approach based on parametric models that can scale to arbitrarily large spaces $\mdlspace$ that are trained using doubly stochastic gradient estimators. 

\subsection{Estimation via surrogate}\label{sec:surrogatemodelweightdensity}
The objective in \autoref{eq:varloss_by_mixture} can be rewritten as a single-variable objective with respect to $\vipp$:
\begin{equation}
    \begin{split}
        \vipp^* &\in \argmin_{\vipp}\min_{\vimkp} \mc{L}(\vimkp, \vipp)=\argmax_{\vipp}\max_{\vim_\vimkp \in \pspace_\vimkpspace} \Ex_{\mdl\sim\vim_\vimkp}[-\ell(\mdl; \vipp) + \log \prior(\mdl)] + \entropy[\vim_\vimkp],\\
    \end{split}
\end{equation}
where $\pspace_\vimkpspace$ denotes the space of probability measures over $\mdlspace$ parameterized by $\vimkp\in\vimkpspace\subseteq\sR^{n_\vimkp}$, and $\entropy$ denotes entropy. If we replace $\pspace_\vimkpspace$ by $\pspace(\mdlspace)$, i.e., the whole space of probability measures over $\mdlspace$, the solution to the inner optimization problem admits a closed-form expression:
\begin{equation}
    \vim_{\ell, \vipp}^*(\mdl) := \frac{\prior(\mdl)\exp(-\ell(\mdl;\vipp))}{\sum_{\mdl'\in\mdlspace}\prior(\mdl')\exp(-\ell(\mdl';\vipp))}\,.
\end{equation}
Computing the expression above within an optimization loop over $\vipp$ in practice would, however, require the evaluation of flow-based densities over the entire model space. We may, instead, follow a cheaper-to-evaluate density $\vim_{\ucb, \vipp}$ which approximates $\vim_{\ell, \vipp}^*$ for a given $\vipp$, by means of learning a surrogate model over $\ell$ within the \emph{same} optimization loop\footnote{We are here assuming that the prior $\prior(\mdl)$ is cheap to evaluate. If not, we can model $-\ell(\mdl;\vipp) + \log \prior(\mdl)$, instead, with a surrogate, which leads to similar theoretical guarantees after minimal adjustments.}. In particular, we derive a Gaussian process (GP) upper confidence bound \citep{SrinivasGaussian2010}, which provides the following approximation to the optimal model probabilities:
\begin{equation}
    \vim_{\ucb, \iterIdx}(\mdl) := \frac{\prior(\mdl) \exp \ucb_\iterIdx(\mdl)}{\sum_{\mdl'\in\mdlspace} \prior(\mdl') \exp \ucb_\iterIdx(\mdl')},
\end{equation}
where $\ucb_\iterIdx(\mdl) := \gpMean_\iterIdx(\mdl, \vipp_\iterIdx) + \beta\sigma_\iterIdx(\mdl, \vipp_\iterIdx)$, with $\gpMean_\iterIdx$ and $\sigma_\iterIdx^2$ representing the posterior mean and variance of a GP model conditioned on all mini-batches of data $\batch_{\iterIdx} := \{\vipp_{\iterIdx-1}, \mdl_{\iterIdx,i}, \log h_{\vipp_{\iterIdx-1}}(\refvars_{\iterIdx,i}|\mdl_{\iterIdx,i})\}_{i=1}^\nbatch$ available at iteration $\iterIdx$ of stochastic gradient descent, and $\vipp_\iterIdx$ denotes the current flow parameters. In this form, $\ucb_\iterIdx$ provides an upper confidence bound (UCB) over $-\ell(\mdl;\vipp_\iterIdx)$ determined by the choice of confidence parameter $\beta \geq 0$. The GP posterior mean and variance can be derived in closed form if the observation noise is Gaussian with, e.g., variance $\sigma_\obsnoise^2$. We, however, show that a sub-Gaussian noise assumption is sufficient to use a conventional GP model. In addition, if $\vipp_\iterIdx$ follows a convergent sequence (e.g., by ensuring diminishing step sizes during gradient-based optimization), we have the following guarantee.
\begin{corollary}
    \label{thr:main-ucb-convergence}
    Let $\ell \sim \gp(0, \kernel)$, where $\kernel: (\mdlspace\times\vpspace)^2\to\sR$ is a bounded, continuous positive-semidefinite kernel over $\mdlspace\times\vpspace$. Assume $\log h_\vipp(\refvars|\mdl) - \ell(\mdl;\vipp)$ is $\sigma_\obsnoise^2$-sub-Gaussian with respect to $\refvars\sim\refdist$. 
    Then, if $\vipp_\iterIdx$ follows a convergent sequence, the following also holds:
    \begin{equation}
        \KL(\vim_{\ucb,\iterIdx} || \vim_{\ell,\vipp_\iterIdx}^*) \in \bigo_{\Prob}(\iterIdx^{-1/2}), 
    \end{equation}
    where $\bigo_{\Prob}$ characterizes convergence in probability.\footnote{$\xi_\iterIdx \in \bigo_{\Prob}(\anyfunction_\iterIdx)$ if $\lim_{\anyconstant\to\infty}\limsup_{\iterIdx\to\infty}\Prob[\xi_\iterIdx\anyfunction_\iterIdx^{-1} > \anyconstant] = 0$.}
\end{corollary}
%
The result above tells us that the UCB-based models distribution approaches the optimal distribution at a rate of $\bigo_{\Prob}(\iterIdx^{-1/2})$ and ultimately converges to it as $\iterIdx\to\infty$. Therefore, a stochastic gradient optimizer using samples from the surrogate density $\vim_{\ucb,\iterIdx}$ should asymptotically converge to the optimization path determined by the optimal $\vim_{\ell, \vipp_\iterIdx}^*$. That is, under appropriate settings for, e.g., its learning rate schedule, the optimization will converge to $\vipp^*$. Lastly, note that the result in \autoref{thr:main-ucb-convergence} is independent of the choice of $\beta$, which can be set to $\beta = 0$. Our analysis is mainly based on obtaining enough samples almost everywhere across the model space, which can be ensured by sampling according to the predictive mean $\gpMean_\iterIdx$ of the surrogate, as $\exp \gpMean_\iterIdx > 0$ under mild assumptions. However, in practice, a non-zero value of $\beta$ helps to accelerate convergence in finite time by encouraging exploration. Corollary \ref{thr:main-ucb-convergence} is a direct application of \autoref{thr:kl-bound}, proved in Appendix \ref{sec:convergence}, where we also contrast it with existing results \citep{oliveiranoregret2021}.

Due to the reliance on GP-based approximations, a naive implementation of this approach would incur a cost of $\bigo(\nbatch^3\iterIdx^3)$ per stochastic gradient step, where $\nbatch$ is the mini-batch size, due to matrix inversions \citep{Rasmussen2006}. However, for model spaces of moderate cardinality $|\mdlspace| = M$, we can keep compute costs linear with the number of optimization steps by applying recursive equations to evaluate the GP posterior mean and covariance over the model space (see Eq. \ref{eq:gp-mean-update} and \ref{eq:gp-cov-update}), 
leading to a cost of $\mc{O}(\nbatch^3+M\nbatch^2+M^2\nbatch)=\mc{O}(M^2 \nbatch)$ per step, as $\nbatch\ll M$, totaling $\mc{O}(TM^2\nbatch)$ over $T$ steps. 
Sparse approximations to GPs can further reduce this cost \citep{Rasmussen2006, Gijsberts2013Realtime}  to make it practical for larger model spaces. For our purposes, we implemented a diagonal Gaussian approximation, which makes the cost linear in the batch size and constant in $\iterIdx$ via a mean-field approximation.

\subsection{Categorical and neural probability mass functions}\label{sec:isomorphismcategoricalmade}

By \cref{prop:unique_categorical_representation}, we may represent probability distributions over the model space $\mdlspace$ 
by arbitrarily parametrized categorical distributions.
A drawback of the surrogate
is the need to maintain and update estimates over the entire model space, which can be impractical for spaces of very large cardinality, 
such as DAG discovery.
Hence, we introduce two parametric alternatives.

\paragraph{Categorical:}
Assume $|\mdlspace| = \nmodels\in\sN$. Then, for $\vimkp\in\sR^\nmodels$, the distribution over $\mdlspace$ is defined by
$\smash{\vim_{\vimkp}(\mdl) := (\sum_{j=1}^\nmodels \exp\vimkp_j)^{-1}\sum_{i=1}^\nmodels\indic{\mdl_i = \mdl}\exp\vimkp_i\,}$.
The logit weights vector $\vimkp$ is unconstrained in $\sR^\nmodels$ and can be jointly optimized with $\vipp$ by gradient methods. 
Density evaluations and the entropy can be 
computed
with memory cost $\mc{O}(|\mdlspace|)$. 

\paragraph{Autoregressive:} 
If the model space is too large
we may use a 
structured sample generation process which allows for the number of parameters to be smaller than
cardinality of the model space
i.e., $\dim(\vimkp) < \nmodels$. For instance, \citet{GermainMADE2015} proposed an autoregressive parametrization for distributions over binary strings $\vct{s} \in \{0,1\}^{d_s}$ via the decomposition $\smash{p_\vimkp(\vct{s}) = \prod_{i=1}^{d_s} p_\vimkp(s_i | s_1, \dots, s_{i-1})}$.
For each $\vct{s}$, we
assign a unique $\mdl\in\mdlspace$ and 
define $\vim_{\vimkp}(\mdl) := p_\vimkp(\vct{s}(\mdl))$. The conditional densities and sampling can be implemented via MADE, allowing us to map the entire model space with fewer parameters
when
$2^{d_s} \geq |\mdlspace|$. The same reasoning can be applied to a DAG via decomposition of its adjacency matrix.
Details of MADE
are 
in Appendix~\ref{apdx:madeplus}, 
and for
DAGs in Appendix~\ref{apdx:dagmdl}.

\subsection{Estimation via Monte Carlo gradients}\label{sec:mcgmodelweightdensity}
When $|\mdlspace|$ is too large to use a surrogate-based approach, 
or to even parameterize an entire vector of categorical weights
in physical memory, we can employ neural-based 
methods that use gradient descent and estimation of the gradients of $\vimkp$ via Monte Carlo estimation of gradients (MCG) \citep{MohamedMonte2020}. 
Using $\nabla_{\vimkp}q_{\vimkp}(\mdl) = q_{\vimkp}(\mdl) \nabla_{\vimkp}\log q_{\vimkp}(\mdl)$, the gradient of the expectation in \eqref{eq:varloss_by_mixture} with respect to $\vimkp$ is
\begin{align}
    \nabla_\vimkp \mc{L}(\vimkp,\vipp) 
     =&\,\Ex_{\mdl\sim\vimkq}
            \left[ \ell(\mdl;\vipp) \nabla_\vimkp \log\vimkq(\mdl) \right] 
    +\Ex_{\mdl\sim\vimkq}\left[\log\dfrac{q_\vimkp(\mdl)}{\prior(\mdl)}\nabla_\vimkp \log\vimkq(\mdl)\right].
\end{align}
In practice, the variance of this estimator can be very high.
However, techniques exist to reduce this variance~\citep{PaisleyVariational2012, Ranganath2014Black, MohamedMonte2020} for general applications. 
We
use a control variate $\controlvariate$ in the form
\begin{align}
    \nabla_\vimkp \mc{L}(\vimkp,\vipp)=
    \Ex_{\mdl\sim\vimkq}
        \left[
            g(\vipp,\vimkp,\controlvariate)
                \nabla_\vimkp\log\vimkq(\mdl)
            \right],
\end{align}
where $g(\vipp,\vimkp,\controlvariate)=\Ex_{\refvars\sim \nu_\dmax}\left[\log h_{\vipp}(\refvars|\mdl)+\log\vimkq(\mdl)-\log\prior(\mdl)-\controlvariate\right]$.
We compute $\controlvariate$ using the method described in Appendix~\ref{apdx:sfecontrolvariate} (full description in Appendix~\ref{apdx:mcgsfe}).

The benefit of using MCG for variational parameter estimation is the flexibility of choice for $\vimkq$. 
We compare two:
(1) MCG of the logits of a standard categorical distribution, and (2) MCG of multi-layer perceptron weights that parameterize a configuration of the MADE neural autoregressive density estimator of \citet{GermainMADE2015} (see Appendix~\ref{apdx:dagmdl}).
When $|\mdlspace|$ is large, such implementations of $\vimkq$ permit an efficient approximate representation of the true model distribution.
\subsection{Information-Limiting the optimization}
The convergence of $\vimkp\rightarrow\vimkp^*$ is dependent on the convergence of $\vipp\rightarrow\vipp^*$, and optimal sample efficiency for the inference of $\vipp$ is achieved when $\vimkp\approx\vimkp^*$. 
Intuitively, $\vipq$ should focus primarily on the higher-probability models that contribute most to estimator variance, but discovering these models requires stable approximation of each $\vipq(\cvars\mid\mdl)$ to inform $\nabla_\vimkp$. This circular dependence motivates practical regulation of the optimization of $\vimkp$ when estimating $\nabla_\vimkp$ via Monte Carlo gradient estimates, addressing an instability similar to that discussed in the reinforcement learning literature \citep{SchulmanProximal2017}, but without modifying the objective. 
Our approach is to reduce the variance of the estimates of $\vim_{\vimkp}$ by bounding the information gain in the transition $\vim_{\vimkp_\iterIdx} \to \vim_{\vimkp_{\iterIdx+1}}$, which determines the step size, thereby stabilizing the optimization (detailed in Appendix~\ref{apdx:ig_threshold}).

\section{Related work}\label{sec:relatedwork}


Conditional normalizing flows \citep{WinklerLearning2019,Durkan2019Neural} have emerged as powerful tools for incorporating conditioning information. Existing methods  use the context variable as a conditioning input, but fewer adapt the flow architecture itself. An
exception is the transport‐based reversible jump MCMC method~\citep{DaviesTransport2023}, which learns proposals for transdimensional moves, but 
does not readily allow its use as an inverse autoregressive flow \citep{KingmaImproving2016}. In contrast, we introduce an identity‐parameterized \method{} transformation without identity-map training. We bypass path-wise approximations to discrete distributions~\citep{JangCategorical2022,MaddisonConcrete2022}, instead comparing Monte Carlo gradient estimation \citep{MohamedMonte2020} with Bayesian optimization \citep{ShahriariTaking2016}. We adopt an  information-based approach to scale gradient steps using ``small steps,'' inspired by reinforcement learning \citep{SchulmanProximal2017}.
Bayesian methods for model selection and optimization have advanced with black‐box variational inference \citep{Ranganath2014Black, Titsias2014Doubly, Wingate2013Automated} and flexible flows \citep{RezendeVariational2015, PapamakariosMasked2017, Durkan2019Neural}. Recent work in amortized Bayesian mixture models \citep{Kucharsky2025Amortized} shows amortization over multiple mixture components using conditional normalizing flows, but not for variable dimensions. Conversely, \citet{Li2020ACFlow} introduces an architecture for learning imputation over transdimensional inputs, but lacks immediate application as a variational density. Our approach  unifies transdimensional inference with flow‐based variational methods, bypassing the need for tailored dimension jumps and broadening applications.
\section{Experiments}\label{sec:examples}


We present experiments involving synthetic and real data on two 
representative applications: robust variable selection and directed acyclic graphs.
To evaluate the quality of the approximation $q_{\vimkp,\vipp}(\mdl,\cvars)$ to the target distribution $\target(\mdl, \cvars)$ for a relatively small $|\mdlspace|<2^{19}$ model space, we use the \textit{average negative log-likelihood} (NLL) computed over a set of samples drawn from \( \target \) via a baseline sampling method, in this case reversible jump 
MCMC~\citep{SissonTransdimensional2005}.
Let \( \{(\mdl^i, \cvars^i)\}_{i=1}^N \) denote \( N \) independent samples from \( \target(\mdl, \cvars) \).
The average NLL 
corresponds to the \textit{cross-entropy} \( H(\target, q_{\vimkp,\vipp}) \) between 
\( \target \) and \( q_{\vimkp,\vipp} \), which quantifies the expected number of bits needed to encode samples from \( \target \) using \( q_{\vimkp,\vipp} \),
and is defined as $\smash{\text{NLL} = \frac{1}{N} \sum_{i=1}^{N} -\log q_{\vimkp,\vipp}(\mdl^i,\cvars^i)}$.
Comparison of VTI DAG inference quality with baseline frequentist and Bayesian approaches use standard metrics \citep{Kummerfeld2019Simulations}.

\subsection{Bayesian misspecified robust variable selection}\label{sec:robustvs}

We study a robust Bayesian variable selection problem where the response $y \in \mathbb{R}$ is related to predictors $\vct{x} \in \mathbb{R}^p$ (including an intercept) through a linear model.
The innovation is a mixture-of-Gaussians noise specification, accommodating outliers via a heavy-tailed component.
A subset indicator $\vct{\gamma} \in \{0,1\}^p$ selects which predictors enter the model.
If $\vct{\beta} \in \mathbb{R}^p$ are the coefficients, only the components where $\gamma_j = 1$ contribute to the linear predictor.
In particular, for data $\{(\vct{x}_i,y_i)\}_{i=1}^n$ the prediction function  is $\vct{\mu}(\vct{x}) = \vct{x}^{\top}\!(\vct{\beta} \odot \vct{\gamma})$, the likelihood is 
\begin{align}
p\bigl(y_i \mid \vct{x}_i,\vct{\beta},\vct{\gamma}\bigr) 
&=
(1 - \alpha)\,\mathcal{N}\!\Bigl(y_i;\vct{\mu}(\vct{x}_i),\sigma_1^2\Bigr)
\;+\;
\alpha\,\mathcal{N}\!\Bigl(y_i;\vct{\mu}(\vct{x}_i),\sigma_2^2\Bigr),\label{eq:robustvs-likelihood}
\end{align}and priors
$\smash{p(\gamma) \;=\; 2^{-p}}$ and
$\smash{p(\vct{\beta}) \;=\;\mathcal{N}(0,\,\sigma_\beta^2\mm{I})}$.
Here, $\alpha$ controls the fraction of outliers, and $(\sigma_1^2,\sigma_2^2)$ encode the 
variances of in-distribution and outliers, respectively. 
To complicate the inference problem, two misspecified data-generating processes were used (medium- and high-misspecification) which
encourages multi-modality in the approximating posterior $\target(\cvars|\mdl)$. 

\begin{figure}[t]
\centering
\setlength{\tabcolsep}{1pt}
\begin{tabular}{@{}c !{\vrule width 0.5pt} c@{}}
\includegraphics[width=0.49\textwidth]{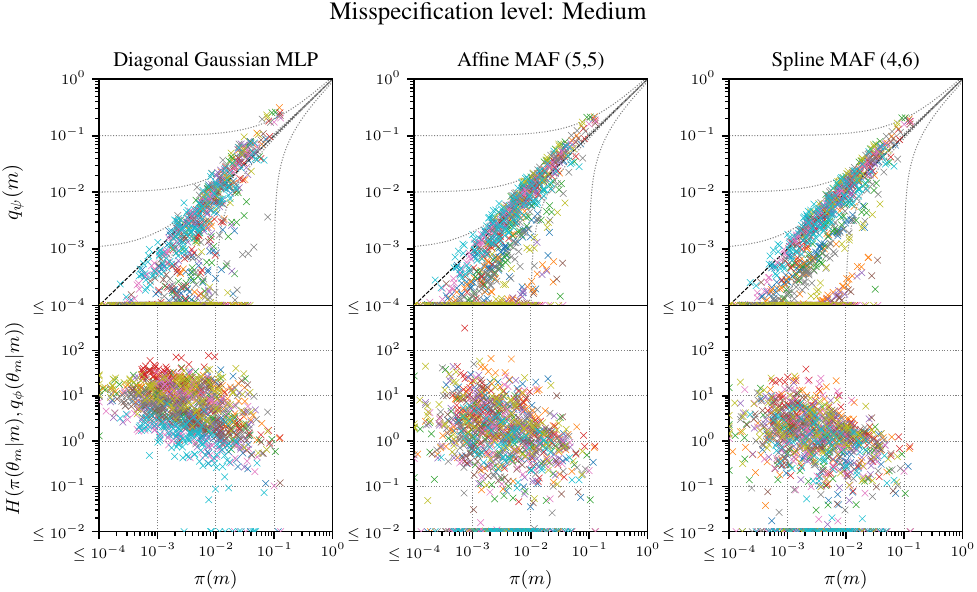} &
\includegraphics[width=0.49\textwidth]{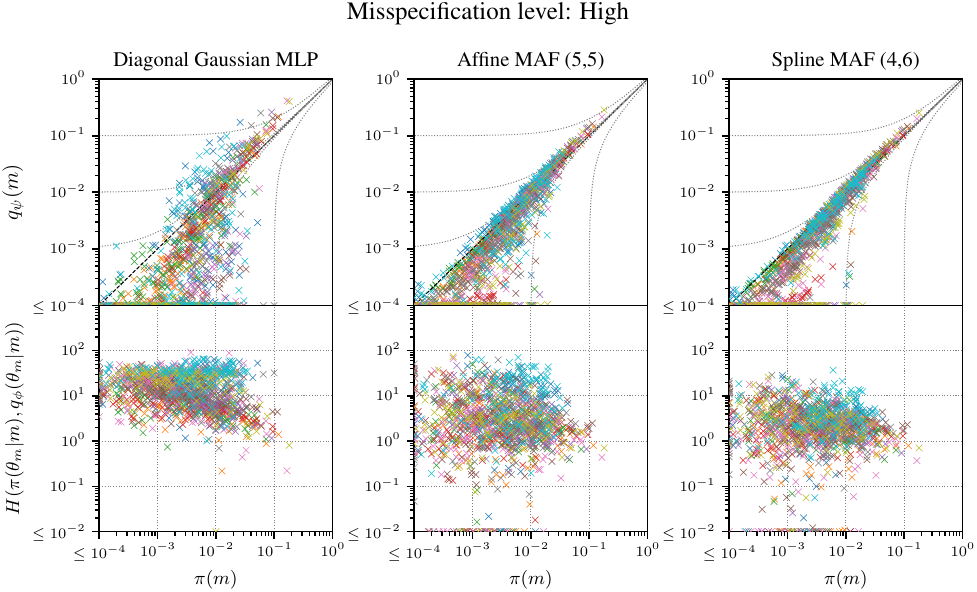}
\end{tabular}
\caption{Quality of VTI approximation for Bayesian misspecified robust variable selection. Outer columns denote medium (left) or high (right) likelihood misspecificaton, inner columns indicate different normalizing flow constructions,  increasing flow expressivity from left to right. 
Flow types are described in Appendix~\ref{apdx:flowarchitecture}.
\textit{Top row:} Estimated model probabilities $q_{\vimkp}(\mdl)$ vs true model probabilities $\target(\mdl)$ on the log scale.
\textit{Bottom row:} Cross entropy between individual model estimates $q_{\vipp}(\cvars|\mdl)$ and true density $\target(\cvars|\mdl)$ versus true model probability.
Colors indicate 10 replicated analyses, each with $|\mdlspace|=2^7$ models.}
    \label{fig:robustvsmodelprob}
\end{figure}

Table \ref{tab:robustvsdgpsetup} in Appendix \ref{apdx:rvsadditionalresults} summarizes the full experiment configuration. Figure \ref{fig:robustvsmodelprob} offers a holistic assessment of inference quality relative to a sampling baseline using RJMCMC, where cross-entropy reduces as flow expressivity increases. It shows two problem settings, mid and high misspecification, and for each setting shows how increasing complexity of the variational density (left-to-right panels) improves the quality of the approximations of both $\target(\cvars|\mdl)$ (bottom row) and estimated model probabilities (top row),
and that the approximation quality of $\target(\cvars|\mdl)$ is higher for higher probability models.

%
\paragraph{Cardinality sweep:} Using the focused prior setup on both the medium and high misspecification level targets, we sweep the cardinality of the model space  $|\mdlspace|$ from $2^9$ to $2^{24}$ and compute the cross entropy \( H(\target, q_{\vimkp,\vipp}) \), where samples $(\mdl,\cvars)\sim\target$ are obtained via RJMCMC (see Appendix \ref{apdx:robustvsrjmcmc}). Figure \ref{fig:apdxrobustvssweepcard} (left) compares the cross entropy between the three $\vimkq(\mdl)$ types discussed in Sections \ref{sec:surrogatemodelweightdensity} and \ref{sec:mcgmodelweightdensity} in simulated problems of increasing $|\mdlspace|$.
%
%
As expected, \( H(\target, q_{\vimkp,\vipp}) \) generally increases with $|\mdlspace|$ when the flow architecture is held fixed. The surrogate method (blue bars) performs comparably with the other methods for the smaller model spaces ($|\mdlspace|=2^9$), whereas the neural density (orange bars) performs consistently as $|\mdlspace|$ increases. Figure \ref{fig:apdxrobustvssweepcard} (right) shows two bivariate plots of selected variables ($\smash{\cvars^{(1)},\cvars^{(5)})}$ from the posterior inferred using RJMCMC and VTI. This qualitative visual comparison shows how well the CoSMIC flow is able to capture non-trivial model distributions versus the sampling approach (for the full multivariate comparison see Figure \ref{fig:robustvsmultivariate}). 
 Appendix \ref{apdx:rvsadditionalresults}
describes the experiments in further detail and demonstrates VTI robustness to diffuse priors.
\begin{figure}[t]
    \centering
  \setlength{\tabcolsep}{1pt}
  \begin{tabular}{@{}c c !{\vrule width 0.5pt} c c@{}}
  \includegraphics[width=0.245\linewidth]{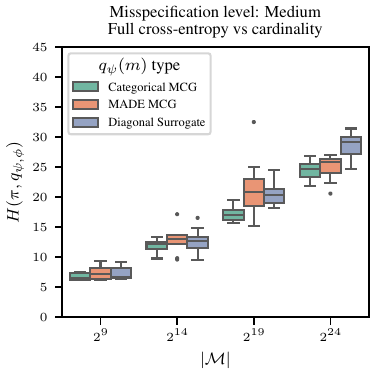} &
    \includegraphics[width=0.245\linewidth]{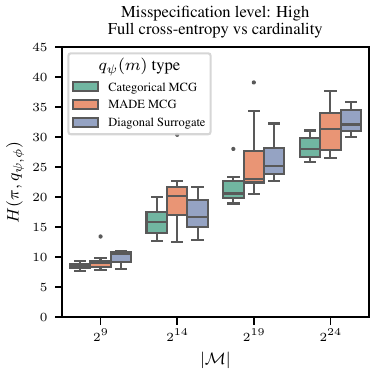} &
    \raisebox{0\height}{\includegraphics[width=0.245\linewidth, trim={0 0 0 0}, clip]{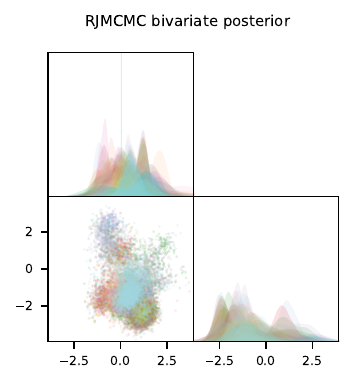}} &
    \raisebox{0\height}{\includegraphics[width=0.245\linewidth, trim={0 0 0 0}, clip]{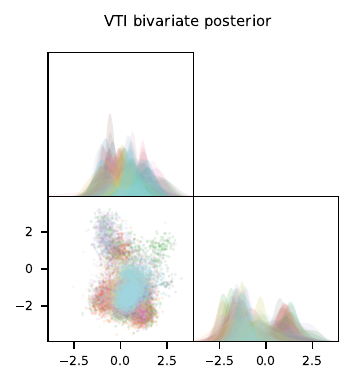}}
    \end{tabular}
    \caption{\textit{Left:} A simulation study of the robust variable selection example showing the cross entropy (NLL) between RJMCMC samples and an flow-based variational transdimensional density using rational quadratic spline CoSMIC flows under a fixed number of iterations (30,000). Each cardinality was run with 10 independently sampled synthetic data sets. \textit{Right:} Comparison of bivariate plots of variables $\smash{\cvars^{(1)},\cvars^{(5)}}$ obtained by RJMCMC and VTI for a single $\smash{|\mdlspace|=2^7}$ problem.}
    \label{fig:apdxrobustvssweepcard}
\end{figure}

\subsection{Bayesian structure learning via non-linear directed acyclic graph discovery}\label{sec:bayesian_dag_problem}

\begin{figure}[t]
    \centering
    \includegraphics[width=1\linewidth, trim={0 0 0 1cm}, clip]{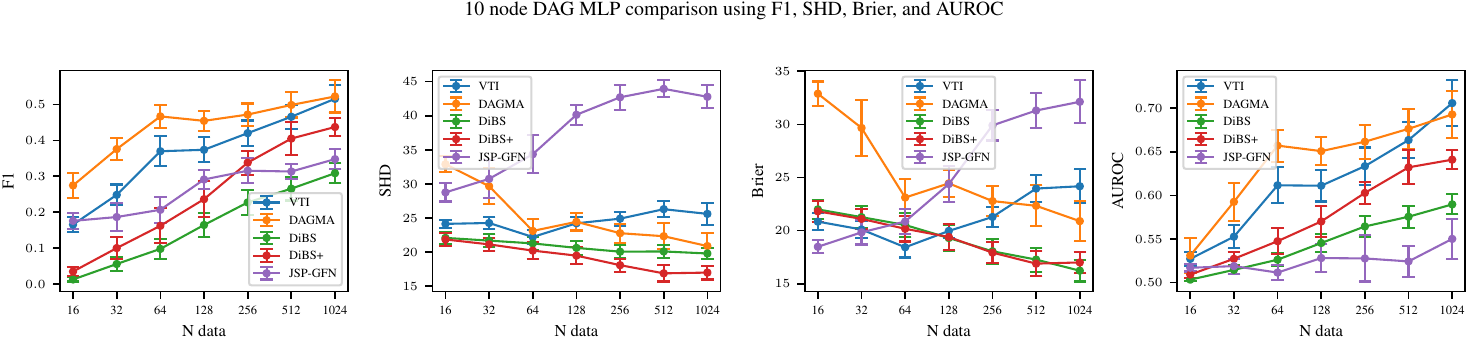}
    \caption{Simulation study comparing VTI to DAGMA~\citep{Bello2022DAGMA}, DiBS/DiBS+~\cite{Lorch2021DiBS}, and JSP-GFlowNets~\cite{Deleu2023Joint} for discovery of a 10-node non-linear DAG visualized 
    using standard metrics (Appendix~\ref{apdx:dagmetrics}, left to right, where better is: higher, lower, lower, higher). Bars display mean and standard error over nine i.i.d. repetitions for each data set size.}
    \label{fig:mlpdagsimstudy}
\end{figure}

We consider a dataset of real-valued observations, denoted by $\MLPDData \in \mathbb{R}^{n \times \MLPDnodes}$, where $n$ is the number of data samples and $\MLPDnodes$ is the number of nodes. Our goal is to perform Bayesian inference over a space of non-linear structural equation models (SEMs) which is isomorphic to a space of directed acyclic graphs (DAGs) and non-linear functions over the active edges. A DAG is represented by a directed adjacency matrix $\MLPDAdj \in \{0,1\}^{\MLPDnodes \times \MLPDnodes}$, where $A_{ij} = 1$ indicates a directed edge from node $i$ to node $j$ and $A_{ij} = 0$ otherwise. The acyclicity constraint requires that the directed edges in $\MLPDAdj$ do not form any directed cycle.
In a \emph{non-linear} SEM, each node $X_j$ depends non-linearly on its parents in the form $\MLPDData = f(\MLPDData) + \epsilon$, $\epsilon_j \sim \mathcal{N}\!\bigl(0,\,\sigma^2\bigr)$, where $f : \mathbb{R}^{\MLPDnodes}\mapsto\mathbb{R}^{\MLPDnodes}$ is a nonlinear function possessing an acyclic Jacobian matrix. 
We follow 
\citep{Bello2022DAGMA, thompson2025prodag}
whereby 
$f$ is a multi-layer-perceptron (MLP) structured as $f(\MLPDData)=(f_1(\MLPDData),\dots,f_{\MLPDnodes}(\MLPDData))^\top$. 
We implement $f$ using a single hidden layer, with rectified linear unit (ReLU) activation functions used to model non-linearity where the bias term can be optionally included (see Appendix~\ref{apdx:mlpdagdescription}).
By introducing a topological ordering of the $\MLPDnodes$ nodes, we simultaneously enforce acyclicity and a consistent mapping of parameters to each graph. Let $\MLPDPerm$ be a permutation matrix that reorders nodes into a valid topological order and define $\MLPDEdges$ to be strictly upper-triangular. By construction, any acyclic adjacency matrix can be represented as $\MLPDAdj \;=\; \MLPDPerm^\top\,\MLPDEdges\,\MLPDPerm$.
Each edge is guaranteed to point from lower-indexed nodes to higher-indexed nodes \emph{in the topological order} \citep{Bonilla2024Variational}. Note that this parametrization does not conform to \cref{prop:unique_categorical_representation}, as the correspondence between $(\MLPDPerm, \MLPDEdges)$ and $\MLPDAdj$ is many-to-one. However, this does not violate the consistent parameter mapping. We use a MADE-based discrete distribution \citep{GermainMADE2015} for $\vimkq$ for inference over a very high cardinality model space (see Appendices \ref{apdx:madeplus} and \ref{apdx:dagmdl} for details).
The simulation study in Figure \ref{fig:mlpdagsimstudy} contrasts VTI with state-of-the-art Bayesian and non-Bayesian baselines (DiBS/DiBS+~\cite{Lorch2021DiBS}, JSP-GFlowNets~\cite{Deleu2023Joint}, and DAGMA~\citep{Bello2022DAGMA}) with the aim of demonstrating that the performance of the generic VTI approach can be competitive with application-specific approaches, where one would expect the latter to have better performance. Evaluation of each method is depicted using the commonly accepted F1 score, structural Hamming distance (SHD), Brier score, and area under receiver operating curve (AUROC) (see Appendix~\ref{apdx:dagmetrics}). A complete description of this study is in Appendix~\ref{apdx:dagsimstudy}.
%
%
\begin{table}[b]
    \caption{Comparison of DAG discovery on flow cytometry data \citep{Sachs2005Causal}: VTI versus baselines}
     \label{tab:dagsachsresults}
    \centering
    \begin{tabular}{cccccc}
        \toprule
         Method & F1 & SHD & Brier & AUROC \\
        \midrule
         VTI non-linear DAG & \textbf{0.44} & \textbf{23.0} & 23.0 & \textbf{0.68} \\
         DAGMA non-linear & 0.32 & 25.0 & 25.0 & 0.60 \\ 
         DiBS+ non-linear & 0.22 & 28.0 & \textbf{17.0} & 0.54 \\
         JSP-GFN non-linear & 0.23 & 54.5 & 44.0 & 0.51 \\
        \bottomrule
    \end{tabular}
\end{table}

\vspace{-3mm}
\paragraph{Real data example in flow cytometry:}
\citet{Sachs2005Causal} use Bayesian networks to analyze multi-parameter single-cell data for deriving causal influences in cellular signaling networks of human immune cells. 
Causal interactions are validated by comparing to a domain-agreed
adjacency matrix representing causality within the data, establishing a baseline for causal prediction accuracy. We use VTI to discover the distribution of non-linear DAGs for these data, comprising $\MLPDsamples=7466$ entries over $\MLPDnodes=11$ nodes, 
and benchmark this against the agreed adjacency. Table \ref{tab:dagsachsresults} shows strong performance of VTI 
compared to state of the art methods. A complete description is in Appendix~\ref{apdx:realdatadagexample}.
%
\section{Discussion}\label{sec:discussion}

We have introduced \method\, normalizing flows as a means to implement variational transdimensional inference (VTI), the approximation of a target density over a transdimensional space with a single variational density.
VTI is broadly applicable to a wide class of transdimensional inference problems.
%
Although the specification of a CoSMIC flow requires augmenting all model dimensions to $\dmax$, VTI is not sensitive to these added dimensions during training and inference due to the construction of the auxiliary variable transforms.
We presented two approaches for simultaneously optimizing the variational parameters $\vimkp,\vipp$. 
The Gaussian surrogate-based approach benefits from our derivation of the approximation error bounds and established convergence guarantees for the marginal models distribution under convergent optimization steps. 
%
%
The two approaches that use Monte Carlo gradient estimation for SGD optimization
benefit from 
recent advances in neural architectures and
neural approximation of very large model spaces. 
The choice of model sampler is dependent on both the cardinality of the model space and the structure of the problem. When $|\mdlspace|$
is small,
the Gaussian process surrogate-based sampler or the categorical sampler using Monte Carlo gradients are both appropriate, although in practice it is usually safe to default to the latter approach.
For high cardinality problems,
we recommend a neural model sampler for approximate inference on the distribution of model weights.




%
%
The quality of the VTI approximation possesses
two notable characteristics.
%
Firstly, those models $\mdl\in\mdlspace$ estimated to have 
large posterior model probabilities will contribute most significantly to the loss. Hence the \method\, flow will 
produce a relatively more accurate (in the KL sense) 
approximation of such models, compared to models with low probabilities.
This effect is seen in Figure \ref{fig:robustvsmodelprob} (bottom row).
While one might prefer greater accuracy on more dominant models,
structured 
changes to $\mc{L}(\vimkp,\vipp)$
could give greater control over where the quality of the variational approximation should focus.
The second characteristic is that when the 
normalizing flow is unable to 
approximate the conditional target $\target(\cvars\gvn\mdl)$ well, 
a smaller loss can be achieved by shrinking the estimated model probability $q_\vimkp(\mdl)$ to zero. This effect is seen in Figure \ref{fig:robustvsmodelprob} (top row), which lessens as flow expressivity increases. Here the question is how to design the 
normalizing flow, i.e.~the flow context $\contextvars$ and the mapping $\ctm(\mdl)$, to 
best
allocate resources to produce good approximations of models likely to be of relatively high posterior model probability.
In many transdimensional problems, two models could be considered adjacent by structural similarity (e.g.~in variable selection where they differ by one included covariate) and so may have similar posterior model probabilities.
This could be achieved by
e.g.~extending the architecture of the context encoder (Appendix~\ref{apdx:flowarchitecture}) to capture similarities between models that generalize over the model space, and learning structural similarity within the surrogate-based model sampler itself via a reward-based criterion.


The left-align permutation used in Proposition \ref{prop:cosmic-factorisation} raises the question of whether the alignment of variables across models in the normalizing flow is important for computational efficiency. More broadly: (a) is there shared information between models? And, if so, (b) would careful manual construction of the flow improve exploitation of this versus allowing the optimization to determine it agnostically?
(a) is answered by the robust variable selection example in Figures \ref{fig:apdxrobustvssweepcard} (right) and \ref{fig:robustvsmultivariate} where high probability mass regions for each model do not overlap and thus there there is no such shared information. To answer (b) would be an avenue for future research.
Future work could also 
derive convergence rates, which will depend on the
choice of optimization algorithm for the flow parameters, and extension of the \method{} architecture to coupling flows, in applications outside variational inference. 
In addition,
our analysis for the surrogate-based approach is general enough to be extended to a variety of methods for approximating 
a distribution over models.
%
Finally, in extending VI to the transdimensional setting we inherit the same strengths and weaknesses of single-model VI, including the challenges of mode-collapse. Users would need to take the same steps to manage it as in the standard setting. Mode collapse in the model distribution is mitigated by the exploration versus exploitation strategies discussed in Section \ref{sec:modelweights}.

\section*{Acknowledgments}
This research was made possible by funding support from 
the CSIRO Machine Learning and Artificial Intelligence Future Science Platform. The authors would also like to thank Ryan Thompson for critical advice in the construction of the non-linear DAG example, and Daniel M. Steinberg and Edwin Bonilla for advice on Monte Carlo gradient estimation methods. SAS is supported by the Australian Research Council.

{
\small

\ifthenelse{\boolean{isarxiv}\OR\boolean{isfinal}}{
\bibliographystyle{plainnat}   
\bibliography{vti_references_natbib_merged}
}{
  \printbibliography
}

\newpage
\appendix

\section{Implementation of \method{} normalizing flows}
\label{apdx:cosmic-details}

\subsection{Inverse autoregressive flow sampling procedure}
Draw reference samples  
\[
   \refvars=(\refvarcoord^{(1)},\dots,\refvarcoord^{(\dmax)})\sim\refdist_{\dmax}.
\]
For a given $\mdl$, define the permutation matrix $P_\mdl$ that groups active coordinates first:
\[
   (\refvars_\mdl,\refvars_{\setminus\mdl}) := P_\mdl \refvars
   \quad\Longrightarrow\quad
   \refvars_\mdl\in\mathbb R^{d_\mdl},\;
   \refvars_{\setminus\mdl}\in\mathbb R^{\setminus d_\mdl},\,\setminus d_\mdl=\dmax-d_\mdl.
\]

Concatenate the coordinate-wise transforms $\indivtransform_{\ap_i}$ into the map $\transform_\vipp$ and bookend with permutations $P_\mdl$ to give the strict \method{} bijection
\[
   T_{\phi}^{(P_m)}(\refvars)
      :=P_m^{-1}\,
        \bigl(\tau_{\rho_{1}^{C(m)}}(\refvars^{(1)})
              ,\cdots,
              \tau_{\rho_{\dmax}^{C(m)}}(\refvars^{(\dmax)})\bigr)
        \,P_m .
\]

\subsection{Experimental \method{} transform compositions}
\label{apdx:flowarchitecture}

\newcommand{\mdlrevperm}{P_{<d_\mdl}^{\operatorname{rev}}}

The experiments use the below compositions of transforms as inverse autoregressive flows $\transform_\vipp(\refvars\mid\mdl)$ where $\refvars$ are the inputs from the reference distribution and $\mdl$ is the context input. All compositions except for the diagonal Gaussian are assumed to have the strict left-align permutations discussed in Appendix~\ref{apdx:cosmic-details}. The term ``block'' is defined in Appendix~\ref{apdx:madedefinitions}.

\paragraph{Context encoder:} Experiments will sometimes use a context encoder that projects the context input to a higher dimensional space. Typically this will take the form of a multi-layered perceptron with hidden layers of increasing size (fixed to powers of 2) and terminating in an activation layer at the largest size, say $2^{12}$ nodes.

\paragraph{Model-specific reverse-permutation:} Flow compositions commonly include reverse permutations to ensure expressibility of an autoregressive-NN-based flow is (approximately) the same for all coordinates. Denoting the generic reverse permutation for all coordinates as $P^{\operatorname{rev}}$, we assume the strict left-right permutation $P_\mdl$ (as per Appendix~\ref{apdx:cosmic-details}) has been applied, and hence define the left-most $d_\mdl$-coordinate reverse permutation $\mdlrevperm$.

\newcommand{\TMAF}[1]{\transform_{\vipp_{#1}}^{\operatorname{Affine}}}
\newcommand{\TRQSMAF}[1]{\transform_{\vipp_{#1}}^{\operatorname{RQ-Spline}}}

\paragraph{Affine(5,5):} The learned component is the affine masked autoregressive transform~\citep{PapamakariosMasked2017}, denoted here as $\TMAF{k}$ for transforms $k=1,\dots,5$, each having 5 blocks. We set $\transform_\vipp:=\TMAF{5}\circ\mdlrevperm\circ\dots\circ\mdlrevperm\circ\TMAF{1}$.

\paragraph{Spline(4,6):} The learned component is the rational quadratic spline masked autoregressive flow architecture \citep{Durkan2019Neural}, denoted here as $\TRQSMAF{k}$. Each $\TRQSMAF{k}$ has 6 blocks. Additionally, we define a fixed global affine  transform $\transform_{\mu_g,\sigma_g}$ that is not dependent on inputs nor context and hence has only two learnable parameters: scale $\mu_g$ and shift $\sigma_g$. We set $\transform_\vipp:=\transform_{\mu_g,\sigma_g}\circ\TRQSMAF{5}\circ\mdlrevperm\circ\dots\circ\mdlrevperm\circ\TRQSMAF{1}$.

\subsection{Autoregressive flow definitions}\label{apdx:madedefinitions}
\newcommand{\madearchinputs}{\mathbf{x}}
\newcommand{\madearchcontext}{\mathbf{z}}
\newcommand{\madearchfeatures}{d}
\newcommand{\madearchactivation}{\sigma}
\newcommand{\madearchdropout}{\mathrm{Dropout}}
\newcommand{\madearchBN}{\mathrm{BN}}
\newcommand{\madearchmaskedlinear}{\mathrm{MaskedLinear}}
\newcommand{\madearchblockoutput}{\mathbf{h}}

We use the residual variant of the Masked Autoencoder for Distribution Estimation (MADE) \citep{GermainMADE2015}, implemented in PyTorch by \citep{Durkan2020Nflows}. Each block maintains the autoregressive property by assigning degrees \(\{1, \dots, \madearchfeatures\}\) to inputs and propagating them forward.

Given input $\madearchinputs\in\mathbb{R}^{\madearchfeatures}$ and optional context $\madearchcontext$, each residual block computes:
\[
\begin{aligned}
\madearchblockoutput &= \madearchinputs + \madearchmaskedlinear_2\left(
    \madearchactivation\left(
      \madearchBN_2\left(
        \madearchmaskedlinear_1\left(
          \madearchactivation\left(
            \madearchBN_1(\madearchinputs)
          \right)
        \right) + \delta(\madearchcontext)
      \right)
    \right)
  \right).
\end{aligned}
\]
Here, \(\madearchmaskedlinear_i\) are masked linear layers respecting the autoregressive structure, \(\madearchBN_i\) are optional batch norm layers, and \(\delta(\madearchcontext)\) is an optional context projection. All layers preserve feature dimensionality and respect degree ordering to ensure autoregressive validity.

\section{Analysis of a \method{} normalizing flow}\label{apdx:cosmic-proof}

\begin{proof}[Proof of \cref{lemma:identity-map}]
Let
$I(\mdl)=\{\,i\in\{1,\dots,\dmax\}:\modeltomask_i(\mdl)=1\,\}$, and $I^c$ be the complement. 
The result holds from \eqref{eqn:cosmic-rho-m} as, for all coordinates $i\in I^c(m)$,
$    \auxvars_{\setminus\mdl}^{(i)}
    = \indivtransform_{\ap_i^\ctm}(\refvars_{\setminus\mdl}^{(i)})
    = 
    \indivtransform_{\staticap}(\refvars_{\setminus\mdl}^{(i)})
    =\refvars_{\setminus\mdl}^{(i)}.$
\end{proof}

\begin{proof}[Proof of \cref{prop:cosmic-factorisation}]

\smallskip
\noindent
\emph{(a)–(b) Density factorization and marginal consistency.}

\smallskip
We aim to prove \begin{align*}
\text{(a)}&\quad\satvipq^{\lrperm}(\cvars,\auxvars_{\setminus\mdl}\mid\mdl)
   \;=\;
   q_\vipp
(\cvars\mid\mdl)\;
   \refdist_{d_{\setminus\mdl}}\!\bigl(\auxvars_{\setminus\mdl}\bigr),\\
\quad
\text{and so (b)}&\quad
\int \satvipq^{\lrperm}(\cvars,\auxvars_{\setminus\mdl}\mid\mdl)\,
      d\auxvars_{\setminus\mdl}
   = q_\vipp
(\cvars\mid\mdl).
\end{align*}
Write $\transform_{\phi}^{\lrperm,-1}=(\transform_{\vipp,\mdl}^{-1},\operatorname{Id})$, where $\operatorname{Id}$ denotes the identity transform, and let the permuted reference vector be
$P_\mdl\refvars=(\refvars_\mdl,\refvars_{\setminus\mdl})
                 \in\mathbb{R}^{\dmdl}\times\mathbb{R}^{d_{\setminus\mdl}}$.
Because the masking function $\ctm$ sets every transform
$\indivtransform_{\ap_i}$ with $\ctm_i(\mdl)=0$ to the identity,
the inverse flow splits as
\[
\transform_{\vipp}^{\lrperm,-1}(\cvars,\auxvars_{\setminus\mdl})
   \;=\;
   \bigl(
        \transform_{\vipp,\mdl}^{-1}(\cvars),
        \;\auxvars_{\setminus\mdl}
    \bigr),
\]
where
$\transform_{\vipp,\mdl}^{-1}: \Theta_\mdl\to\mathbb{R}^{\dmdl}$
is the active block and the dummy block is exactly the identity.
Consequently, the Jacobian matrix of
$\transform_{\vipp}^{\lrperm,-1}$ is block upper‑triangular with
\(
\det\nabla\transform_{\vipp}^{\lrperm,-1}
  = \det\nabla\transform_{\vipp,\mdl}^{-1}\times 1.
\)

%

Apply change‑of‑variables with
$\refdist_{\dmax}=\refdist_{d_m}\otimes\refdist_{\setminus d_m}$ to obtain
\[
\begin{aligned}
\satvipq^{\lrperm}(\cvars,\auxvars_{\setminus\mdl}\!\mid\!\mdl)
 &=
 \refdist_{\dmax}\!\bigl(
     \transform_{\vipp}^{\lrperm,-1}(\cvars,\auxvars_{\setminus\mdl})
 \bigr)
 \bigl|\det\nabla\transform_{\vipp}^{\lrperm,-1}\bigr|
\\
 &=
 \refdist_{\dmdl}\!\bigl(
     \transform_{\vipp,m}^{-1}(\cvars)
 \bigr)
 \,\refdist_{d_{\setminus\mdl}}(\auxvars_{\setminus\mdl})\,
 \bigl|\det\nabla\transform_{\vipp,m}^{-1}(\cvars)\bigr|
\\
 &=
 q_{\vipp}(\cvars\!\mid\!\mdl)\;
 \refdist_{d_{\setminus\mdl}}(\auxvars_{\setminus\mdl}),
\end{aligned}
\]
which proves the factorization (a).

Integrating the right‑hand side over
$\auxvars_{\setminus\mdl}$ recovers (b)
$q_{\vipp}(\cvars\!\mid\!\mdl)$, completing the proof.%
\end{proof}

\label{apdx:cancellation-proof}

\begin{proof}[Proof of \cref{prop:cancellation}]
It is sufficient to show $\dkl(\tilde{q}_{\vimkp,\vipp}||\sattargetunnorm) = \dkl(q_{\vimkp,\vipp}||\targetunnorm) := \mc{L}(\vimkp,\vipp)$. Note by \eqref{eq:saturatedtarget}, $\sattargetunnorm(\mdl,\cvars,\auxvars_{\setminus\mdl})=\prior(\mdl)\sattargetunnorm(\cvars, \auxvars_{\setminus\mdl}\mid\mdl)=\prior(\mdl)\targetunnorm(\cvars\mid\mdl)\refdist_{d_{\setminus\mdl}}(\auxvars_{\setminus\mdl})$.
\[
\begin{aligned}
    \dkl(\tilde{q}_{\vimkp,\vipp}||\sattargetunnorm)
    &= \Ex_{\mdl\sim\vimkq}\left[
            \Ex_{(\cvars,\auxvars_{\setminus\mdl})\sim\satvipq}\left[
                \log\left(
                    \dfrac
                        {\vimkq(\mdl)\satvipq(\cvars,\auxvars_{\setminus\mdl}\mid\mdl)}
                        {\prior(\mdl)\sattargetunnorm(\cvars,\auxvars_{\setminus\mdl}\mid\mdl)}
                \right)
            \right]
        \right] \\
        &= \Ex_{\mdl\sim\vimkq}\left[
            \Ex_{(\cvars,\auxvars_{\setminus\mdl})\sim\satvipq}\left[
                \log\left(
                    \dfrac
                        {\satvipq(\cvars,\auxvars_{\setminus\mdl}\mid\mdl)}
                        {\sattargetunnorm(\cvars,\auxvars_{\setminus\mdl}\mid\mdl)}
                \right)
            \right]
            +\log\left(\vimkq(\mdl)\right) - \log\left(\prior(\mdl)\right)
        \right]\\
        &= \Ex_{\mdl\sim\vimkq}\left[
        \log\left(\vimkq(\mdl)\right) - \log\left(\prior(\mdl)\right)
        \right]+\\
        &\quad
        \Ex_{\mdl\sim\vimkq}\left[
            \Ex_{\refvars\sim\refdist_{\dmax}}\left[
                \log\left(
                    \dfrac
                        {\refdist_{d_\mdl}(\refvars_{d_\mdl})\cancel{\refdist_{d_{\setminus\mdl}}(\refvars_{\setminus\mdl})}|\det \nabla\transform_\vipp (\refvars)|^{-1}}
                        {\targetunnorm(\cvars\mid\mdl)\cancel{\refdist_{d_{\setminus\mdl}}(\auxvars_{\setminus\mdl})}}
                \right)
            \right]
        \right]\,\,\text{by Proposition \ref{prop:cosmic-factorisation}}\\  
        &=   \dkl(q_{\vimkp,\vipp}||\targetunnorm)
        =\mc{L}(\vimkp,\vipp).
\end{aligned}
\]
\end{proof}

\begin{corollary}[Computational complexity]\label{apdx:cosmic-complexity}
\smallskip
\noindent
\begin{itemize}
      \item \emph{Sampling (forward IAF)}:
            all coordinates can be updated in parallel
            $\;\Rightarrow\;$ $\mc{O}(1)$ wall‑time depth.
      \item \emph{Evaluation (inverse direction)}:
            must populate $\refvars^{(<i)}$ sequentially
            $\;\Rightarrow\;$ $\mc{O}(\dmax)$ arithmetic operations,
            identical to a standard IAF.
      \end{itemize}
\end{corollary}
\begin{proof}
    The forward IAF updates $\theta_m$ via closed‑form $\tau_i$
that read \emph{previous outputs}—all available after one pass through
the network— which are thereby fully parallelizable.
Conversely, evaluating $T_{\phi}^{(m),-1}$ at an arbitrary point in
$\Theta_m\times\mathcal M$ must reconstruct $z$ sequentially,
exactly as for any IAF, giving $O(d_{\max})$ time.
\end{proof}

\newpage\section{Theoretical analysis of the model weights distribution}\label{apdx:theoreticalanalysis}
We consider the following bi-level stochastic optimization problem over a function $\objective: \mdlspace \times \vpspace \to \sR$ as:
\begin{equation}
    \vipp^* \in \argmax_{\vipp \in \vpspace} \max_{\vim_\objective \in \pspace(\mdlspace)} \Ex_{\mdl \sim \vim} [\objective(\mdl, \vipp) + \log\prior(\mdl)] + \entropy[\vim_\objective] = \argmax_{\vipp\in\vpspace} \Ex_{\mdl \sim \vim_{\objective,\vipp}^*}[\objective(\mdl, \vipp) + \log\prior(\mdl)]\,,
    \label{eq:prob-general}
\end{equation}
where $\pspace(\mdlspace)$ denotes the space of probability measures over $\mdlspace$, $\entropy$ is the entropy, and the optimal $\vim_\objective$ for a given $\vipp$ can be shown to be:
\begin{equation}
    \vim_{\objective,\vipp}^*(\mdl) := \frac{\prior(\mdl) \exp \objective(\mdl,\vipp)}{\sum_{\mdl'\in \mdlspace} \prior(\mdl') \exp \objective(\mdl', \vipp)}\,, \quad \mdl \in \mdlspace\,.
\end{equation}
This formulation corresponds to a stochastic optimization problem over two variables $\vipp$ and $\vim_\objective$, where the optimum for $\vim_\objective$ has a closed-form expression $\vim_{\objective,\vipp}^*$ for every given $\vipp\in\vpspace$. To solve this problem, we will follow a sequential optimization process over $\vipp$ (e.g., stochastic gradient descent).
However, sampling from the optimal model distribution $\vim_{\objective,\vipp}^*$ (above) requires evaluating the summation in the normalization constant, which is expensive. 
Therefore, we will instead approximate each $\vim_{\objective, \vipp_\iterIdx}^*$ with a distribution $\vim_{\ucb, \iterIdx}$ composed of a cheaper-to-evaluate surrogate $\ucb_\iterIdx$ based on noisy observations $\observation_{\iterIdx-1, i} = \observable(\refvars_i, \mdl_i, \vipp_{\iterIdx-1})$, where $\refvars_i\sim\refdist$ and $\mdl_i \sim \vim_{\ucb, \iterIdx-1}$, $i \in \{1, \dots, \nbatch\}$, such that $\Ex_{\refvars\sim\refdist}[\observable(\refvars, \mdl, \vipp)] = \objective(\mdl, \vipp)$.
If we ensure that $\vim_{\ucb, \iterIdx}$ approaches $\vim_{\objective, \vipp_\iterIdx}^*$ over time, optimization steps based on $\vim_{\ucb, \iterIdx}$ will eventually follow $\vim_{\objective, \vipp_\iterIdx}^*$ and allow for the optimum $\vipp^*$ to be reached.

\subsection{Regularity assumptions}
We make the following assumptions about the function $\objective$ and the observation noise.

\begin{assumption}
    \label{a:k-continuous}
    The objective $\objective$ is a sample from a zero-mean Gaussian process prior with a bounded, positive-semidefinite covariance function $\kernel: (\mdlspace \times \vpspace)^2\to\R$, which is continuous over $\vpspace$.
\end{assumption}

The GP assumption allows us to derive closed-form expressions for predictions over $\objective$ and their associated uncertainty. The continuity assumption on $\kernel$ is easily satisfied by most practical covariance functions and ensures that, if $\vipp_\iterIdx$ converges to some $\vipp^*$, GP-based estimates $\objective(\mdl, \vipp^*)$ will also converge for every $\mdl\in\mdlspace$. To model predictions over $\objective$ with closed-form GP updates, we also need Gaussian assumptions about the observation noise, which is given by:
\begin{equation}
    \obsnoise_{\mdl, \vipp} := \observable(\refvars, \mdl, \vipp) - \objective(\mdl, \vipp), \quad \refvars\sim\refdist, \quad \mdl \in \mdlspace, \vipp\in\vpspace\,.
\end{equation}
However, as we will show in our analysis, sub-Gaussian tails are enough for GP modeling, which we formalize next.

\begin{assumption}
    \label{a:sub-g-noise}
    The observation noise is $\sigma_\obsnoise^2$-sub-Gaussian, i.e., given any $\mdl\in\mdlspace$ and $\vipp\in\vpspace$, we have: 
    \begin{equation}
        \forall s\in\sR, \quad \Ex[\exp(s\obsnoise_{\mdl, \vipp})] \leq \exp\left(\frac{1}{2}s^2\sigma_\obsnoise^2\right).
    \end{equation}
\end{assumption}

This mild assumption is satisfied, for example, when $\refdist$ is a zero-mean Gaussian distribution and $\observable$ is Lipschitz continuous on its first argument, in which case $\sigma_\obsnoise$ only depends on $\observable$ through its Lipschitz constant \citep{Boucheron2013, Pisier2016}.


\subsection{Gaussian process model}
Under the GP assumption $\objective\sim\gp(0, \kernel)$, the posterior over $\objective$ is again a Gaussian process. Suppose at each iteration $\iterIdx\geq 1$ of stochastic gradient descent we sample a mini-batch $\{\mdl_{\iterIdx, i}\}_{i=1}^\nbatch$ from a variational posterior approximating $\vim_{\objective, \vipp}^*$ 
at $\vipp = \vipp_{\iterIdx-1}$. Given a batch of observations $\batch_\iterIdx := \{\vipp_{\iterIdx-1}, \mdl_{\iterIdx, i}, \observation_{\iterIdx,i} \}_{i=1}^\nbatch$, the GP posterior $\objective|\batch_{1, \dots, \iterIdx} \sim \gp(\gpMean_\iterIdx, \kernel_\iterIdx)$ has its mean and covariance described by the following recursive equations:
\begin{align}
    \gpMean_\iterIdx(\mdl, \vipp) &= \gpMean_{\iterIdx-1}(\mdl, \vipp) + \vct\kernel_{\iterIdx-1}(\mdl, \vipp)^\transpose(\mKernel_{\iterIdx-1} + \sigma_\obsnoise^2\eye)^{-1}(\vct\observation_\iterIdx - \vct{\gpMean}_{\iterIdx-1}) \label{eq:gp-mean-update}\\
    \kernel_\iterIdx(\mdl, \vipp, \mdl', \vipp') &= \kernel_{\iterIdx-1}(\mdl, \vipp, \mdl',\vipp') - \vct\kernel_{\iterIdx-1}(\mdl, \vipp)^\transpose(\mKernel_{\iterIdx-1} + \sigma_\obsnoise^2\eye)^{-1}\vct\kernel_{\iterIdx-1}(\mdl', \vipp), \label{eq:gp-cov-update}
\end{align}
where $\vct\kernel_{\iterIdx-1}(\mdl, \vipp) := [\kernel_{\iterIdx-1}(\mdl, \vipp, \mdl_{\iterIdx, i}, \vipp_{\iterIdx-1})]_{i=1}^\nbatch \in \sR^\nbatch$, $\mKernel_{\iterIdx-1} := [\kernel_{\iterIdx-1}(\mdl_{\iterIdx,i}, \vipp_{\iterIdx-1}, \mdl_{\iterIdx,j}, \vipp_{\iterIdx-1})]_{i,j=1}^\nbatch \in \sR^{\nbatch\times\nbatch}$, and $\vct\gpMean_{\iterIdx-1} := [\gpMean_{\iterIdx-1}(\mdl_{\iterIdx,i}, \vipp_{\iterIdx-1})]_{\iterIdx=1}^\nbatch \in \sR^\nbatch$, with $\gpMean_0 = 0$ and $\kernel_0 = \kernel$. Any pointwise prediction is then modeled as $\objective(\mdl, \vipp)|\batch_{1, \dots, \iterIdx} \sim \normal(\gpMean_\iterIdx(\mdl, \vipp), \sigma_\iterIdx^2(\mdl, \vipp))$, where $\sigma_\iterIdx^2(\mdl, \vipp) := \kernel_\iterIdx(\mdl, \vipp, \mdl, \vipp)$, for $(\mdl, \vipp) \in \mdlspace\times\vpspace$.


\subsection{Upper confidence bound (UCB) algorithm}
Given the GP posterior, we formulate an upper confidence bound algorithm \citep{SrinivasGaussian2010} with:
\begin{equation}
    \ucb_\iterIdx(\mdl) := \gpMean_\iterIdx(\mdl, \vipp_\iterIdx) + \beta_\iterIdx\sigma_\iterIdx(\mdl, \vipp_\iterIdx), \quad \mdl \in \mdlspace\,,
\end{equation}
where $\beta_\iterIdx > 0$ is a parameter controlling the size of the confidence bound, which we will discuss in our analysis.
We then derive a sampling distribution based on using the UCB as a surrogate for $\objective$ as:
\begin{equation}
    \vim_{\ucb, \iterIdx} \in \argmax_{\vim \in \pspace(\mdlspace)}  \Ex_{\mdl\sim\vim}[\ucb_\iterIdx(\mdl) + \log \prior(\mdl) - \log \vim(\mdl)]\,.
\end{equation}
The solution to this optimization is available in closed form as the UCB softmax:
\begin{equation}
    \vim_{\ucb, \iterIdx}(\mdl) = \frac{\prior(\mdl) \exp \ucb_\iterIdx(\mdl)}{\sum_{\mdl'\in\mdlspace} \prior(\mdl') \exp\ucb_\iterIdx(\mdl')}, \quad \mdl\in\mdlspace\,.
\end{equation}
Equipped with this UCB-based sampling distribution, we follow the generic procedure outlined in \autoref{alg:ucb}. The algorithm starts by sampling from the current UCB distribution. A sample-based estimate of the optimization objective $\Ex_{\mdl\sim\vipp_\iterIdx}[\objective(\mdl, \vipp_{\iterIdx-1})] \approx \frac{1}{\nbatch} \sum_{i=1}^\nbatch \observable(\refvars_{\iterIdx, i}, \mdl_{\iterIdx,i}, \vipp_{\iterIdx-1})$ is then passed to the algorithm responsible for updating the parameters $\vipp_\iterIdx$, e.g., a stochastic gradient descent update.
Once the parameters are updated, we reevaluate the objective and update our GP. The procedure then repeats up to a given number of iterations $\nIterations \in \sN$.

\begin{algorithm}[t]
    \caption{Stochastic optimization with UCB sampling}
    \label{alg:ucb}
    \begin{algorithmic}
        \FOR{$\iterIdx\in\{1,\dots,\nIterations\}$}
           \STATE $\{\mdl_{\iterIdx, i}\}_{i=1}^\nbatch \sim \vim_{\ucb, \iterIdx-1}$
           \STATE $\{\refvars_{\iterIdx, i}\}_{i=1}^\nbatch \sim \refdist$
           \STATE $\observation_{\iterIdx, i} = \observable(\refvars_{\iterIdx, i}, \mdl_{\iterIdx, i}, \vipp_{\iterIdx-1})$, for $i \in \{1, \dots, \nbatch\}$
           \STATE $\vipp_{\iterIdx} \leftarrow$ \textsc{UpdateParameters}$\left(\vipp_{\iterIdx-1}, \{\observable(\refvars_{\iterIdx, i}, \mdl_{\iterIdx,i}, \vipp_{\iterIdx-1})\}_{i=1}^\nbatch, \vim_{\ucb,\iterIdx-1} \right)$
           \STATE $\gpMean_\iterIdx, \kernel_\iterIdx \leftarrow$ \textsc{UpdateSurrogate}$(\{\mdl_{\iterIdx, i}, \observation_{\iterIdx, i}\}_{i=1}^\nbatch, \vipp_{\iterIdx-1}, \gpMean_{\iterIdx-1}, \kernel_{\iterIdx-1})$
        \ENDFOR
    \end{algorithmic}
\end{algorithm}

\subsection{Approximation errors under sub-Gaussian noise}
In the following, we derive generic concentration bounds for GP predictions under sub-Gaussian observation noise. We start by showing that the approximation error between the GP mean and the true function is sub-Gaussian.
\begin{lemma}
    \label{thr:sub-g-process}
    Let $\objective \sim \gp(0, \kernel)$ be a zero-mean Gaussian process with a given positive-definite covariance function $\kernel: \domain \times \domain \to \sR$. Assume we are given a sequence of observations $\observation_\obsIdx = \objective(\location_\obsIdx) + \obsnoise_\obsIdx$, where $\location_\obsIdx \in \domain$ and $\obsnoise_\obsIdx$ is $\sigma_\obsnoise^2$-sub-Gaussian noise, for all $\obsIdx \in \sN$. Let $\gpMean_\obsIdx$ and $\sigma_\obsIdx^2$ denote the predictive mean and variance, respectively, of the GP posterior under the assumption that the noise is zero-mean Gaussian with variance given by $\sigma_\obsnoise^2$. Then, for all $\obsIdx\geq 0$ and all $\location\in\domain$, we have that $\objective(\location) - \gpMean_\obsIdx(\location)$ is $\sigma_\obsIdx^2(\location)$-sub-Gaussian.
\end{lemma}
\begin{proof}
	For $\obsIdx=0$, the proof is trivial as, without observations, we only have the prior with $\gpMean_0(\location) = 0$ and $\sigma_0^2(\location) = \kernel(\location,\location)$. Now let $\locations_\obsIdx := \{\location_i\}_{i=1}^\obsIdx \subset\domain$ denote a set of $\obsIdx \geq 1$ observed locations. For any given $\location\in\domain$, expanding the GP posterior mean from its definition, the approximation error can be decomposed as:
	\begin{equation}
		\begin{split}
			\Delta_\obsIdx(\location) := \objective(\location) - \gpMean_\obsIdx(\location) &= \objective(\location) - \kernel(\location, \locations_\obsIdx)(\Kernel_\obsIdx + \sigma_\obsnoise^2\eye)^{-1}(\vct\objective_\obsIdx + \vct\obsnoise_\obsIdx)\\
            &= \objective(\location) - \kernel(\location, \locations_\obsIdx)(\Kernel_\obsIdx + \sigma_\obsnoise^2\eye)^{-1}\vct\objective_\obsIdx - \kernel(\location, \locations_\obsIdx)(\Kernel_\obsIdx + \sigma_\obsnoise^2\eye)^{-1}\vct\obsnoise_\obsIdx\,,
		\end{split}
		\label{eq:sub-g-process-approximation-components}
	\end{equation}
    where $\kernel(\location, \locations_\obsIdx) := [\kernel(\location,\location_1), \dots, \kernel(\location,\location_\obsIdx)]$, $\mKernel_\obsIdx := [\kernel(\location_i, \location_j)]_{i,j=1}^\obsIdx$, $\vct\objective_\obsIdx := [\objective(\location_i)]_{i=1}^\obsIdx$, and $\vct\obsnoise_\obsIdx := [\obsnoise_i]_{i=1}^\obsIdx$.
	The last term on the right-hand side above is sub-Gaussian, since $\Ex[\vct\obsnoise_\obsIdx] = 0$ and, letting $\vct \alpha_\obsIdx := (\Kernel_\obsIdx + \sigma_\obsnoise^2\eye)^{-1} \kernel(\locations_\obsIdx, \location)$, we have a sum of independent sub-Gaussian random variables, see e.g.~\citep{Pisier2016}, Lemma 1.1:
	\begin{equation}
		\Ex[\exp(\vct\alpha_\obsIdx^\transpose \vct\obsnoise_\obsIdx)] = \Ex\left[\exp\left(\sum_{i=1}^\obsIdx \alpha_{\obsIdx,i}\obsnoise_{\obsIdx,i} \right)\right] = \prod_{i=1}^\obsIdx\Ex[\exp(\alpha_{\obsIdx,i}\obsnoise_{\obsIdx,i})] \leq \exp \left(\frac{1}{2} \sigma_\obsnoise^2\sum_{i=1}^\obsIdx\alpha_{\obsIdx,i}^2\right), 
        \label{eq:sub-g-noise-component}
	\end{equation}
    which follows from the definition of sub-Gaussian noise (cf. \autoref{a:sub-g-noise}).
    The remaining term on the right-hand side of \eqref{eq:sub-g-process-approximation-components} is a zero-mean Gaussian random variable with
    variance given by:
	\begin{equation}
		\begin{split}
			&\Var[\objective(\location) - \kernel(\location, \locations_\obsIdx)(\Kernel_\obsIdx + \sigma_\obsnoise^2\eye)^{-1}\vct\objective_\obsIdx]\\
			 &= \kernel(\location,\location) - 2 \kernel(\location, \locations_\obsIdx)(\Kernel_\obsIdx + \sigma_\obsnoise^2\eye)^{-1}\kernel(\locations_\obsIdx,\location) +  \kernel(\location, \locations_\obsIdx)(\Kernel_\obsIdx + \sigma_\obsnoise^2\eye)^{-1}\Kernel_\obsIdx(\Kernel_\obsIdx + \sigma_\obsnoise^2\eye)^{-1}\kernel(\locations_\obsIdx,\location)\\
			 &= \kernel(\location,\location)
			 - 2 \kernel(\location, \locations_\obsIdx)(\Kernel_\obsIdx + \sigma_\obsnoise^2\eye)^{-1}\kernel(\locations_\obsIdx,\location)
			 +  \vct\alpha_\obsIdx^\transpose \Kernel_\obsIdx \vct\alpha_\obsIdx\,.
             \label{eq:sub-g-objective-component}
		\end{split}
	\end{equation}
	   As \eqref{eq:sub-g-process-approximation-components} describes the sum of two independent sub-Gaussian random variables, we can follow similar reasoning to the one applied in \eqref{eq:sub-g-noise-component} to show that $\Delta_\obsIdx(\location)$ is $s_\obsIdx^2(\location)$-sub-Gaussian for some $s_\obsIdx^2(\location) > 0$. The resulting sub-Gaussian parameter $s_\obsIdx^2(\location)$ is then bounded by the sum of the individual sub-Gaussian parameters in equations \ref{eq:sub-g-noise-component} and \ref{eq:sub-g-objective-component} as:
	\begin{equation}
		\begin{split}
			s_\obsIdx^2(\location) &\leq \kernel(\location,\location)
			- 2 \kernel(\location, \locations_\obsIdx)(\Kernel_\obsIdx + \sigma_\obsnoise^2\eye)^{-1}\kernel(\locations_\obsIdx,\location)
			+  \vct\alpha_\obsIdx^\transpose \Kernel_\obsIdx \vct\alpha_\obsIdx
			+ \sigma_\obsnoise^2 \vct\alpha_\obsIdx^\transpose \vct\alpha_\obsIdx\\
            &= \kernel(\location,\location) - 2 \kernel(\location, \locations_\obsIdx)(\Kernel_\obsIdx + \sigma_\obsnoise^2\eye)^{-1}\kernel(\locations_\obsIdx,\location) +  \vct\alpha_\obsIdx^\transpose(\Kernel_\obsIdx+\sigma_\obsnoise^2\eye)\vct\alpha_\obsIdx\\
			&= \kernel(\location,\location) - 2 \kernel(\location, \locations_\obsIdx)(\Kernel_\obsIdx + \sigma_\obsnoise^2\eye)^{-1}\kernel(\locations_\obsIdx,\location) +  \kernel(\location, \locations_\obsIdx)(\Kernel_\obsIdx + \sigma_\obsnoise^2\eye)^{-1}\kernel(\locations_\obsIdx,\location)\\
			&=\kernel(\location,\location) - \kernel(\location, \locations_\obsIdx)(\Kernel_\obsIdx + \sigma_\obsnoise^2\eye)^{-1}\kernel(\locations_\obsIdx,\location)\\
			&= \sigma_\obsIdx^2(\location)\,,
		\end{split}
	\end{equation}
	which concludes the proof.
\end{proof}

\subsection{Convergence guarantees}
\label{sec:convergence}
Now we apply the error bounds above to the general optimization problem in \eqref{eq:prob-general}.

\begin{assumption}
	\label{a:param-seq}
	The sequence of parameters $\{\vipp_\iterIdx\}_{\iterIdx=1}^\infty$ is a Cauchy sequence, i.e.:
	\begin{equation}
		\forall \stepsize > 0, \quad \exists \nIterations_\stepsize \in \sN:\quad \norm{\vipp_{\iterIdx+1} - \vipp_{\iterIdx}} \leq \stepsize, \quad \forall \iterIdx \geq \nIterations_\stepsize\,.
	\end{equation}
\end{assumption}
The assumption above can be guaranteed by, e.g., diminishing step sizes during (stochastic) gradient descent. It essentially means that $\vipp_\iterIdx$ will converge to some $\hat\vipp\in\vpspace\subseteq\sR^{n_\vipp}$, though not requiring it to be the optimum.

\begin{assumption}
    \label{a:prior}
    The prior $\prior(\mdl)$ has full support over $\mdlspace$. 
\end{assumption}
Such assumption ensures that the prior would not wrongly assign zero probability to plausible models.

\begin{lemma}
	\label{thr:variance-convergence}
	Let assumptions \ref{a:k-continuous} to \ref{a:prior} hold, and set $\beta_\iterIdx= \beta > 0$, for all $\iterIdx\in \{0, 1, 2, \dots\}$. Then the following almost surely holds:
	\begin{equation}
		\sigma_\iterIdx^2(\mdl, \vipp_\iterIdx) \in \bigo(\iterIdx^{-1}), \quad \forall \mdl \in \mdlspace\,.
	\end{equation}
\end{lemma}
\begin{proof}
	Consider the following upper bound on the predictive variance of a GP model \cite[Lem. D.3]{Steinberg2024Variational}:
	\begin{equation}
		\forall \iterIdx\in \sN, \quad \sigma_\iterIdx^2(\mdl, \vipp) \leq \frac{\sigma_\obsnoise^2\sigma_0^2(\mdl, \vipp)}{\sigma_\obsnoise^2 + \sigma_0^2(\mdl,\vipp)\obscount_\iterIdx(\mdl, \vipp)}, \quad \forall (\mdl, \vipp) \in \mdlspace\times\vpspace\,,
		\label{eq:variance-bound}
	\end{equation}
	where $\obscount_\iterIdx(\mdl, \vipp)$ denotes the number of observations collected at $(\mdl, \vipp) \in \mdlspace\times\vpspace$ up to time $\iterIdx\geq 1$. In addition, letting $\history_\iterIdx$ denote the $\sigma$-algebra generated by the history of all random variables measurable at time $\iterIdx$, and setting $\hat\vipp := \lim_{\iterIdx\to\infty} \vipp_\iterIdx$, the second Borel-Cantelli lemma \citep{Dubins1965Sharper} tells us that:\footnote{More precisely, the second Borell-Cantelli lemma shows that the two sides of \autoref{eq:count-limit} are proportional to each other, while equality holds if the right-hand side diverges.}
	\begin{equation}
		\forall\mdl \in \mdlspace, \quad \lim_{\iterIdx\to\infty} \obscount_\iterIdx(\mdl, \hat\vipp) = \lim_{\iterIdx\to\infty} \sum_{i=1}^\iterIdx \Prob\left[ \mdl_i = \mdl \mid \history_{i-1} \right]\,.
		\label{eq:count-limit}
	\end{equation}
	Therefore, for $\sigma_\iterIdx^2(\mdl, \hat\vipp)\to 0$, we need the series above to diverge. To ensure the latter, we can show that the conditional probabilities in \cref{eq:cond-probs} have a nonzero lower bound or, if they converge to zero, that they do so slowly enough.

	We now derive a lower bound on the sampling probabilities. First, observe that:
	\begin{equation}
		\forall\iterIdx\in \sN, \quad \Ex\left[\norm{\gpMean_\iterIdx(\cdot, \vipp_\iterIdx)}_\infty\right]
		= \Ex[\norm{\Ex[\objective(\cdot, \vipp_\iterIdx) \mid \history_\iterIdx]}_\infty] 
		\leq \Ex[\Ex[\norm{\objective(\cdot, \vipp)}_\infty \mid \history_\iterIdx ] ],
		\quad \forall \vipp \in \vpspace\,,
	\end{equation}
	where  $\norm{\objective(\cdot, \vipp)}_\infty = \sup_{\mdl\in\mdlspace} |\objective(\mdl, \vipp)|$ denotes the supremum norm of $\objective(\cdot,\vipp)$, and we applied Jensen's inequality in the last step. Since the kernel $\kernel$ is continuous and bounded, the sub-Gaussian parameter $\sigma_\iterIdx^2(\cdot,\vipp_\iterIdx)$ has a maximum in $\mdlspace$, which is finite. As the expected value of the maximum of a finite collection of sub-Gaussian random variables is bounded \cite[see, e.g.,][Thr. 2.5]{Boucheron2013}, it follows that the GP mean $\gpMean_\iterIdx$ is almost surely bounded at all times (by, e.g., Markov's inequality). Considering the model sampling probabilities and that $\prior_{\min} := \min_{\mdl \in \mdlspace} \prior(\mdl) > 0$ by \autoref{a:prior}, we then have that the following almost surely holds:
	\begin{equation}
		\begin{split}
			\forall\iterIdx \geq 0, \quad \Prob\left[ \mdl_{\iterIdx+1}
			= \mdl \mid \history_{\iterIdx} \right] = \vim_{\ucb, {\iterIdx}}(\mdl)
			&\geq \frac{
				\prior_{\min} \exp(-\norm{\gpMean_\iterIdx(\cdot, \vipp_\iterIdx)}_\infty + \beta\sigma_\iterIdx(\mdl, \vipp_\iterIdx))
			}{
				\sum_{\mdl'\in\mdlspace} \exp(\norm{\gpMean_\iterIdx(\cdot, \vipp_\iterIdx)}_\infty + \beta\sigma_\iterIdx(\mdl', \vipp_\iterIdx))
			}\\
			&\geq \frac{
				\prior_{\min} \exp(-2\norm{\gpMean_\iterIdx(\cdot, \vipp_\iterIdx)}_\infty + \beta\sigma_\iterIdx(\mdl, \vipp_\iterIdx))
			}{
				|\mdlspace| \max_{\mdl'\in\mdlspace}\exp(\beta\sigma_\iterIdx(\mdl', \vipp_\iterIdx)) 
			}\,.
		\end{split}
		\label{eq:cond-probs}
	\end{equation}
	As, for every $\mdl\in\mdlspace$, the sequence $\{\sigma_\iterIdx^2(\mdl,\vipp_\iterIdx)\}_{\iterIdx=0}^\infty$ is non-negative and non-increasing, it has a limit by the monotone convergence theorem. 
	Let $\sigma_* :=  \lim_{\iterIdx\to\infty} \max_{\mdl\in\mdlspace}\sigma_\iterIdx(\mdl, \vipp_\iterIdx)$, and let $\mdl_* \in \mdlspace$ be one of the maximizers of $\lim_{\iterIdx\to\infty} \sigma_\iterIdx(\cdot, \vipp_\iterIdx)$. If $\sigma_* > 0$, by \autoref{eq:cond-probs}, we have for $\mdl_*$ that:
	\begin{equation}
		\begin{split}
			\lim_{\iterIdx\to\infty} \Prob\left[ \mdl_{\iterIdx+1} = \mdl_* \mid \history_{\iterIdx} \right]
			&\geq
			\lim_{\iterIdx\to\infty} 
			\frac{
				\prior_{\min} \exp(-2\norm{\gpMean_\iterIdx(\cdot, \vipp_\iterIdx)}_\infty+\beta\sigma_\iterIdx(\mdl_*, \vipp_\iterIdx))
			}{
				|\mdlspace| \max_{\mdl\in\mdlspace}\exp(\beta\sigma_\iterIdx(\mdl, \vipp_\iterIdx))
			}\\
			&= 
			\lim_{\iterIdx\to\infty} 
			\frac{
				\prior_{\min} \exp(-2\norm{\gpMean_\iterIdx(\cdot, \vipp_\iterIdx)}_\infty+\beta\sigma_*)
			}{
				|\mdlspace| \exp(\beta\sigma_*)
			}\\
			&=
			\lim_{\iterIdx\to\infty} \frac{\prior_{\min} \exp(-2\norm{\gpMean_\iterIdx(\cdot, \vipp_\iterIdx)}_\infty)}{|\mdlspace|}\\
			&\geq \frac{\prior_{\min} \exp(-2\Ex[\norm{\objective(\cdot, \hat\vipp)}_\infty \mid \history_\infty])}{|\mdlspace|}\\
			&=: \bound_\mdl > 0\,,
		\end{split}
		\label{eq:cond-probs-lb}
	\end{equation}
	which implies $\obscount_\iterIdx(\mdl_*, \hat\vipp)\to\infty$ by \autoref{eq:count-limit}. However, in that case, we must have $\sigma_*^2 = \lim_{\iterIdx\to\infty} \sigma_\iterIdx^2(\mdl_*, \vipp_\iterIdx) = 0$ by \autoref{eq:variance-bound}, which is a contradiction. Therefore, $\sigma_* = 0$, and consequently $\lim_{\iterIdx\to\infty} \sigma_\iterIdx(\mdl, \vipp_\iterIdx) \leq \sigma_* = 0$, for all $\mdl\in\mdlspace$.
	
	Finally, we show that $\sigma_\iterIdx^2(\cdot,\vipp_\iterIdx)\in\bigo(\iterIdx^{-1})$. As we have seen that $\lim_{\iterIdx\to\infty} \sigma_\iterIdx(\cdot, \vipp_\iterIdx) = 0$ above, applying the limit to \eqref{eq:cond-probs}, we see that $\Prob\left[ \mdl_\iterIdx = \mdl \mid \history_{\iterIdx-1} \right] \to \bound_\mdl > 0$,  for each $\mdl\in\mdlspace$. Hence, $\obscount_\iterIdx(\mdl,\vipp_\iterIdx)^{-1} \in \bigo(\iterIdx^{-1})$, implying that $\sigma_\iterIdx^2(\cdot, \vipp_\iterIdx)$ is $\mc{O}(\iterIdx^{-1})$ asymptotically by \autoref{eq:count-limit}, which concludes the proof.
\end{proof}

\begin{definition}
    Let $\{\xi_\iterIdx\}_{\iterIdx\in\sN}$ be a real-valued stochastic process. We say that $\xi_\iterIdx \in \bigo_{\Prob}(\anyfunction(\iterIdx))$, for a positive function $\anyfunction:\sN\to(0,\infty)$, if:
    \begin{equation}
    	\forall \varepsilon > 0, \quad \exists \anyconstant_\varepsilon \in (0, \infty), \obscount_\varepsilon\in\sN: \quad \Prob\left[\frac{|\xi_\iterIdx|}{\anyfunction(\iterIdx)} > \anyconstant_\varepsilon\right] \leq \varepsilon, \quad \forall \iterIdx \geq \obscount_\varepsilon\,,
    \end{equation}
    or equivalently that:
    \begin{equation}
    	\lim_{\anyconstant\to\infty} \limsup_{\iterIdx\to\infty} \Prob\left[\frac{|\xi_\iterIdx|}{\anyfunction(\iterIdx)} > \anyconstant\right] = 0\,.
    \end{equation}
\end{definition} 

\begin{theorem}
    \label{thr:kl-bound}
    Under the assumptions in \cref{thr:variance-convergence}, we have that the following holds in probability: 
    \begin{equation}
        \dkl(\vim_{\ucb, \iterIdx} || \vim_{\objective, \vipp_\iterIdx}^*) \in \bigo_{\Prob}(\iterIdx^{-1/2})\,. 
    \end{equation}
\end{theorem}
\begin{proof}
    Expanding from the definition of the KL divergence and the variational distributions, we have that:
    \begin{equation}
    	\begin{split}
    		 \iterIdx\geq 0, \quad \dkl(\vim_{\ucb, \iterIdx} || \vim_{\objective, \vipp_\iterIdx}^*) &= \Ex_{\mdl\sim\vim_{\ucb, \iterIdx}}\left[\log \vim_{\ucb, \iterIdx}(\mdl) - \log \vim_{\objective,\vipp_\iterIdx}^*(\mdl)\right]\\
    		 &=\Ex_{\mdl\sim\vim_{\ucb, \iterIdx}} [ \ucb_\iterIdx(\mdl) - \objective(\mdl, \vipp_\iterIdx) ]\\
             &\quad + \log \left(\sum_{\mdl'\in\mdlspace}  \prior(\mdl') \exp \objective(\mdl', \vipp_\iterIdx) \right) - \log \left(\sum_{\mdl'\in\mdlspace}  \prior(\mdl') \exp \ucb_\iterIdx(\mdl') \right) \,.
    		 \label{eq:kl-error-expansion}
    	\end{split}
    \end{equation}
    Under assumptions \ref{a:k-continuous} and \ref{a:sub-g-noise}, 
    given any $\beta>0$, applying standard sub-Gaussian concentration results \citep{Boucheron2013} and a union bound, we have that, for all $\iterIdx\geq 0$:
    \begin{equation}
    	\begin{split}
	    	\Prob[ \exists \mdl\in\mdlspace:\:|\objective(\mdl, \vipp_\iterIdx) - \gpMean_\iterIdx(\mdl, \vipp_\iterIdx)| > \beta\sigma_\iterIdx(\mdl, \vipp_\iterIdx) ]
	    	&\leq \sum_{\mdl\in\mdlspace} \Prob[|\objective(\mdl, \vipp_\iterIdx) - \gpMean_\iterIdx(\mdl, \vipp_\iterIdx)| > \beta\sigma_\iterIdx(\mdl, \vipp_\iterIdx)]\\
	    	&\leq 2|\mdlspace|\exp\left(-\frac{\beta^2}{2}\right)\\
	    	&=: \delta_\beta\,.
	    \end{split}
    \end{equation}
    With probability at least $1-\delta_\beta$, it then follows that:
    \begin{align}
    	 \ucb_\iterIdx(\mdl) - \objective(\mdl, \vipp_\iterIdx) = \gpMean_\iterIdx(\mdl, \vipp_\iterIdx) + \beta\sigma_\iterIdx(\mdl, \vipp_\iterIdx) - \objective(\mdl, \vipp_\iterIdx) &\leq 2\beta\sigma_\iterIdx(\mdl,\vipp_\iterIdx)\\
    	 \log \left(\sum_{\mdl'\in\mdlspace}  \prior(\mdl') \exp \objective(\mdl', \vipp_\iterIdx) \right) - \log \left(\sum_{\mdl'\in\mdlspace}  \prior(\mdl') \exp \ucb_\iterIdx(\mdl') \right) &\leq 0
	\end{align}
    for all $\mdl\in\mdlspace$. Hence, with the same probability, it holds that:
    \begin{equation}
    	\forall\iterIdx\geq 0, \quad \dkl(\vim_{\ucb, \iterIdx} || \vim_{\objective, \vipp_\iterIdx}^*) \leq 2\beta \Ex_{\mdl\sim\vim_{\ucb,\iterIdx}}[\sigma_\iterIdx(\mdl, \vipp_\iterIdx)] \leq 2\beta \norm{\sigma_\iterIdx(\cdot, \vipp_\iterIdx)}_\infty\,.
    \end{equation}
    By \cref{thr:variance-convergence}, we know that $\sigma_\iterIdx(\mdl, \vipp_\iterIdx) \in \bigo(\iterIdx^{-1/2})$, so that there exists $\anyconstant > 0$ such that $\sigma_\iterIdx(\mdl, \vipp_\iterIdx) \leq \anyconstant\iterIdx^{-1/2}$, for all $\mdl\in\mdlspace$. We then have that:
    \begin{equation}
    	\begin{split}
    		\lim_{\beta\to\infty} \limsup_{\iterIdx\to\infty} \Prob\left[\frac{\dkl(\vim_{\ucb, \iterIdx} || \vim_{\objective, \vipp_\iterIdx}^*)}{\anyconstant\iterIdx^{-1/2}} > 2\beta\right] &\leq \lim_{\beta\to\infty} \limsup_{\iterIdx\to\infty}\Prob\left[\dkl(\vim_{\ucb, \iterIdx} || \vim_{\objective, \vipp_\iterIdx}^*) > 2\beta\norm{\sigma_\iterIdx(\cdot, \vipp_\iterIdx)}_\infty\right]\\
    		&\leq \lim_{\beta\to\infty} 2|\mdlspace|\exp\left(-\frac{\beta^2}{2}\right) = 0\,,
    	\end{split}
    \end{equation}
    which concludes the proof.
\end{proof}

\begin{remark}
    The result in \autoref{thr:kl-bound} is similar to Corollary 1 in \citet{oliveiranoregret2021}, which also derives the concentration bound for the KL divergence between a surrogate-based approximation of a posterior and the true posterior. However, \citeauthor{oliveiranoregret2021}'s result only provides an asymptotic convergence rate requires an upper bound on the information gain of the surrogate model of order $o(\sqrt{t})$ and an appropriately scaled UCB parameter $\beta_\iterIdx$, whereas our result shows that we do not need either of these assumptions whenever a sampling lower bound can be guaranteed, i.e., $\inf_{\iterIdx\in\sN, \mdl\in\mdlspace} \Prob[\mdl_{\iterIdx+1} = \mdl | \history_\iterIdx] > 0$. In addition, \citet{oliveiranoregret2021} only deals with the static setting where the target posterior does not change over time, while in our case we have a changing $\vipp_\iterIdx$ that leads to different targets per optimization step. This non-stationarity requires additional care with the convergence analysis.
\end{remark}

\newpage\section{Bijective equivalence between discrete distributions}\label{apdx:categoricaldiscreteproof}
\begin{proposition}
    \label{prop:unique_categorical_representation}
    Every finite discrete distribution over a finite support \( \mdlspace = \{\mdl_1, \mdl_2, \ldots, \mdl_k\} \) has a unique representation as a categorical distribution. Specifically, there exists a bijective mapping between the set of all finite discrete distributions on \( \mdlspace \) and the set of categorical distributions parameterized by probability vectors \( \boldsymbol{\catmdlparam} \) over \( \mdlspace \).
\end{proposition}
\begin{proof}
    Let \( \mathcal{P} \) denote the set of all finite discrete distributions over \( \mdlspace \), and let \( \mathcal{C} \) denote the set of categorical distributions parameterized by \( \boldsymbol{\theta} \).

    \textbf{Injectivity:} Suppose \( \boldsymbol{\theta} \) and \( \boldsymbol{\phi} \) are two distinct probability vectors in \( \mathcal{C} \). Then, there exists at least one index \( i \) such that \( \theta_i \neq \phi_i \). Consequently, the corresponding distributions assign different probabilities to \( \mdl_i \), implying \( \boldsymbol{\theta} \neq \boldsymbol{\phi} \).

    \textbf{Surjectivity:} For any finite discrete distribution \( \mathbf{p} \in \mathcal{P} \), define \( \boldsymbol{\theta} = \mathbf{p} \). Since \( \mathbf{p} \) satisfies \( \theta_i \geq 0 \) and \( \sum_{i=1}^{k} \theta_i = 1 \), \( \boldsymbol{\theta} \) is a valid parameterization in \( \mathcal{C} \). Thus, every \( \mathbf{p} \) corresponds to some \( \boldsymbol{\theta} \).

    Since the mapping is both injective and surjective, it is bijective. Therefore, every finite discrete distribution has a unique categorical distribution representation.
\end{proof}
\newpage\section{Monte Carlo gradients via score function estimation}\label{apdx:mcgsfe}

An alternative to the reparameterization trick is the score function estimator (SFE), which circumvents the issue of non-differentiable samples from discrete distributions by using the log trick to compute the gradients of a function with respect to variational parameters. In the case of the distribution of models, we have the identity
\begin{align}
    \nabla_{\vimkp}q_{\vimkp}(\mdl) = q_{\vimkp}(\mdl) \nabla_{\vimkp}\log q_{\vimkp}(\mdl).
\end{align}
By the Leibniz integral rule, the gradient of the expectation in \eqref{eq:varloss_by_mixture} with respect to the parameters of the discrete distribution is
\begin{align*}
    \nabla_\vimkp \mc{L}(\vipp,\vimkp)
    &=
    \nabla_\vimkp \Ex_{\mdl\sim\vimkq}
            \left[\ell(\mdl) \right] +
    \nabla_\vimkp\Ex_{\mdl\sim\vimkq}\left[\log\dfrac{q_\vimkp(\mdl)}{\prior(\mdl)}\right]\\
    &=
    \Ex_{\mdl\sim\vimkq}
            \left[ \ell(\mdl) \nabla_\vimkp \log\vimkq(\mdl) \right]
    +
    \Ex_{\mdl\sim\vimkq}\left[\log\dfrac{q_\vimkp(\mdl)}{\prior(\mdl)}\nabla_\vimkp \log\vimkq(\mdl)\right].
\end{align*}
In practice, the variance of this estimator can be very high when the batch size is not large. However, there are techniques to reduce this variance for general applications. The simplest of which is to use a control variate $\controlvariate$ in the form
\begin{align}
    \nabla_\vimkp \mc{L}(\vipp,\vimkp)
    &=
    \Ex_{\mdl\sim\vimkq}
        \left[
            \Ex_{\vct{z}\sim \nu_\dmax}
                \left[
                    h(\vimkp,\vipp,\refvars)-\varsigma
                \right]
                \nabla_\vimkp\log\vimkq(\mdl)
            \right].
\end{align}
By simply choosing $\controlvariate=\Ex_{t\in\{1,\dots,T\}}\left[\mc{L}(\vipp,\vimkp)\right]$, where the expectation is estimated online over the iterations of the optimizer, we can reduce variance of $\nabla_\vimkp\mc{L}(\vipp,\vimkp)$. See Appendix~\ref{apdx:sfecontrolvariate} for implementation details.

\subsection{Control variate for score function estimator}\label{apdx:sfecontrolvariate}

\newcommand{\mcgcvbeta}{\beta}              
\newcommand{\mcgcvB}{B}                     
\newcommand{\mcgcvIter}{t}                  
\newcommand{\mcgcvLoss}{\ell}               
\newcommand{\mcgcvBiased}{\tilde{\mu}}        
\newcommand{\mcgcvUnbiased}{\mu}              
\newcommand{\mcgcvGrad}{\tilde{\nabla}_{\vimkp}}                  
\newcommand{\mcgcvGradIter}[1]{\tilde{\nabla}_{\vimkp,#1}}                  
\newcommand{\mcgcvpsi}{\vimkp}                
\newcommand{\mcgcvk}{\mdl}                     

We adopt the approach used in \citet{KingmaAdam2017} for obtaining an unbiased running first moment of the loss function. 
At iteration $\mcgcvIter$ we draw $\mcgcvB$ samples
$\{\mcgcvk_{\mcgcvIter,n}\}_{n=1}^{\mcgcvB}$ and compute
\[
  \mcgcvLoss_{\mcgcvIter,n} \;=\; -\widehat{\mc{L}}_{\mcgcvIter,n},
  \qquad n=1,\dots,\mcgcvB.
\]

With fixed decay $\mcgcvbeta\in(0,1)$, update the (biased) first moment exactly as in the approach of the Adam optimizer~\citep{KingmaAdam2017}:
\[
  \mcgcvBiased_{\mcgcvIter}
  \;\gets\;
  \mcgcvbeta\,\mcgcvBiased_{\mcgcvIter-1}
  \;+\;
  (1-\mcgcvbeta)\,
  \bar{\mcgcvLoss}_{\mcgcvIter},
  \qquad
  \bar{\mcgcvLoss}_{\mcgcvIter}
  =\frac{1}{\mcgcvB}\sum_{n=1}^{\mcgcvB}\mcgcvLoss_{\mcgcvIter,n}.
\]

To remove the initialization bias,
\[
  \mcgcvUnbiased_{\mcgcvIter}
  \;\gets\;
  \frac{\mcgcvBiased_{\mcgcvIter}}
       {1-\mcgcvbeta^{\mcgcvIter}}.
\]

Using $\controlvariate_{\mcgcvIter}:=\mcgcvUnbiased_{\mcgcvIter}$ as a baseline,
the Monte Carlo gradient estimator becomes
\[
  \mcgcvGradIter{\mcgcvIter}
  \;\gets\;
  \frac{1}{\mcgcvB}
  \sum_{n=1}^{\mcgcvB}
  \bigl(\mcgcvLoss_{\mcgcvIter,n}-\controlvariate_{\mcgcvIter}\bigr)
  \nabla_{\mcgcvpsi}\log q_{\mcgcvpsi}\!\bigl(\mcgcvk_{\mcgcvIter,n}\bigr).
\]
Because $\controlvariate_{\mcgcvIter}$ is independent of each
$\mcgcvk_{\mcgcvIter,n}$, the estimator remains unbiased while the baseline substantially
reduces its variance.

\subsection{Controlling learning rate via the information gain}\label{apdx:ig_threshold}


When using stochastic gradient descent for optimization over parameters of both $\vimkq$ and $\vipq$, it is necessary to use careful scaling of the estimated gradients to ensure the optimizer does not ``drop off a cliff'' into a local minimum. Such phenomena has been observed in related fields such as proximal policy gradients~\citep{SchulmanProximal2017} where the authors demonstrate empirically such a necessity in reinforcement learning problems. In essence, we want to control the learning rate of $\vimkp$ with respect to the convergence of $\vipp\rightarrow\vipp^*$. We show empirical results for controlling this rate and leave any mathematical properties for the optimal scaling to future research.

One approach is to control the rate of \textit{information gain} (IG) of $\vimkq$ during the simultaneous optimization over both $\vimkp$ and $\vipp$. By assuming a bounded rate of information gain for $\vipq$ (achieved via gradient clipping) we only need to consider computing the IG over successive $q_{\vimkp}^{(t)}$ for steps $t=1,\dots,T$. Defining the IG in terms of entropy, we have
\begin{align}
    \mathrm{IG}(q_{\vimkp}^{(t+1)}\gvn q_{\vimkp}^{(t)}) &= \mathrm{H}(q_{\vimkp}^{(t)}) - \mathrm{H}(q_{\vimkp}^{(t+1)}).
\end{align}
When $\vimkq$ is a categorical distribution, this quantity is available analytically.
However, in general this is not available, but it can be estimated via Monte Carlo integration and importance sampling using available quantities (see Appendix~\ref{apdx:montecarloinformation}). 
We choose to set a threshold for the IG between steps, denoted $\igthreshold$, and then at each successive step $t$ we scale $\nabla_\vimkp$ using an iterative method such as bisection\footnote{In preliminary investigations, other approaches for implementation of this threshold such as constrained optimization and computation of Lagrange multipliers were trialed without success, possibly due to the geometry of the optimization landscape.}.

\subsection{Monte Carlo estimation of information}
\label{apdx:montecarloinformation}

\newcommand{\MADEFN}{\operatorname{NN}_\vimkp}
\newcommand{\MADEFNPRIME}{\operatorname{NN}_{\vimkp'}}
The below procedure assumes $\vimkq$ represents a distribution over strings of Bernoulli variables. Let $\vimkp\in\R^{n_\vimkp}$ parameterize a masked autoencoder that determines logits for a product Bernoulli distribution
\[
\vimkq(\mdl)
=\prod_{i=1}^{d}\sigma\!\bigl(\MADEFN^{(i)}(\mdl)\bigr)^{\mdl^{(i)}}
          \!\bigl[1-\sigma\!\bigl(\MADEFN^{(i)}(\mdl)\bigr)\bigr]^{1-\mdl^{(i)}},
\qquad
\mdl\in\{0,1\}^{d},
\]
with MADE logits $\MADEFN^{(i)}(\,\cdot\,)$ and
$\sigma(\MADEFN)=(1+e^{-\MADEFN})^{-1}$.  
After an SGD proposal $\vimkp'=\vimkp-\alpha\,\nabla_\vimkp$,  
we estimate the entropy reduction
\begin{align}
\Delta\mathcal I(\alpha)
=\mathrm{H}\bigl(\vimkq\bigr)
-\mathrm{H}\bigl(q_{\vimkp'}\bigr),
\qquad
\mathrm{H}(p)=-\!\!\sum_{\mdl}p(\mdl)\log p(\mdl).
\label{eq:apdxigthreshold}
\end{align}

To reduce computation at the expense of introducing some estimation bias, we employ importance weights to re-use the current sample of model indicators in an iterative search to scale the gradient step. Draw a mini-batch
$\{\mdl^{(n)}\}_{n=1}^{N}\overset{\text{i.i.d.}}{\sim}\vimkq$
\emph{once}; no re-sampling is needed afterwards.  
Because the expectation in \eqref{eq:apdxigthreshold} switches from
$\vimkq$ to $q_{\vimkp'}$, rewrite
\begin{align}
\mathrm{H}\bigl(q_{\vimkp'}\bigr)=
-\E_{\mdl\sim \vimkq}
\!\bigl[w_{\vimkp',\vimkp}(\mdl)\,
        \log q_{\vimkp'}(\mdl)\bigr],
\quad
w_{\vimkp',\vimkp}(\mdl):=
\frac{q_{\vimkp'}(\mdl)}{\vimkq(\mdl)}.
\end{align}
For a Bernoulli product the weight factorizes:
\begin{align}
w_{\vimkp',\vimkp}(\mdl)
=\prod_{i=1}^{d}
\frac{\sigma(\MADEFNPRIME^{(i)})^{\mdl^{(i)}}[1-\sigma(\MADEFNPRIME^{(i)})]^{1-\mdl^{(i)}}}
     {\sigma(\MADEFN^{(i)} )^{\mdl^{(i)}}[1-\sigma(\MADEFN^{(i)} )]^{1-\mdl^{(i)}}}
\quad\bigl(\MADEFN^{(i)}:=\MADEFN^{(i)}(\mdl),~
      \MADEFNPRIME^{(i)}:=\MADEFNPRIME^{(i)}(\mdl)\bigr),
\end{align}
implemented stably via
$\frac{\sigma(\MADEFNPRIME^{(i)})}{\sigma(\MADEFN^{(i)})}
=\exp\bigl[\log(1+e^{-\MADEFN^{(i)}})-\log(1+e^{-\MADEFNPRIME^{(i)}})\bigr]$.

The mini-batch estimator is therefore
\begin{align}
\widehat{\mathrm{H}}_{N}(\vimkp')
=-\frac{1}{N}\sum_{n=1}^{N}
w_{\vimkp',\vimkp}\!\bigl(\mdl^{(n)}\bigr)
\log q_{\vimkp'}\!\bigl(\mdl^{(n)}\bigr),\qquad
\widehat{\mathrm{H}}_{N}(\vimkp)
=-\frac{1}{N}\sum_{n=1}^{N}\log \vimkq\!\bigl(\mdl^{(n)}\bigr).
\end{align}

Given a tolerance $\varepsilon>0$, reduce $\alpha\leftarrow0.5\,\alpha$
until
\begin{align}
\bigl|\widehat{\mathrm{H}}_{N}(\vimkp)
       -\widehat{\mathrm{H}}_{N}(\vimkp')\bigr|\le\varepsilon.
\label{eq:apdxhalvestep}
\end{align}
If no $\alpha>10^{-20}$ satisfies \eqref{eq:apdxhalvestep}, discard the update by
setting the gradient to $\vct 0$.
Otherwise, apply the accepted scaled gradient.

\newpage
\section{Robust variable selection example details and additional results}\label{apdx:rvsadditionalresults}

The likelihood  is
\begin{align}
p\bigl(y_i \mid \vct{x}_i,\vct{\beta},\vct{\gamma}\bigr) 
&= 
(1 - \alpha)\,\mathcal{N}\!\Bigl(y_i;\vct{\mu}(\vct{x}_i),\sigma_1^2\Bigr)
\;+\;
\alpha\,\mathcal{N}\!\Bigl(y_i;\vct{\mu}(\vct{x}_i),\sigma_2^2\Bigr),\label{eq:robustvs-likelihood-adpx}
\end{align}
with priors
$p(\gamma) \;=\; 2^{-p}$ and
$p(\vct{\beta}) \;=\;\mathcal{N}(0,\,\sigma_\beta^2\mm{I})$. Each of the parameters in the likelihood are described in Table \ref{tab:robustvsdgpsetup} under the Misspecification:None column. The data generating setup in Table \ref{tab:robustvsdgpsetup} describes three levels of misspecification to induce poor identifiability and thus a posterior that is challenging to fit using simple variational density families, such as mean field inference. This exemplifies the use of normalizing flows for this experiment. While many parameters are shared, some differ strongly between the likelihood and DGP. In particular, notice the difference in $\sigma_1,\sigma_2$. Also, for the highly misspecified DGP, correlation between included covariates $i$ and excluded covariates $j$ is induced by a factor of 
$\rho_{i,j}=0.1$
for a proportion of $j$, making the recovery of the DGP using any inference method a challenging and improbable task. For every data set, $\vct{\beta}$ will be either $\vct{\beta}_1$ or $\vct{\beta}_2$ with probability $0.5$.

\begin{table}[bht]
\centering
\caption{Data generating setup}
\label{tab:robustvsdgpsetup}
\begin{tabular}{|c|c|c|c|}
\hline
\textbf{Parameter} & \multicolumn{3}{c|}{\textbf{Misspecification to likelihood}} \\
\cline{2-4}
& \textbf{None} & \textbf{Mid} & \textbf{High}  \\ 
\hline
Number of data points $|\vct{x}|$ & \multicolumn{3}{c|}{50}   \\ 
 Dimension of $\vct{\beta}$ & \multicolumn{3}{c|}{8}   \\ 
 Dimension of $\vct{\gamma}$ &
\multicolumn{3}{c|}{7}\\ 
 $|\mdlspace|$ &
\multicolumn{3}{c|}{$2^7=128$}  \\
 Probability of inclusion $\mathbb{P}(\gamma_i=1)$ & \multicolumn{3}{c|}{$0.4$} \\
 \cline{2-4}
 Non-outlier $\sigma_1$ & 1 & 2 & 4 \\
 Outlier $\sigma_2$ & 10 & 5 & 4 \\
 \cline{2-3}
 Probability of correlation $\mathbb{P}(\rho_{i,j}>0|\gamma_i=1,\gamma_j=0)$ & \multicolumn{2}{c|}{0} & 0.4 \\
 Total correlation factor $\sum_j \rho_{i,j}$ & \multicolumn{2}{c|}{0} & 0.1 \\
 \cline{2-4}
 $\vct{\beta}_1$ & \multicolumn{3}{c|}{0.5} \\
 \cline{2-4}
 $\vct{\beta}_2$ & 0.5 & \multicolumn{2}{c|}{1.5} \\
 \cline{2-4}
 Outlier probability $\alpha$ & \multicolumn{3}{c|}{0.1} \\
\hline
\end{tabular}
\end{table}

Lastly, during the inference process, we consider two separate experiments for each DGP: a ``focused-prior'' experiment where $\sigma_\beta=1.5$, and a ``wide-prior''  experiment where $\sigma_\beta=10$. These two scenarios cause a significant difference between the inferred reversible jump MCMC model probabilities and the inferred VTI model probabilities, as can be seen in the subsequent figures. 

VTI inference was conducted on a cluster of GPU nodes with mixed Nvidia RTX3090 and H100 cards. On the former we used float32 precision for MLP architectures, the latter used float64.

\subsection{Focused versus wide priors}\label{apdx:robustvsprioreffects}

Each of Figures \ref{fig:robustvssweep_noms_fprior}--\ref{fig:robustvssweep_highms_wprior} is a replicate of Figure \ref{fig:robustvsmodelprob} in the main text, showing a sweep of 10 randomly generated data sets (indicated by different colours) according to the corresponding setup in Table \ref{tab:robustvsdgpsetup} using three different variational families: diagonal Gaussian MLP (a CoSMIC mean-field variational family), a composition of 5 affine masked autoregressive flows each with 5 hidden blocks, and a composition of 4 rational quadratic spline masked autoregressive flows each with 6 hidden blocks. The expressiveness of each variational family increases from left to right in each figure.

In the $\sigma_\beta=1.5$ focused prior setting (Figures \ref{fig:robustvssweep_noms_fprior}, 
\ref{fig:robustvssweep_midms_fprior},
\ref{fig:robustvssweep_highms_fprior}) 
performance is generally good, as per Figure \ref{fig:robustvsmodelprob} in the main text: (i) the model probability estimates (top row) tend to move closer to the $y=x$ line as the expressiveness of the variational family increases (left to right plots); (ii) the slight S-shape of the model probability estimates around the $y=x$ line is easily interpretable as the the variational objective $\mc{L}(\vimkp,\vipp)$ (\eqref{eq:varloss_by_mixture}) will naturally favour models with higher posterior model probability over those with lower probabilities; (iii) the true data generating process models (triangles) are generally given high posterior model probabilities; and (iv) individual model posteriors are better estimated for higher probability models (negative slope on the bottom rows).

For the $\sigma_\beta=10$ wide prior setting (Figures \ref{fig:robustvssweep_noms_wprior},
\ref{fig:robustvssweep_midms_wprior},
\ref{fig:robustvssweep_highms_wprior}) performance at first glance appears much worse, particularly in terms of estimating model probabilities. However, on closer inspection this is not the case. It is well known (e.g. \cite{Gelman2020Holes}) that the marginal likelihood (a.k.a. model evidence; a component of the posterior model probability) can be highly sensitive to diffuse priors. In such cases (as with $\sigma_\beta=10$) the posterior will tend to unreasonably favour those models with fewer parameters, and particularly (in the case of regression models) the null model with no predictors, even in the presence of a very clear relationship between predictors and response. This effect can be clearly seen in Figures \ref{fig:robustvssweep_noms_wprior},
\ref{fig:robustvssweep_midms_wprior},
\ref{fig:robustvssweep_highms_wprior} (top row), where the null model (indicated by a circle) is given far higher posterior model probability on the $\pi(\mdl)$ axis than the actual data generating process (triangles). In contrast, the true data generating process (triangles) is generally given a high posterior model probability (comparable with the focused prior setting in Figures \ref{fig:robustvssweep_noms_fprior}, 
\ref{fig:robustvssweep_midms_fprior},
\ref{fig:robustvssweep_highms_fprior}) under the VTI approximation. From these results we conclude that: (i) the posterior model probabilities that depend on the marginal likelihood (i.e., the estimates of $\pi(\mdl)$ on the $x$-axis) have been affected by the wide prior to unreasonably favour models with less parameters; (ii) the VTI-based posterior model probability estimates suggest that they are less sensitive to the undesirable effects of this prior; and (iii) in combination the resulting plots in Figures \ref{fig:robustvssweep_noms_wprior},
\ref{fig:robustvssweep_midms_wprior},
\ref{fig:robustvssweep_highms_wprior} (top row) only appear to indicate worse performance of VTI compared to the gold standard than is actually the case.

\subsection{Within model comparison}

Figure \ref{fig:robustvsmultivariate} illustrates a typical comparison between the reversible jump MCMC estimated posterior distribution and the VTI approximation. The figure shows the posterior of the data generating process model from the first  high misspecification dataset in Figure \ref{fig:robustvsmodelprob} (main text). While there are some small differences, the main features of the posterior appear to be well captured.

\begin{figure}[H]
    \centering
    \begin{overpic}[width=\linewidth]{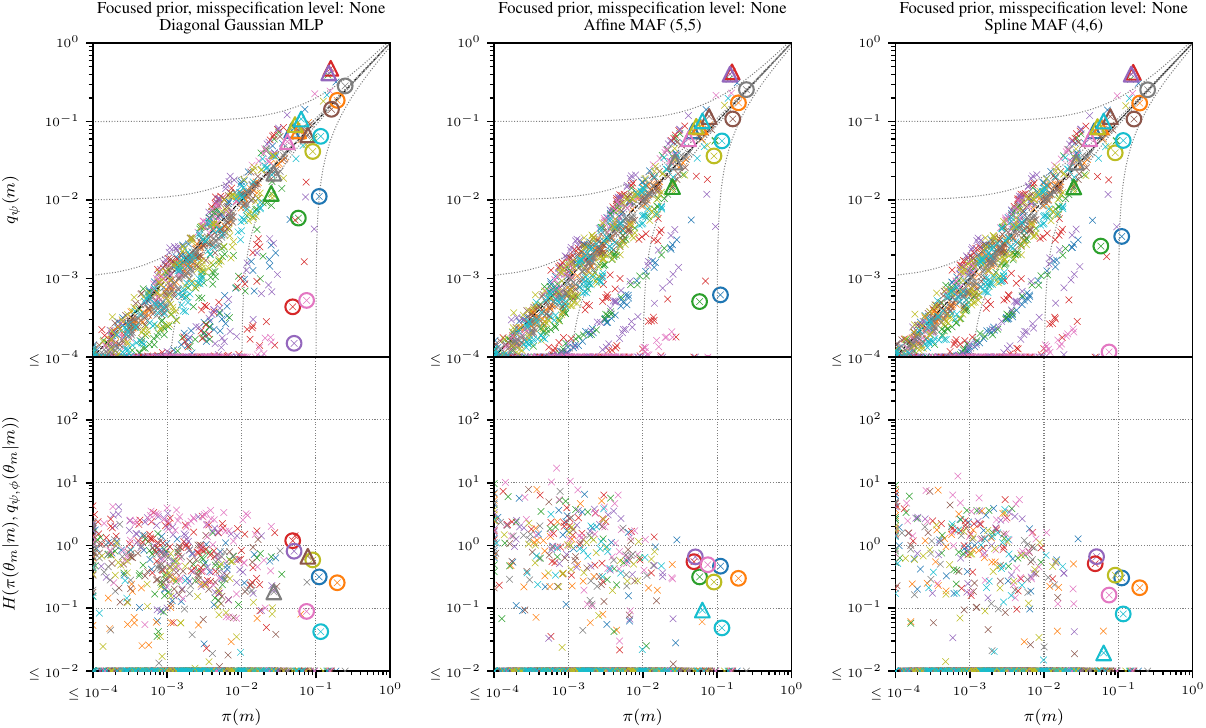}
    \put(0,7){%
      \colorbox{white}{%
        \rotatebox{90}{\fontsize{5}{5}\selectfont\bfseries $H(\target(\theta_\mdl\mid\mdl),\vipq(\theta_\mdl\mid\mdl))$}%
      }%
    }
    \end{overpic}
    \caption{As Figure \ref{fig:robustvsmodelprob} (main text), but under:
    {\em no misspecification ($\sigma_1=1,\sigma_2=10$), focused prior ($\sigma_\beta=1.5$)}. 
    Circles indicate the null model (constant only, no predictors); triangles indicate the data generating process.}
    \label{fig:robustvssweep_noms_fprior}
\end{figure}

\begin{figure}[H]
    \centering
    \begin{overpic}[width=\linewidth]{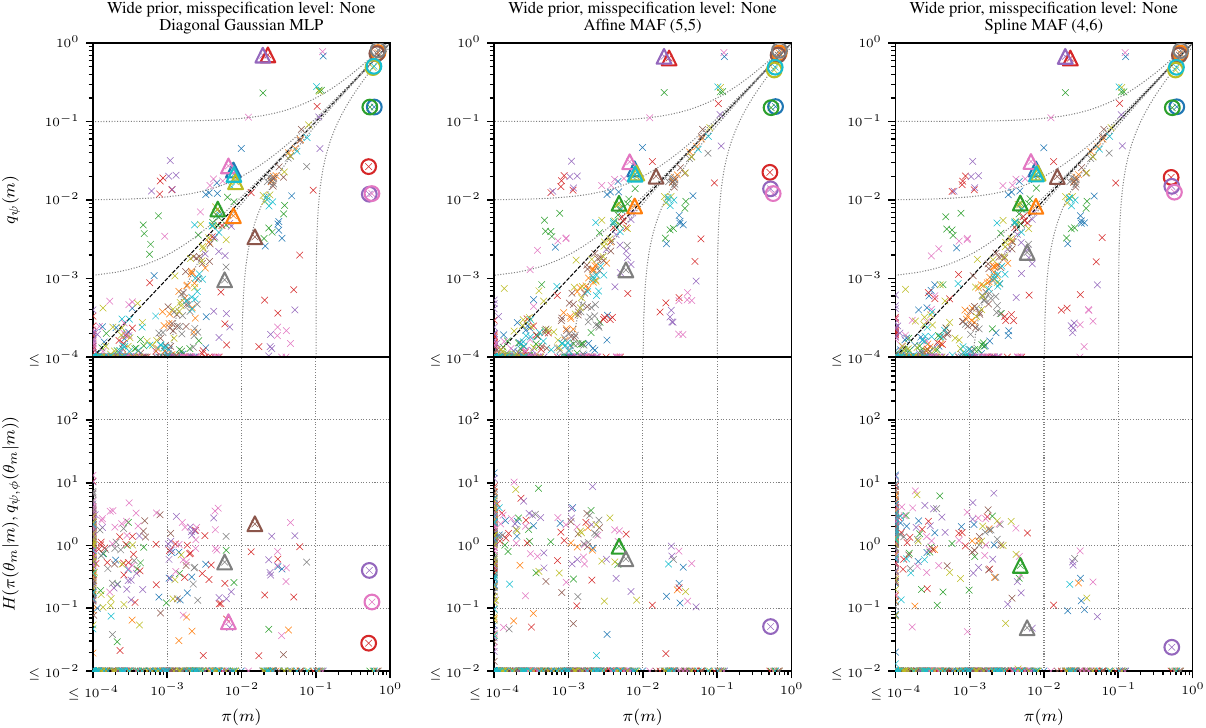}
    \put(0,7){%
      \colorbox{white}{%
        \rotatebox{90}{\fontsize{5}{5}\selectfont\bfseries $H(\target(\theta_\mdl\mid\mdl),\vipq(\theta_\mdl\mid\mdl))$}%
      }%
    }
    \end{overpic}
    \caption{As Figure \ref{fig:robustvsmodelprob} (main text), but under:
    {\em no misspecification ($\sigma_1=1,\sigma_2=10$), wide prior ($\sigma_\beta=10$)}. 
    Circles indicate the null model (constant only, no predictors); triangles indicate the data generating process.
    }
    \label{fig:robustvssweep_noms_wprior}
\end{figure}

\begin{figure}[H]
    \centering
    \begin{overpic}[width=\linewidth]{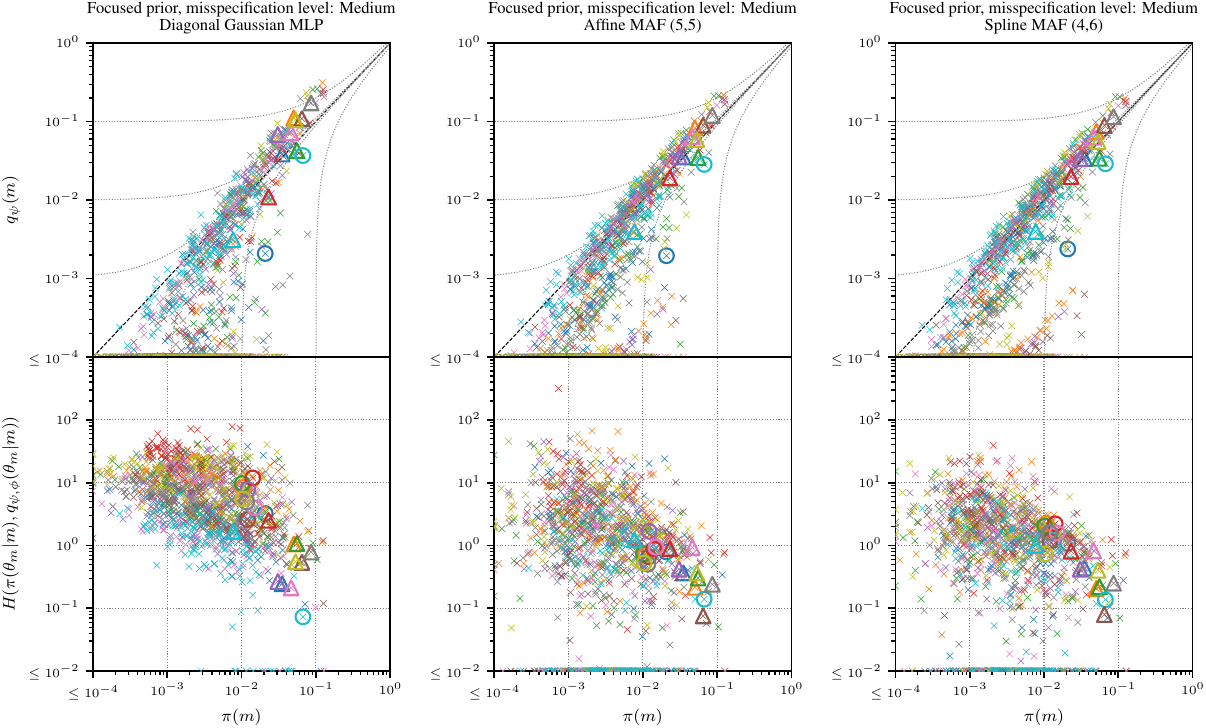}
    \put(0,7){%
      \colorbox{white}{%
        \rotatebox{90}{\fontsize{5}{5}\selectfont\bfseries $H(\target(\theta_\mdl\mid\mdl),\vipq(\theta_\mdl\mid\mdl))$}%
      }%
    }
    \end{overpic}
    \caption{As Figure \ref{fig:robustvsmodelprob} (main text), but under:
    {\em mid misspecification ($\sigma_1=2,\sigma_2=5$), focused prior ($\sigma_\beta=1.5$)}. 
    Circles indicate the null model (constant only, no predictors); triangles indicate the data generating process.
    }
    \label{fig:robustvssweep_midms_fprior}
\end{figure}

\begin{figure}[H]
    \centering
    \begin{overpic}[width=\linewidth]{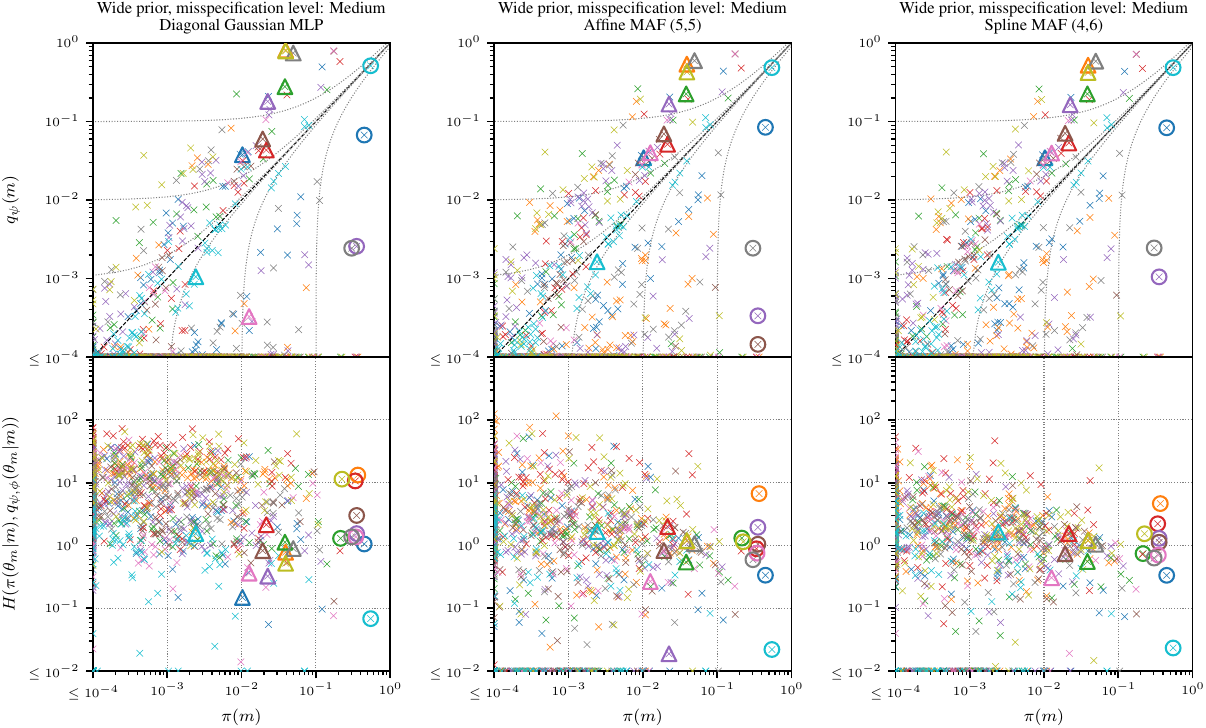}
    \put(0,7){%
      \colorbox{white}{%
        \rotatebox{90}{\fontsize{5}{5}\selectfont\bfseries $H(\target(\theta_\mdl\mid\mdl),\vipq(\theta_\mdl\mid\mdl))$}%
      }%
    }
    \end{overpic}
    \caption{
    As Figure \ref{fig:robustvsmodelprob} (main text), but under:
    {\em mid misspecification ($\sigma_1=2,\sigma_2=5$), wide prior ($\sigma_\beta=10$)}. 
    Circles indicate the null model (constant only, no predictors); triangles indicate the data generating process.
    }
    \label{fig:robustvssweep_midms_wprior}
\end{figure}

\begin{figure}[H]
    \centering
    \begin{overpic}[width=\linewidth]{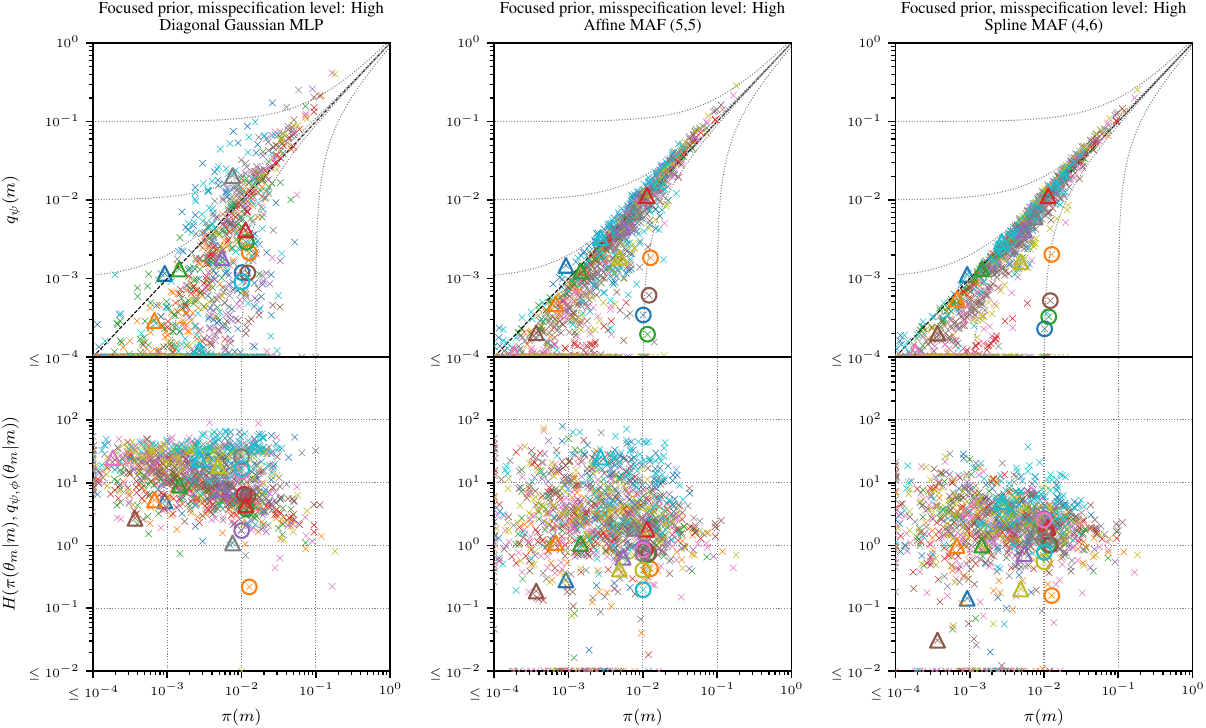}
    \put(0,7){%
      \colorbox{white}{%
        \rotatebox{90}{\fontsize{5}{5}\selectfont\bfseries $H(\target(\theta_\mdl\mid\mdl),\vipq(\theta_\mdl\mid\mdl))$}%
      }%
    }
    \end{overpic}
    \caption{
    As Figure \ref{fig:robustvsmodelprob} (main text), but under:
    {\em high misspecification ($\sigma_1=4,\sigma_2=4$), focused prior ($\sigma_\beta=1.5$)}. 
    Circles indicate the null model (constant only, no predictors); triangles indicate the data generating process.
    }
    \label{fig:robustvssweep_highms_fprior}
\end{figure}

\begin{figure}[H]
    \centering
    \begin{overpic}[width=\linewidth]{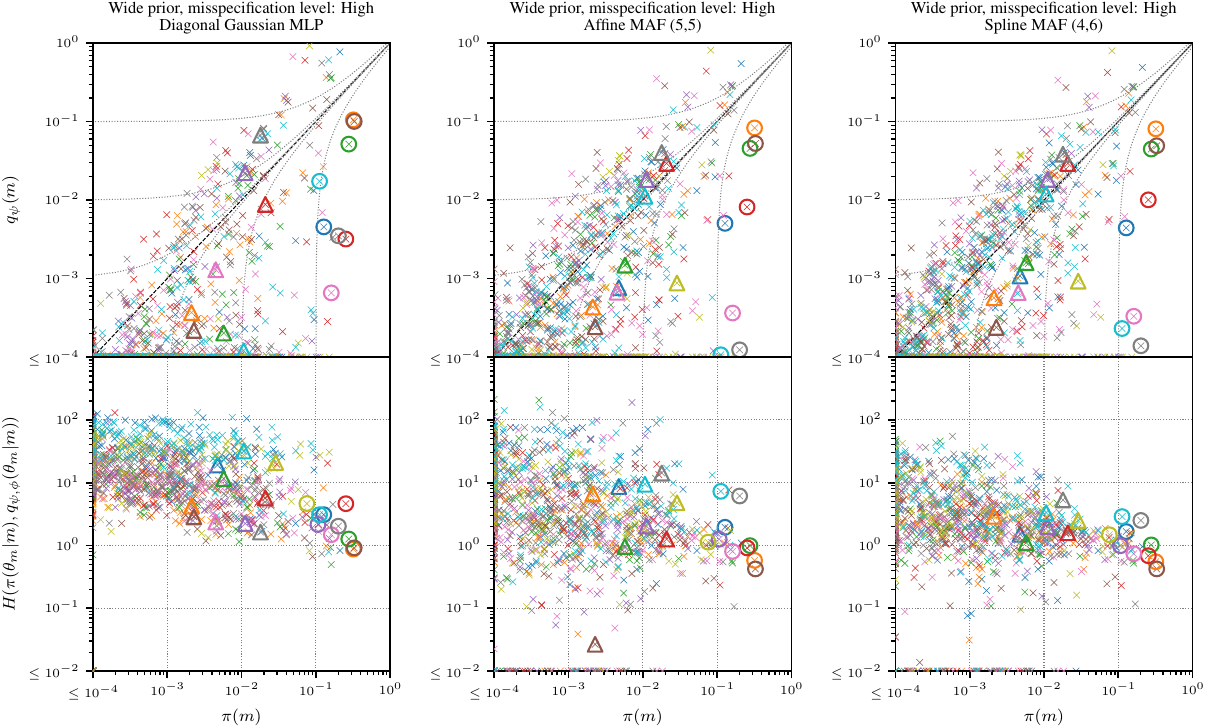}
    \put(0,7){%
      \colorbox{white}{%
        \rotatebox{90}{\fontsize{5}{5}\selectfont\bfseries $H(\target(\theta_\mdl\mid\mdl),\vipq(\theta_\mdl\mid\mdl))$}%
      }%
    }
    \end{overpic}
    \caption{As Figure \ref{fig:robustvsmodelprob} (main text), but under:
    {\em high misspecification ($\sigma_1=4,\sigma_2=4$), wide prior ($\sigma_\beta=10$)}. 
    Circles indicate the null model (constant only, no predictors); triangles indicate the data generating process.
    }
    \label{fig:robustvssweep_highms_wprior}
\end{figure}

\begin{figure}[H]
    \centering
    \includegraphics[width=0.49\linewidth, trim={0 0 0 1cm}, clip]{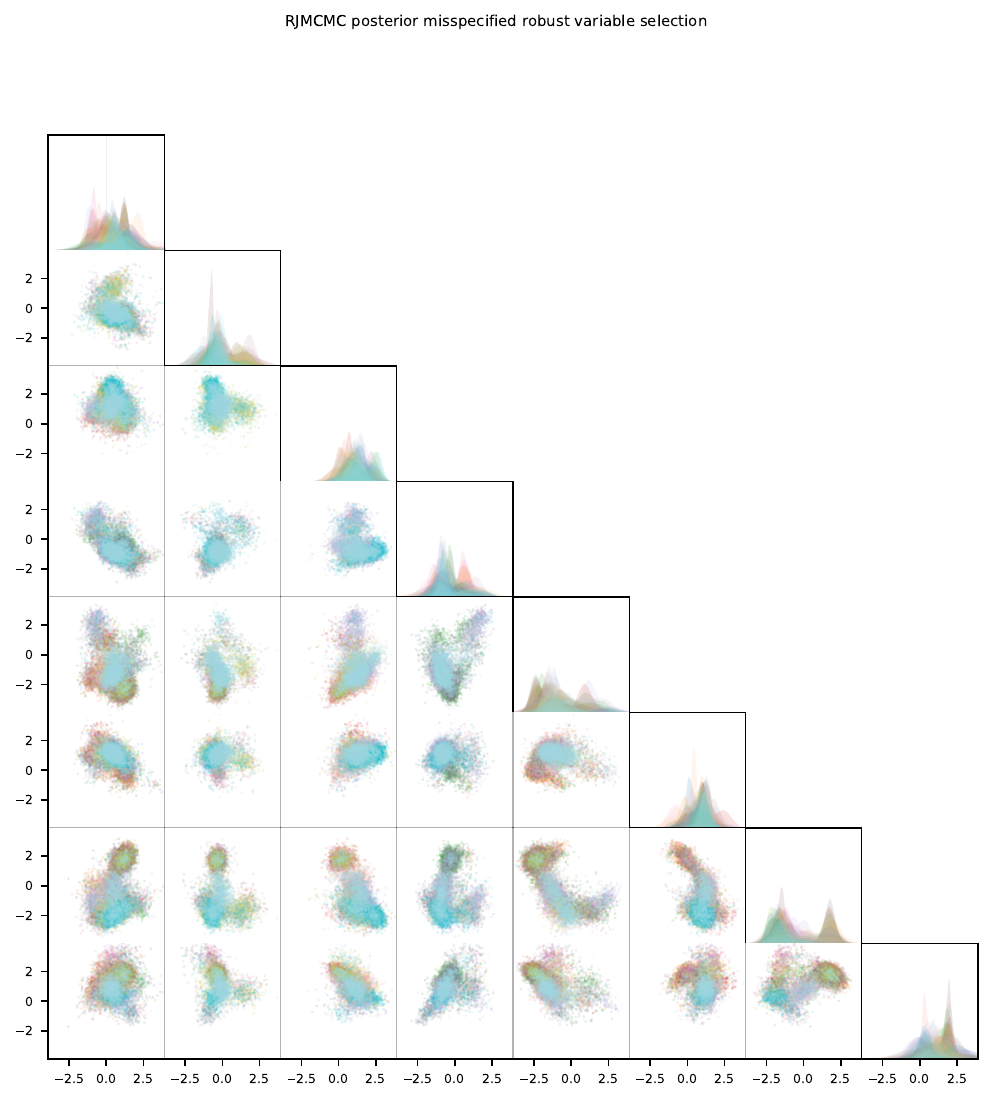}
    \includegraphics[width=0.49\linewidth, trim={0 0 0 1cm}, clip]{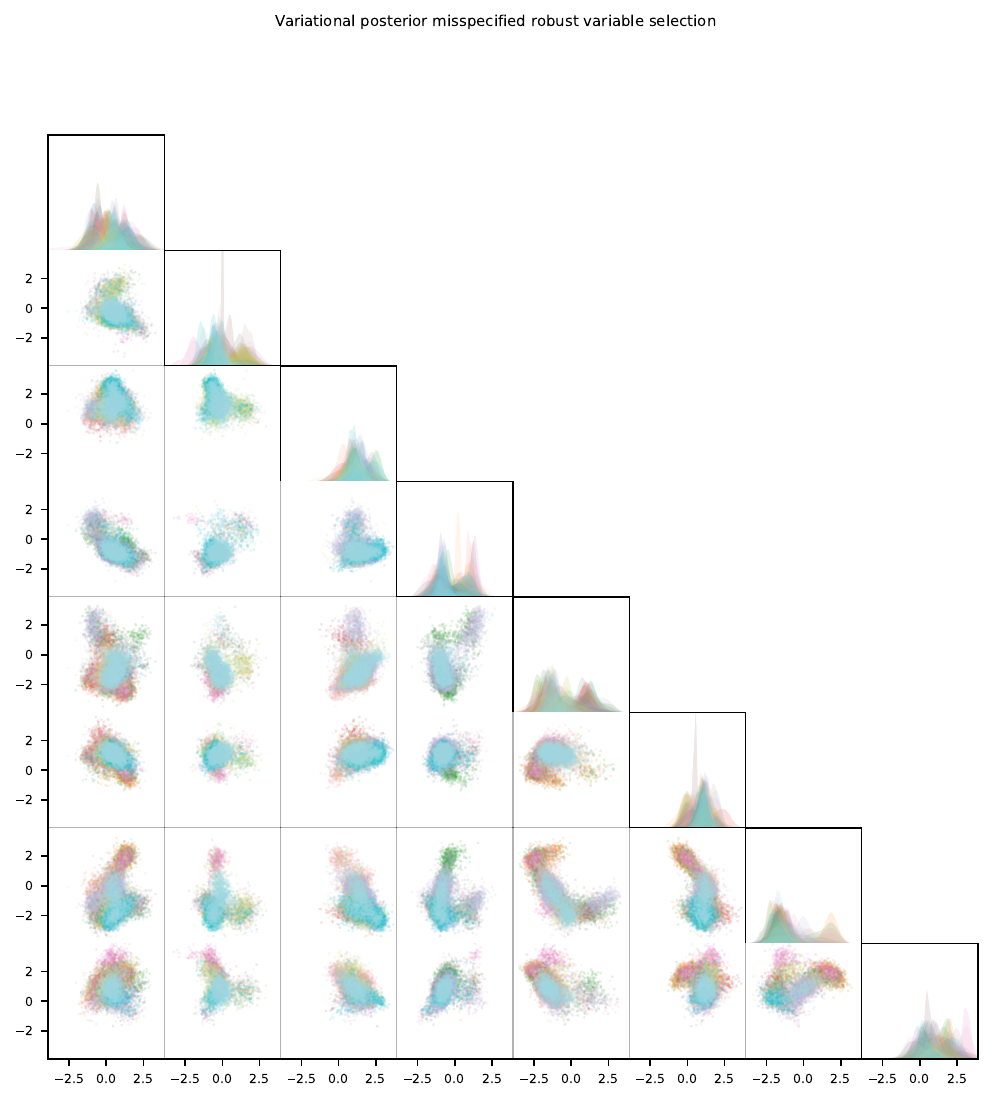}
    \caption{Multivariate plot comparison between reversible jump MCMC (left) and VTI (right) using spline flow composition of four layers and six blocks on the first synthetic narrow-prior high-misspecification data set from the Figure \ref{fig:robustvsmodelprob} (main text) example.} 
    \label{fig:robustvsmultivariate}
\end{figure}
\newpage\subsection{Baseline reversible jump MCMC for robust variable selection}\label{apdx:robustvsrjmcmc}





Consider the linear model \( y = X \beta + \varepsilon \) with \( \varepsilon \sim \mathcal{N}(0, \sigma^2 I) \). We introduce a binary mask \( \mdl \in \{0,1\}^p \) to indicate active coefficients in \( \beta \in \mathbb{R}^p \). The reversible jump MCMC algorithm explores the model space by proposing bit-flips in \( \mdl \), corresponding to adding (birth) or removing (death) predictors.

\textbf{Jacobian Determinant}: For bit-flipping moves in a saturated space where the dimensionality remains constant (\( \dim(\mdl') = \dim(\mdl) \)), the transformation is bijective with a Jacobian determinant of 1:
\[
\left| \frac{\partial(\mdl', \beta')}{\partial(\mdl, \beta)} \right| = 1.
\]
Thus, the Jacobian does not affect the acceptance probability.

\textbf{Birth Move} \((\mdl, \beta) \rightarrow (\mdl', \beta)\): A birth move flips a bit in $\mdl\rightarrow\mdl'$ from 0 to 1. Given the birth/death move ratio
\[
r_{\text{b-d}}(\mdl,\mdl',\beta)  = \frac{p(y \mid \beta, \mdl') \, p(\beta \mid \mdl') \, \pi(\mdl')}{p(y \mid \beta, \mdl) \, p(\beta \mid \mdl) \, \pi(\mdl)},
\]
the acceptance probability is:
\[
\alpha_{\text{birth}}(\mdl,\mdl',\beta) = \min\left\{ 1, r_{\text{b-d}}(\mdl,\mdl',\beta) \right\}.
\]
\textbf{Death Move} \((\mdl, \beta) \rightarrow (\mdl', \beta)\):
A death move flips a bit in $\mdl\rightarrow\mdl'$ from 1 to 0. Using the same birth/death move ratio, the acceptance probability is: 
\[\alpha_{\text{death}}(\mdl,\mdl',\beta)  = \min\left\{ 1, r_{\text{b-d}}(\mdl,\mdl',\beta)\right\}.
\]

\textbf{Within-Model Gaussian Proposal} \(\beta \rightarrow \beta'\): Within a fixed model \( \mdl \), propose a new \( \beta' \) using a symmetric random-walk:
\[
\alpha_{\text{within}}(\mdl,\beta,\beta') = \min\left\{ 1, \frac{p(y \mid \beta', \mdl) \, p(\beta' \mid \mdl)}{p(y \mid \beta, \mdl) \, p(\beta \mid \mdl)} \right\}
\]
Since the proposal is symmetric, the proposal densities cancel out in the acceptance probability.


\newpage

\section{Example description: Bayesian inference of multi-layer-perceptron
         directed acyclic graph discovery}\label{apdx:mlpdagdescription}

\paragraph{Notation:}
\begin{align*}
\MLPDnodes & && \text{number of nodes in graph}\\
\MLPDsamples & && \text{number of data samples}\\
\MLPDData        &\in \mathbb{R}^{\MLPDsamples\times\MLPDnodes}
             &&\text{(rows are i.i.d.\ samples)}\\
\MLPDPerm        &\in\mathcal P_{\MLPDnodes}
             &&\text{permutation matrix (node order)}\\
\MLPDEdges       &\in\{0,1\}^{\MLPDnodes\times\MLPDnodes}
             &&\text{strictly upper–triangular edge mask}\\
\MLPDAdj         &=\MLPDPerm^{\top}\MLPDEdges\MLPDPerm
             &&\text{adjacency in canonical order (code default)}\\
\MLPDparents{j}  &=\{\,i<j : \MLPDEdges_{ij}=1\,\}
             &&\text{parents of node $j$ in the sorted order.}
\end{align*}
\newcommand{\MLPDsubweight}{\mathbf{W}}
\paragraph{Node-wise conditional mean:}
Fix hidden width $\MLPDhid$ and a model indicator $\MLPDModel=(\MLPDPerm,\MLPDEdges)$.
For each \emph{non-root} node $j=2,\dots,\MLPDnodes$ define parameters
\[
\MLPDThetaJ{j}
   =\bigl(\MLPDsubweight^{(1)}_j,b^{(1)}_j,\MLPDsubweight^{(2)}_j,b^{(2)}_j\bigr)
   \in\mathbb{R}^{(j+2)\MLPDhid + 1},
\]
with
\(
\MLPDsubweight^{(1)}_j\!\in\!\mathbb{R}^{\MLPDhid\times(j-1)},\;
b^{(1)}_j\!\in\!\mathbb{R}^{\MLPDhid},\;
\MLPDsubweight^{(2)}_j\!\in\!\mathbb{R}^{1\times\MLPDhid},\;
b^{(2)}_j\!\in\!\mathbb{R}.
\)
Let $\MLPDuvec{j}:=\MLPDEdges_{1:(j-1),\,j}$ be the \((j-1)\)-vector of active parents.  
Writing $\MLPDData_{1:j-1}$ to denote the $1,\dots,j-1$ columns of $\MLPDData$,
\begin{align}
\MLPfunc{j}(\MLPDData_{1:j-1};\MLPDThetaJ{j},\MLPDEdges)
   = \MLPDsubweight^{(2)}_j\,
     \MLPDrelu\!\bigl(\MLPDsubweight^{(1)}_j(\MLPDData_{1:j-1}\odot\MLPDuvec{j})+b^{(1)}_j\bigr)
     +b^{(2)}_j,
\qquad
\MLPfunc{1}(\,\cdot\,)\equiv 0 .\label{eq:apdxmlpdagmapping}
\end{align}

\paragraph{Gaussian likelihood:}
Let $\MLPDPermmap$ be the permutation associated with $\MLPDPerm$
(so $\MLPDData_{\MLPDPermmap(j)}$ is column $j$ after sorting). 
With
homoscedastic noise $\MLPDNoiseVar$,
\[
\boxed{%
\log p\!\bigl(\MLPDData\!\mid\!\MLPDPerm,\MLPDEdges,\MLPDParamVec\bigr)
   =-\frac{\MLPDsamples\!\MLPDnodes}{2}\log\!\bigl(2\pi\MLPDNoiseVar\bigr)
   -\frac{1}{2\MLPDNoiseVar}
     \sum_{s=1}^{\MLPDsamples}\sum_{j=1}^{\MLPDnodes}
       \Bigl(
         \MLPDData_{\MLPDPermmap(j)}^{(s)}
        -\MLPfunc{j}\!\bigl(
            \MLPDData_{\MLPDPermmap(1:j-1)}^{(s)};
            \MLPDThetaJ{j},\MLPDEdges
          \bigr)
       \Bigr)^{2}}
\]

\paragraph{Parameter prior (masked i.i.d.\ Gaussian):}
Let $\MLPDctxmask(\MLPDModel)\subseteq\{1,\dots,\dim\MLPDParamVec\}$ be the index set
that survives the mask.  Then
\[
\boxed{%
\prior(\MLPDParamVec\!\mid\!\MLPDPerm,\MLPDEdges)
   =\!\!\prod_{k\in\MLPDctxmask(\MLPDModel)}
     \mathcal{N}\!\bigl(\MLPDParamVec_k;0,\MLPDPriorScale^{2}\bigr)}
\]
(parameters outside $\MLPDctxmask(\MLPDModel)$ are handled by a reference density).

\paragraph{Structural prior:}
\[
\boxed{%
\prior(\MLPDPerm,\MLPDEdges)\;\propto\;
  \exp\!\bigl(-\MLPDSparsityPen\,\|\MLPDEdges\|_{1}\bigr)},
  \qquad \MLPDSparsityPen\ge 0 ,
\]
with $\MLPDPerm$ a permutation matrix and $\MLPDEdges$ strictly upper triangular.

The un-normalised log-posterior is the sum of the three boxed terms above.

\subsection{Data generating process}\label{apdx:mlpdagdgp}
The data generating procedure generally follows the simulation design in \citet{thompson2025prodag}.
\paragraph{Global hyper-parameters:}
\begin{align*}
\MLPDnodes :\text{ number of nodes},\quad
\MLPDhid :\text{ hidden width},\quad
\MLPDNoiseVar :\text{ noise variance},\quad\\
\MLPDEdgeProb\in(0,1):\text{ edge probability},\quad
\MLPDPriorScale>0:\text{ parameter prior scale}.
\end{align*}

\paragraph{Sample graph structure:}
\begin{align*}
\MLPDPerm &\sim \text{Uniform}\bigl(\mathcal P_{\MLPDnodes}\bigr),\\
\MLPDEdges_{ij} &\stackrel{\text{iid}}{\sim} \operatorname{Bernoulli}(\MLPDEdgeProb),
\qquad 1\le i<j\le\MLPDnodes,\\
\MLPDAdj &= \MLPDPerm^{\top}\MLPDEdges\MLPDPerm .
\end{align*}

\paragraph{Sample node parameters:}
Let the \emph{bias flag} $\MLPDBiasFlag\in\{0,1\}$
($\MLPDBiasFlag=1$ keeps both bias vectors, $\MLPDBiasFlag=0$ sets them to $0$).
For each non-root node $j=2,\dots,\MLPDnodes$ draw independently
\[
\MLPDThetaJ{j}
   =\bigl(
      \MLPDsubweight^{(1)}_j,\;\MLPDBiasFlag\,b^{(1)}_j,\;
      \MLPDsubweight^{(2)}_j,\;\MLPDBiasFlag\,b^{(2)}_j
     \bigr),
\qquad
\bigl[\MLPDThetaJ{j}\bigr]_k\stackrel{\text{iid}}{\sim}
[-0.7,-0.3]\cup[0.3,0.7],
\]
while the root node has $\MLPDThetaJ{1}=\varnothing$. Note the active parameters are drawn uniformly from a non-zero range rather than from the prior.

\paragraph{Context–to–mask map:}
For $\MLPDModel=(\MLPDPerm,\MLPDEdges)$, $\MLPDctxmask(\MLPDModel)=\MLPDctxmask(\MLPDEdges)\subseteq\{1,\ldots,\dim\MLPDParamVec\}$
keeps exactly the coordinates satisfying the conditions:
\begin{enumerate}\itemsep2pt
\item Column $i$ of $\MLPDsubweight^{(1)}_j$ is \emph{active} iff $\MLPDEdges_{ij}=1$;
\item If $\sum_{i<j}\MLPDEdges_{ij}=0$ then \emph{all} parameters in $\MLPDThetaJ{j}$ are masked.
\end{enumerate}
(The permutation $\MLPDPerm$ has no effect on the mask.)

\paragraph{Data generation (topological order):}
Let $\MLPDPermmap$ be the permutation induced by $\MLPDPerm$.
For each sample $s=1,\dots,\MLPDsamples$ generate sequentially
\begin{align*}
\MLPDData_{\MLPDPermmap(1)}^{(s)} &= \MLPDNoiseStd\,\varepsilon_{1s},\\
\MLPDData_{\MLPDPermmap(j)}^{(s)} &=
      \MLPfunc{j}\!\bigl(
        \MLPDData_{\MLPDPermmap(1:j-1)}^{(s)};
        \MLPDThetaJ{j},\MLPDEdges
      \bigr)
      +\MLPDNoiseStd\,\varepsilon_{js},
      \quad j=2,\dots,\MLPDnodes,
\end{align*}
where $\varepsilon_{js}\stackrel{\text{iid}}{\sim}\mathcal{N}(0,1)$ and
\[
\MLPfunc{j}(\vct z;\MLPDThetaJ{j},\MLPDEdges)
   = \MLPDsubweight^{(2)}_j\,
     \MLPDrelu\!\bigl(
        \MLPDsubweight^{(1)}_j(\vct z\odot\MLPDuvec{j}) + \MLPDBiasFlag\,b^{(1)}_j
     \bigr)
     +\MLPDBiasFlag\,b^{(2)}_j,
\quad
\MLPDuvec{j}:=\MLPDEdges_{1:(j-1),\,j}.
\]

Collecting the $\MLPDsamples$ draws gives
\[
\MLPDData =
\begin{bmatrix}
\MLPDData^{(1)}\\[-2pt]\vdots\\[-2pt]\MLPDData^{(\MLPDsamples)}
\end{bmatrix}
\in\mathbb{R}^{\MLPDsamples\times\MLPDnodes},
\quad
\text{stored in topological order } \bigl(
  \MLPDData_{\MLPDPermmap(1)},\dots,\MLPDData_{\MLPDPermmap(\MLPDnodes)}
\bigr).
\]



\subsection{Comparison metrics}\label{apdx:dagmetrics}
Given knowledge of a ``true'' adjacency matrix $A$, each experiment uses four scores for comparison with the estimated posterior: F1, structured Hamming distance (SHD), Brier score, and area under the receiver operating characteristic curve (AUROC). This follows the experiment setup in \citet{thompson2025prodag}.

\subsection{Common inference setup}

For each data set in both the simulation study and real data example, VTI is run a total of 10 replicates using different random seeds, and the posterior is selected where the terminal loss is minimized. For DAGMA, the sparsity hyperparameter is swept from $\lambda^{\operatorname{min}}=10^{-3}$ to $\lambda^{\operatorname{max}}=1$ over 10 logarithmically spaced values. For the autoregressive flow, we use Affine(5,5) (see Appendix~\ref{apdx:flowarchitecture}) with a context encoder designed as follows:
\[
\delta(\MLPDPerm,\MLPDEdges) = \sigma^{\lceil\times2\rceil}\circ\dots\circ\sigma^{\lceil\times2\rceil}\circ(\MLPDPerm^\top\MLPDEdges\MLPDPerm),
\]
where $\sigma^{\lceil\times2\rceil}(x):=Wx+b$ broadcasts from $|x|$ to the first power of 2 greater than or equal to $2|x|$. The final dimension of $\delta(\MLPDPerm,\MLPDEdges)$ is 4096.

\subsection{Simulation design}\label{apdx:dagsimstudy}

In the simulation study, the configuration of the MLP is as follows. We set the hidden layer width to $\MLPDhid=10$. We set the number of nodes to $\MLPDnodes=10$. We omit the bias parameters $b_j^{(1)}, b_j^{(2)}$ for all edges, i.e. set $\MLPDBiasFlag=0$. The edge inclusion probability is set to $\MLPDEdgeProb=0.5$. For VTI, the model prior $\prior(\mdl)$ is uniform (i.e. the sparsity parameter is set to $\MLPDSparsityPen=0$). 

We generate 10 i.i.d. complete data sets of length $\MLPDsamples_{\operatorname{max}}=2^{10}$ from the above process. The experiment compares data size against the metrics from Appendix~\ref{apdx:dagmetrics}. The range of data sizes are $\MLPDsamples=16,32,4,128,256,512,1024$, where $\MLPDsamples<\MLPDsamples_{\operatorname{max}}$ simply takes the first $\MLPDsamples$ samples.

VTI inference was conducted on a cluster of GPU nodes with mixed Nvidia RTX3090 and H100 cards. On the former we used float32 precision for MLP architectures, the latter used float64.

In the DAGMA setup, a sweep of the regularization tuning parameter $\lambda$ was conducted for each dataset. The resulting adjacency matrix with the closest number of active edges to the data-generating graph was selected. This resulted in a higher-than-usual score for DAGMA results in the simulation study when compared to other methods. For DiBS/DiBS+, the inference ran for $5,000$ steps over $10$ ``particles'' (each an individual Stein variational gradient descent optimization, see \cite{Lorch2021DiBS}). JSP-GFN was configured to use a batch size of $1024$ over $50,000$ iterations.

\subsection{Real data example}\label{apdx:realdatadagexample}

For VTI, we chose to use a penalized structural model prior $\prior(\mdl)$ that induces ``extra'' sparsity via further down-weighting the probability of graphs with more edges in order to reach an acceptable level of closeness to the ``consensus'' graph in \citet{Sachs2005Causal}. It should be noted that in no other experiment do we use sparsity-inducing priors. We set$\MLPDSparsityPen=200$ and set the number of hidden nodes per edge to $\MLPDhid=5$ and include the bias terms, i.e. $\MLPDBiasFlag=1$.

For DAGMA non-linear, DiBS/DiBS+, and JSP-GFN, we use 10 hidden nodes per edge and no bias term. 

\subsection{DAG Model indicator construction: Lehmer Code Decoding}
%
%
%
A permutation of the ordered set
\(\{1,2,\ldots,\MLPDnodes\}\)%
is represented by a \emph{Lehmer code}
\(\vct c=(c_1,c_2,\ldots,c_{\MLPDnodes})\),
where
\(c_i\in\{0,1,\ldots,\MLPDnodes-i\}\).
At step \(i\,(1\le i\le\MLPDnodes)\) we choose the
\((c_i\!+\!1)\)-th \emph{unused} index in the remaining ascending list.

\smallskip
\noindent
\textit{Example.}\;
For \(\MLPDnodes=5\) and
\(\vct c=(2,1,0,0,0)\)
\[
\begin{array}{@{}l@{\quad}l@{}}
c_1=2 :& \{1,2,3,4,5\}\!\to\!3,\\
c_2=1 :& \{1,2,4,5\}\!\to\!2,\\
c_3=0 :& \{1,4,5\}\!\to\!1,\\
c_4=0 :& \{4,5\}\!\to\!4,\\
c_5=0 :& \{5\}\!\to\!5 .
\end{array}
\]

\paragraph{Permutation-matrix representation.}
The permutation \(\MLPDPermmap\) is stored as a one-hot
\(\MLPDPerm\in\{0,1\}^{\MLPDnodes\times\MLPDnodes}\) with
\(
\MLPDPerm_{r,i}=1
\)
iff row \(r\) is chosen at column \(i\).


Algorithm\,\ref{alg:vectorised-lehmer} decodes each column in parallel.
For column \(i\) the code
\(k\in[0,\,\MLPDnodes-i]\)%
specifies “pick the \((k{+}1)\)-th leftover row.”
The Boolean mask marks currently unused rows; broadcasting the
flattened one-hot vector onto the corresponding
\((\text{batch},\text{row})\) pairs writes the unit entries.
Column \(\MLPDnodes\) is filled by the single row that remains unassigned.
%
This implementation gives a compact \((\MLPDbsize,\MLPDnodes)\) tensor,
expanded by the decoder to
\((\MLPDbsize,\MLPDnodes,\MLPDnodes)\)
for efficient batched linear algebra in our DAG-inference pipeline.

\begin{algorithm}[H]
\caption{Vectorized Lehmer decode via leftover mask}
\label{alg:vectorised-lehmer}
\begin{algorithmic}[1]

\REQUIRE $P_{\text{code}} \in \mathbb{N}^{\MLPDbsize\times\MLPDnodes}$  \COMMENT{batch of Lehmer codes}
\ENSURE  $\MLPDPerm \in \{0,1\}^{\MLPDbsize\times\MLPDnodes\times\MLPDnodes}$

\STATE $\text{bs} \gets \MLPDbsize$
\STATE $\MLPDPerm \gets \mathbf 0_{\,\text{bs}\times\MLPDnodes\times\MLPDnodes}$

\FOR{$i = 1$ \textbf{to} $\MLPDnodes - 1$}
    \STATE $k \gets P_{\text{code}}[:,\,i]$
    \STATE $\text{OneHot} \gets \text{one\_hot}\!\bigl(k,\,\MLPDnodes - i + 1\bigr)$
           \COMMENT{shape $=\text{bs}\times(\MLPDnodes-i+1)$}
    \STATE $\text{Used} \gets \displaystyle\sum_{c=1}^{i-1}\MLPDPerm[:,:,c]$
    \STATE $\text{Mask} \gets (\text{Used}=0)$
    \STATE $\text{Idx}  \gets \operatorname{nonzero}(\text{Mask})$
    \STATE $\MLPDPerm\bigl[\text{Idx}_{[:,0]},\,\text{Idx}_{[:,1]},\,i\bigr]
           \gets \text{reshape}\!\bigl(\text{OneHot},\,{-}1\bigr)$
\ENDFOR

\STATE $\text{Used} \gets \displaystyle\sum_{c=1}^{\MLPDnodes-1}\MLPDPerm[:,:,c]$
\STATE $\text{Last} \gets \operatorname{nonzero}(\text{Used}=0)$
\STATE $\MLPDPerm\bigl[\text{Last}_{[:,0]},\,\text{Last}_{[:,1]},\,\MLPDnodes\bigr] \gets 1$

\RETURN $\MLPDPerm$

\end{algorithmic}
\end{algorithm}


\bigskip
\subsection{Model identifier for directed acyclic graphs}
\label{apdx:dagmdl}

\newcommand{\madecatlogit}[1]{\rho^{\operatorname{cat}}_{#1}}


We encode a permutation matrix
\(\MLPDPerm \in \{0,1\}^{\MLPDnodes \times \MLPDnodes}\) 
using a \emph{compressed Lehmer code} consisting of
\(\MLPDnodes-1\) categorical variables                           
\(\{\madecatlogit{1},\ldots,\madecatlogit{\MLPDnodes-1}\}\).
Here \(\madecatlogit{i}\) has \(\MLPDnodes-i+1\) outcomes.          

Concretely,
\(\madecatlogit{1}\in\{0,1,\ldots,\MLPDnodes-1\}\),                
\(\madecatlogit{2}\in\{0,1,\ldots,\MLPDnodes-2\}\),                
\(\ldots\),
\(\madecatlogit{\MLPDnodes-1}\in\{0,1\}\).                         
Once the first \(\MLPDnodes-1\) columns are fixed, the last column is forced.

Each \(\madecatlogit{i}=k\) is mapped to a one-hot vector of length
\(\MLPDnodes\).  The value \(k\) selects the \((k{+}1)\)-st \emph{available}
row for the \(i\)-th column; previously taken rows remain zero, preserving the
permutation property.
%

Given \(\MLPDPerm\) we form an upper-triangular mask
\(\MLPDEdges\in\{0,1\}^{\MLPDnodes\times\MLPDnodes}\) 
with zero diagonal.  Each entry above the diagonal
(\(i<j\)) is a Bernoulli variable, so \(\MLPDEdges\) flattens to
\(\tfrac{\MLPDnodes(\MLPDnodes-1)}{2}\) bits.               
The adjacency matrix is
\(\MLPDAdj=\MLPDPerm^\top\MLPDEdges\MLPDPerm\), giving a DAG.


We concatenate the \(\MLPDnodes-1\) categorical codes with the
\(\tfrac{\MLPDnodes(\MLPDnodes-1)}{2}\) Bernoulli bits,
yielding a vector \(\vct z\) of length
\((\MLPDnodes-1)+\tfrac{\MLPDnodes(\MLPDnodes-1)}{2}\).
\textsc{\madeplus} consumes \(\vct z\) together with a
\(\mathrm{multiplier\_fn}\) specifying the parameter count for each entry.

Let \(z_j\) denote the \(j\)-th component of \(\vct z\):
\[
\mathrm{multiplier\_fn}(j)=
\begin{cases}
\MLPDnodes-j, & j=1,\ldots,\MLPDnodes-1,\\ 
1, & j=\MLPDnodes,\ldots,
      \MLPDnodes-1+\tfrac{\MLPDnodes(\MLPDnodes-1)}{2}.  
\end{cases}
\]
The architecture yields the autoregressive factorization
\[
p(\vct z)=
\prod_{j=1}^{\,\MLPDnodes-1+\frac{\MLPDnodes(\MLPDnodes-1)}{2}}
p\bigl(z_j \mid z_{<j}\bigr).                                  
\]
The identifier
\(\{\madecatlogit{1},\ldots,\madecatlogit{\MLPDnodes-1},
  \MLPDEdges_{\mathsf{binary}}\}\)                            
is modelled autoregressively by a single \textsc{\madeplus}
network, yielding \(\MLPDAdj=\MLPDPerm^\top\MLPDEdges\MLPDPerm\)
upon sampling.

We employ a structural prior over the space of models with the edge-penalty term \(\gamma\):
\begin{align}
\prior(\MLPDPerm,\MLPDEdges\mid\gamma)
&=
\frac{1}{\MLPDnodes!}\;
\frac{1}{2^{\frac{\MLPDnodes(\MLPDnodes-1)}{2}}}\;
\exp\!\bigl(-\gamma\,\mathrm{nEdges}(\MLPDEdges)\bigr),\\
\mathrm{nEdges}(\MLPDEdges)
&=\sum_{i<j}\MLPDEdges_{ij},\\
\log\prior
&=-\log(\MLPDnodes!)
  -\frac{\MLPDnodes(\MLPDnodes-1)}{2}\log 2
  -\gamma\,\mathrm{nEdges}(\MLPDEdges).
\end{align}
Note that when $\gamma=0$, the prior is uniform.
\subsection{Neural probability mass function for model indicators over large spaces: \madeplus}\label{apdx:madeplus}

To represent a distribution over binary strings, we use the Masked Autoencoder for Density Estimation (MADE) \citep{GermainMADE2015} implementation found in the \citet{Durkan2020Nflows} repository. To represent a more complex discrete distribution such as that required by the $\MLPDPerm,\MLPDEdges$ representation of a directed acyclic graph, we apply a simple extension to this architecture to allow us to vary the output dimension multiplier. For 
presentational clarity we call
this extension \madeplus. The key change in \madeplus is the introduction of a per-dimension output multiplier function
$\,r(i)\,$ that determines how many parameters are emitted for the $i$-th input dimension in the 
autoregressive factorization. 

In the original MADE, all features share a common multiplier $k$, yielding an output 
dimensionality of $\,k \times d\,$ when there are $\,d\,$ input features. Mathematically, 
if $\,\mathbf{x} \in \mathbb{R}^d\,$, the network outputs 
$\,(h_1, h_2, \dots, h_{kd}) \in \mathbb{R}^{kd}\,$. 

In \madeplus, a function $r: \{0, 1, \dots, d-1\} \to \mathbb{N}$ is provided, and the final output dimension is $\sum_{i=0}^{d-1} r(i)$. For each input dimension $\,x_i\,$, the network outputs $\,r(i)\,$ parameters. Concretely, 
where $\,d\,$ is the number of input features, the final output dimension becomes $\textstyle\mathrm{total\_out\_features} = \sum_{i=0}^{d-1} r(i)$. In other words, each input $\,x_i\,$ can be associated with a custom number of distributional 
parameters (e.g., to handle discrete variables of different cardinalities). The masking logic
is preserved by replicating each degree, 
$\mathrm{deg}(x_i)$, 
exactly $\,r(i)\,$ times in 
the final layer.


Below is a simplified, side-by-side pseudocode comparing MADE (left) and \madeplus (right).
Changes in \madeplus are highlighted in \textcolor{green}{green}.

\begin{minipage}[t]{0.49\textwidth}
\begin{algorithm}[H]
    \caption{Original MADE\\ (Final Layer Construction)}
    \label{alg:orig_made}
\begin{algorithmic}
\STATE \quad
\STATE out\_features = features * output\_multiplier
\STATE final\_layer = MaskedLinear(
\STATE \quad in\_degrees = prev\_out\_degrees,
\STATE \quad out\_features = out\_features,
\STATE \quad autoregressive\_features = features,
\STATE \quad is\_output = True
\STATE \quad
\STATE )
\end{algorithmic}
\end{algorithm}
\end{minipage}
\hfill
\begin{minipage}[t]{0.49\textwidth}
\begin{algorithm}[H]
    \caption{\madeplus\\ (Final Layer Construction)}
    \label{alg:madeplus}
\begin{algorithmic}
\STATE total\_out\_features = \(\sum_{i=0}^{features-1}\) \textcolor{green}{$r(i)$}
\STATE final\_layer = MaskedLinear(
\STATE \quad in\_degrees = prev\_out\_degrees,
\STATE \quad out\_features = \textcolor{green}{total\_out\_features},
\STATE \quad autoregressive\_features = features,
\STATE \quad is\_output = True,
\STATE \quad \textcolor{green}{output\_multiplier\_fn = $r(i)$}
\STATE )
\end{algorithmic}
\end{algorithm}
\end{minipage}


By allowing each input dimension $\,X_i\,$ to have its own output multiplier $\,r(i)\,$, 
the \madeplus architecture provides a more flexible autoregressive decomposition:
\[
p(\mathbf{x})
\;=\;
\prod_{i=1}^{d} p\bigl(x_i \,\bigm\vert\, x_{1}, \dots, x_{i-1}\bigr),
\]
where now the conditional distribution for $x_i$ can be parameterized by $r(i)$ parameters 
(e.g., logits for a categorical variable of size $r(i)$, or a mean/variance pair, etc.). 

Hence, one can naturally combine discrete variables of varying dimensions such as Bernoulli and categorical 
variables. For example, if $x_1$ is categorical with 10 categories and 
$x_2$ is a Bernoulli variable, one can specify $r(0) = 10$ and $r(1) = 1$, so that the overall 
conditional densities (or probability mass functions) multiply to form a richer joint model
adapting precisely to each variable's nature.


\ifthenelse{\boolean{isarxiv}}{}{%
\newpage
\section*{NeurIPS Paper Checklist}

\begin{enumerate}

\item {\bf Claims}
    \item[] Question: Do the main claims made in the abstract and introduction accurately reflect the paper's contributions and scope?
    \item[] Answer: \answerYes{} 
    \item[] Justification: Claims made in the abstract and introduction/contributions are met in Sections \ref{sec:method}, \ref{sec:modelweights}, and demonstrated in numerical experiments in Section \ref{sec:examples}.

\item {\bf Limitations}
    \item[] Question: Does the paper discuss the limitations of the work performed by the authors?
    \item[] Answer: \answerYes{} 
    \item[] Justification: While there is no explicitly-titled Limitations section, (i) the effects of using normalising flows with poorer/stronger degrees of expressivity on the quality of the VTI approximation are explicitly discussed in Section \ref{sec:examples} (Experiments) and Appendix \ref{apdx:rvsadditionalresults} (Additional Results); the nature of the VTI approximation over transdimensional model space tending to focus more on high posterior model probability models is discussed in Section \ref{sec:discussion} (Discussion); and the limitations of particular methods of approximating $q_\psi(m)$, depending on the cardinality of the model space $|\cal{M}|$ are discussed in Sections \ref{sec:isomorphismcategoricalmade}, \ref{sec:mcgmodelweightdensity} and \ref{sec:discussion} (Discussion).

\item {\bf Theory assumptions and proofs}
    \item[] Question: For each theoretical result, does the paper provide the full set of assumptions and a complete (and correct) proof?
    \item[] Answer: \answerYes{} 
    \item[] Justification: The methods presented in \cref{sec:method} and \cref{sec:modelweights} are formally analyzed in \cref{apdx:cosmic-details} and \cref{apdx:theoreticalanalysis} respectively. The authors are unaware of any unaddressed assumptions.

    \item {\bf Experimental result reproducibility}
    \item[] Question: Does the paper fully disclose all the information needed to reproduce the main experimental results of the paper to the extent that it affects the main claims and/or conclusions of the paper (regardless of whether the code and data are provided or not)?
    \item[] Answer: \answerYes{} 
    \item[] Justification: Every effort was made to disclose all of the information needed to reproduce the experimental results. \cref{apdx:rvsadditionalresults} and \cref{apdx:mlpdagdescription} contain details of the setup of experiments that are not included in \cref{sec:examples}.

\item {\bf Open access to data and code}
    \item[] Question: Does the paper provide open access to the data and code, with sufficient instructions to faithfully reproduce the main experimental results, as described in supplemental material?
    \item[] Answer: \answerYes{} 
    \item[] Justification: Code will be provided in the supplementary materials.

\item {\bf Experimental setting/details}
    \item[] Question: Does the paper specify all the training and test details (e.g., data splits, hyperparameters, how they were chosen, type of optimizer, etc.) necessary to understand the results?
    \item[] Answer: \answerYes{} 
    \item[] Justification: Almost all details are provided in the main paper or supplementary material appendices. Any smaller details not provided in these sources can be determined from the supplementary code itself. These include the hyperparameters of the optimizer, learning rate schedulers, and so on. 

\item {\bf Experiment statistical significance}
    \item[] Question: Does the paper report error bars suitably and correctly defined or other appropriate information about the statistical significance of the experiments?
    \item[] Answer: \answerYes{} 
    \item[] Justification: The experiment design runs multiple data sets and optimizer seeds. Figures  \ref{fig:mlpdagsimstudy} and \ref{fig:apdxrobustvssweepcard} show error bars. No statistical tests are performed.

\item {\bf Experiments compute resources}
    \item[] Question: For each experiment, does the paper provide sufficient information on the computer resources (type of compute workers, memory, time of execution) needed to reproduce the experiments?
    \item[] Answer: \answerYes{} 
    \item[] Justification: Compute time and resources were not considered to be the main focus of the paper's contribution (which is the development of the first approach to extend variational inference to the transdimensional setting), however \cref{apdx:rvsadditionalresults} and \cref{apdx:dagmdl} mention the GPU architectures used for each experiment.
    
\item {\bf Code of ethics}
    \item[] Question: Does the research conducted in the paper conform, in every respect, with the NeurIPS Code of Ethics \url{https://neurips.cc/public/EthicsGuidelines}?
    \item[] Answer: \answerYes{} 
    \item[] Justification: To the best knowledge of the authors, the paper conforms, in every respect, with the NeurIPS Code of Ethics.

\item {\bf Broader impacts}
    \item[] Question: Does the paper discuss both potential positive societal impacts and negative societal impacts of the work performed?
    \item[] Answer: \answerNA{} 
    \item[] Justification: To the best knowledge of the authors, the paper describes a novel methodology that has no direct societal impact.

\item {\bf Safeguards}
    \item[] Question: Does the paper describe safeguards that have been put in place for responsible release of data or models that have a high risk for misuse (e.g., pretrained language models, image generators, or scraped datasets)?
    \item[] Answer: \answerNA{} 
    \item[] Justification: The experiments use only freely-available data sources.

\item {\bf Licenses for existing assets}
    \item[] Question: Are the creators or original owners of assets (e.g., code, data, models), used in the paper, properly credited and are the license and terms of use explicitly mentioned and properly respected?
    \item[] Answer: \answerYes{} 
    \item[] Justification: All data sources are referenced and the licence and terms are properly respected. Only open source libraries were used in the code.

\item {\bf New assets}
    \item[] Question: Are new assets introduced in the paper well documented and is the documentation provided alongside the assets?
    \item[] Answer: \answerNA{} 
    \item[] Justification: No new assets are released apart from the eventual release of the code used for experiments with the appropriate CC license.

\item {\bf Crowdsourcing and research with human subjects}
    \item[] Question: For crowdsourcing experiments and research with human subjects, does the paper include the full text of instructions given to participants and screenshots, if applicable, as well as details about compensation (if any)? 
    \item[] Answer: \answerNA{} 
    \item[] Justification: No crowdsourcing experiments were conducted.

\item {\bf Institutional review board (IRB) approvals or equivalent for research with human subjects}
    \item[] Question: Does the paper describe potential risks incurred by study participants, whether such risks were disclosed to the subjects, and whether Institutional Review Board (IRB) approvals (or an equivalent approval/review based on the requirements of your country or institution) were obtained?
    \item[] Answer: \answerNA{} 
    \item[] Justification:  No crowdsourcing experiments were conducted.

\item {\bf Declaration of LLM usage}
    \item[] Question: Does the paper describe the usage of LLMs if it is an important, original, or non-standard component of the core methods in this research? Note that if the LLM is used only for writing, editing, or formatting purposes and does not impact the core methodology, scientific rigorousness, or originality of the research, declaration is not required.
    \item[] Answer: \answerNA{} 
    \item[] Justification: LLM technology was not used for any purpose other than grammar and minor textual edits.

\end{enumerate}

}

\end{document}